\documentclass[sn-mathphys,Numbered]{sn-jnl}

\usepackage{graphicx}%
\usepackage{multirow}%
\usepackage{amsmath,amssymb,amsfonts}%
\usepackage{amsthm}%
\usepackage{mathrsfs}%
\usepackage[title]{appendix}%
\usepackage{xcolor}%
\usepackage{textcomp}%
\usepackage{manyfoot}%
\usepackage{booktabs}%
\usepackage{algorithm}%
\usepackage{algorithmicx}%
\usepackage{algpseudocode}%
\usepackage{listings}%
\usepackage{caption}
\usepackage{subcaption}
\usepackage{hyperref}
\usepackage{xcolor}
\newcommand{\ch}[1]{{\color{black}{#1}}}

\begin{document}

\title[Article Title]{DMF-TONN: Direct Mesh-free Topology Optimization using Neural Networks}




\author[1]{\fnm{Aditya} \sur{Joglekar}}
\equalcont{These authors contributed equally to this work.}

\author[1]{\fnm{Hongrui} \sur{Chen}}
\equalcont{These authors contributed equally to this work.}

\author*[1]{\fnm{Levent Burak} \sur{Kara}}\email{lkara@cmu.edu}

\affil[1]{\orgdiv{Department of Mechanical Engineering}, \orgname{Carnegie Mellon University}, \orgaddress{\street{5000 Forbes Ave}, \city{Pittsburgh}, \postcode{15213}, \state{PA}, \country{USA}}}



\abstract{We propose a direct mesh-free method for performing topology optimization by integrating a density field approximation neural network with a displacement field approximation neural network. We show that this direct integration approach can give comparable results to conventional topology optimization techniques, with an added advantage of enabling seamless integration with post-processing software, and a potential of topology optimization with objectives where meshing and Finite Element Analysis (FEA) may be expensive or not suitable. Our approach (DMF-TONN) takes in as inputs the boundary conditions and domain coordinates and finds the optimum density field for minimizing the loss function of compliance and volume fraction constraint violation. The mesh-free nature is enabled by a physics-informed displacement field approximation neural network to solve the linear elasticity partial differential equation and replace the FEA conventionally used for calculating the compliance. We show that using a suitable Fourier Features neural network architecture and hyperparameters, the density field approximation neural network can learn the weights to represent the optimal density field for the given domain and boundary conditions, by directly backpropagating the loss gradient through the displacement field approximation neural network, and unlike prior work there is no requirement of a sensitivity filter, optimality criterion method, or a separate training of density network in each topology optimization iteration.}


\maketitle

\section{Introduction}\label{sec1}
Topology optimization approaches like SIMP (Solid Isotropic Material with Penalisation) (\cite{bendsoe1989optimal,zhou1991coc}) find the optimum structure for a given set of boundary conditions by meshing the design domain and using an iterative process where each iteration involves an  FEA calculation for computing the objectives such as compliance. Therefore, removing
these iterations completely or creating a new class of solvers with a reparameterization of the design variables in this optimization problem is highly desirable. Advances in neural networks, both in learning from large amounts of data and in learning implicit
representations of complex signals show great promise to bring about this transformation, and hence many new approaches trying to utilize neural networks for topology optimization have been recently developed.

Data-driven approaches perform instant optimal topology generation during inference time. However,  they require a large training database generation, a long training time and face generalization issues. Online training approaches use a neural network to represent the density field of designs for an alternative parameterization. They do not face any generalization issues. However, meshing and FEA is still required.

One of the first online training topology optimization approaches, TOuNN, was proposed by \citet{Chandrasekhar2021}. The neural network takes in as inputs the domain coordinates and outputs the density value at each of these coordinates. The loss function consists of the compliance and volume fraction constraint violation. This loss gradient is backpropagated and used for updating the weights of the neural network such that it learns the optimal density distribution for minimizing the loss. The compliance for the density field is calculated as in the traditional SIMP method using FEA. For removing the meshing requirement of FEA and creating a new class of solvers for various partial differential equations (PDEs) in computational mechanics, there have been recent advances and promising results in using physics-informed neural networks (PINNs). \citet{samaniego2020energy} propose an energy approach for solving the linear elasticity PDEs. The displacement field is parameterized by a neural network which takes as input the domain coordinates and outputs the displacements in each direction at each of these coordinates. The loss function consists of the potential energy, which when minimized, will give static equilibrium displacements.

Though the computational time for the neural network PDE approximation frameworks is worse than the current state of the art FEA solvers, there are several potential advantages of this approach, including the mesh-free nature and an easy modelling of non-linear PDEs. Incorporation of these neural network PDE approximation frameworks in online training topology optimization enables mesh-free topology optimization and a new class of solvers for this complex inverse design problem.

\citet{zehnder2021ntopo} were the first to propose such a mesh-free framework for topology optimization with compliance as an objective, where in addition to the density field, the displacement field is also parameterized using a neural network. However, they conclude that connecting the two neural networks directly leads to bad local minima. Hence, they propose  using the optimality criterion method and sensitivity filtering  for calculating target densities. As such,  the density neural network needs to be trained for estimating these target densities in every topology optimization iteration.

In this work, we show that using directly connected displacement field estimation and density field estimation neural networks is indeed an effective approach for mesh-free topology optimization. In particular, we argue that using just one gradient descent step of the density network in each topology optimization iteration without any sensitivity  or density filtering leads to comparable results to conventional topology optimization. Moreover, after the initial run of the displacement network, we significantly reduce the number of iterations in each topology optimization iteration. We show that transfer learning applies here and in this high dimensional and non-convex optimization problem setting, approximate loss and gradients can work well.

We devise DMF-TONN as a method not for replacing SIMP, but for adding to and improving the current class of mesh-free solvers for topology optimization, using the advancements in neural networks. We use Fourier Features and a fully connected layer as the architecture of both our neural networks. We verify the effectiveness of our approach with case studies with different boundary conditions and volume fractions. The implementation of this work is available at: 
\url{https://github.com/AdityaJoglekar/DMF-TONN}.\\

Our main contributions include:
\begin{itemize}
  \item  A framework for achieving mesh-free topology optimization by directly combining displacement field approximation and density field approximation neural networks, with just one gradient descent step of density approximation network in each topology optimization iteration.
\item Extensive analysis and comparison with SIMP for 3D examples, unlike prior related works which focus on 2D examples
  \item A method with an improved computational time compared to prior related works.
\end{itemize}

\begin{figure*}
\centering

\includegraphics[width=\textwidth]{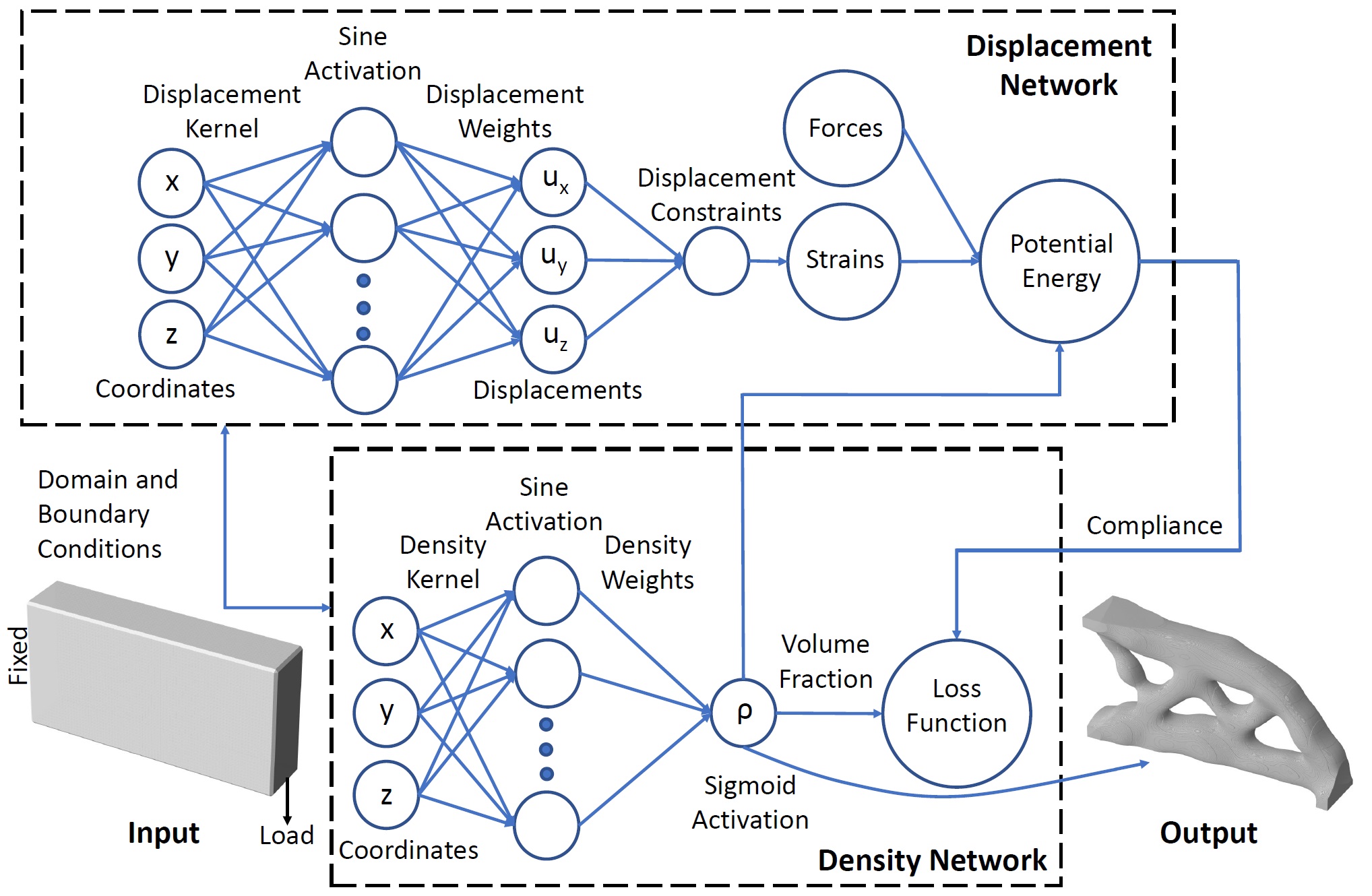}

\caption{Our proposed framework. Each topology optimization iteration consists of: 1) Training the displacement network with current density field, randomly sampled domain coordinates \ch{in each of its iterations} and boundary conditions to obtain static equilibrium displacements. 2) Sampling domain coordinates and performing a forward pass through the density network to obtain current topology output, and displacement network to obtain current compliance, which are passed to the density network loss function 3) Backpropagating density network loss and performing a gradient descent step on density network weights.
}

\label{fig:method}
\end{figure*}

\section{Literature Review}\label{sec2}
\textit{Topology optimization}:
\citet{bendsoe1988generating} introduced the homogenization approach for topology optimization. The SIMP method (\cite{bendsoe1989optimal,zhou1991coc})  considers the relative material density in each element of the Finite Element (FE) mesh as design variables, allowing for a simpler interpretation and optimised designs with more clearly defined features. Other common approaches to topology optimization include the level-set method (\cite{allaire2002level,wang2003level}) and evolutionary algorithms. Improving the optimization results and speed of these approaches using neural networks has seen a lot of development recently and \citet{woldseth2022use} provide an extensive overview on this topic.

\textit{Neural Networks for solving PDEs}: Driven by a neural network's ability to approximate functions, there have been several recent works proposing novel solvers for PDEs. \citet{raissi2019physics} propose PINNs, neural networks that are trained to solve supervised learning tasks while respecting the laws of physics described by general nonlinear partial differential equations.
\citet{samaniego2020energy} propose an approach using neural networks which does not require labelled data points, and just uses domain coordinates and boundary conditions as input to solve computational mechanics PDEs. \citet{nguyen2020deep} develop a deep energy method for finite deformation hyperelasticity. \citet{sitzmann2020implicit} leverage periodic activation functions for implicit neural representations and demonstrate that these networks are ideally suited for representing complex natural signals and their derivatives and solving PDEs. \citet{tancik2020fourier} show that passing input points through a simple Fourier feature mapping enables a multilayer perceptron to learn high-frequency functions in low-dimensional problem domains. We utilize this concept of Fourier Feature mapping for finding good approximations of the displacement field and density field in the low-dimensional coordinate domain. 

\textit{Neural networks for topology optimization}:
Several data-driven methods for topology optimization using neural networks \cite{banga20183d,yu2019deep,nakamura2020deep, nie2021topologygan,behzadi2021real, maze2022diffusion, white2019multiscale} have been proposed. In this review we focus on the online training topology optimization methods, i.e. those methods which do not use any prior data, rather train a neural network in a self-supervised manner for learning the optimal density distribution and topology. \citet{Chandrasekhar2021} explore an online approach where the density field is parameterized using a neural network. Fourier projection based neural network for length scale control (\cite{Chandrasekhar2021Fourier}) and application for multi-material topology optimization (\cite{Chandrasekhar2021MM}) has also been explored. \citet{deng2020topology} propose topology optimization with Deep Representation Learning, with a similar concept of re-parametrization, and demonstrate the effectiveness of their method on  compliance minimization and stress-constrained problems. \citet{zhang2023topology} also employ a design reparameterization approach and improve accuracy of physical response and sensitivity analysis by proposing the BA-MFEA as a substitution for conventional FEA. \citet{hoyer2019neural} use CNNs for density parameterization and directly enforce the constraints in each iteration, reducing the loss function to compliance only. \citet{chen2023concurrent} propose a neural network based approach to topology optimization that aims to reduce the use of support structures in additive manufacturing. \citet{raychen2023idetc} demonstrate that by using a prior initial field on
the unoptimized domain, the efficiency of neural network based
topology optimization can be improved. \citet{he2022deep} and \citet{jeong2023physics} approximate displacement fields using PINNs, but a continuous density field is not learned and the frameworks are not mesh-free. \citet{lu2021physics} demonstrate the effectiveness of hard constraints over soft constraints for solving PDEs in various topology optimization problems. \citet{zehnder2021ntopo} effectively leverage neural representations in the context of mesh-free topology optimization and use multilayer perceptrons to parameterize both the density and displacement fields.  \citet{mai2023physics} develop a similar approach for optimum design of truss structures. We show that unlike in \citet{zehnder2021ntopo}, sensitivity filtering, optimality criterion method and separate training of density network in each topology optimization epoch is not necessary for mesh-free topology optimization using neural networks.

\section{Proposed Method}\label{sec3}
We parameterize the displacement field as well as the density field using neural networks and integrate them as shown in Figure \ref{fig:method}.
\begin{algorithm}
\caption{DMF-TONN}\label{alg:cap}
\begin{algorithmic}[1]
\State Initialize neural networks: $Den_{W_{den}}$, $Disp_{W_{disp}}$
\State Initialize Adam optimizers: $Opt_{den}$, $Opt_{disp}$
\State Initialize domain $\rho_{init}$
\For {$n_{init disp}$ iterations}
\State Sample domain coordinates $X_{disp}$
\State $u_{temp} \leftarrow Disp_{W_{disp}}(X_{disp})$
\State $W_{disp} \leftarrow 
Opt_{disp}.step(W_{disp},\frac{\partial L_{disp}(u_{temp},\rho_{init})}{\partial W_{disp}})$
\EndFor
\For {$n_{opt}$ iterations}

\For {$n_{disp}$ iterations}
\State Sample domain coordinates $X_{disp}$
\State $\rho_{temp} \leftarrow Den_{W_{den}}(X_{disp})$
\State $u_{temp} \leftarrow Disp_{W_{disp}}(X_{disp})$
\State $W_{disp}\leftarrow Opt_{disp}.step(W_{disp},\frac{\partial L_{disp}(u_{temp},\rho_{temp})}{\partial W_{disp}})$
\EndFor
\State Sample domain coordinates $X_{den}$
\State $\rho \leftarrow Den_{W_{den}}(X_{den})$
\State $u \leftarrow Disp_{W_{disp}}(X_{den})$
\State $c \leftarrow L_{disp}(u,\rho) + EW$
\State $W_{den} \leftarrow Opt_{den}.step(W_{den},\frac{\partial L_{den}(\rho,c)}{\partial W_{den}})$
\EndFor
\end{algorithmic}
\end{algorithm}



\subsection{Density Neural Network}
The density neural network $\textit{Den}(\mathbf{X}_{den})$ can be represented as follows:
\begin{equation}
    \textit{Den}(\mathbf{X}_{den}) = \sigma(\sin(\mathbf{X}_{den}\mathbf{K}_{den} +  \mathbf{b})\mathbf{W}_{den})
\end{equation}
The input is a batch of domain coordinates $\mathbf{X}_{den(\text{batchsize} \times 3)}$. We use the domain center as the origin for the coordinates, and the coordinates are normalized with the longest dimension coordinates ranging from -0.5 to 0.5. We use the concept proposed in \citet{tancik2020fourier} and a neural network architecture similar to the one used in \citet{Chandrasekhar2021Fourier}. The first layer weights (kernel $\mathbf{K}_{den(3 \times \text{kernelsize})}$) are fixed, which create Fourier features after passing through the sine activation. We add a bias term $\mathbf{b}$ consisting of ones before applying the sine activation to break off the symmetry about the origin. The kernel is created using a grid of number of dimensions same as the number of domain dimensions, and then reshaping the grid coordinates to the matrix $\mathbf{K}_{den(3 \times \text{kernelsize})}$. The grid size in each dimension dictates how good it can represent topological features, and the grid's range of values control the frequency of the output topology, with higher ranges of values giving a topology with more intricate features. Note that this grid is not a mesh structure, and consists solely of coordinates. We find that making the kernel trainable can slightly improve  compliance. However, we keep it fixed for all the experiments in this paper assuming that the slight increase in performance may not be preferable to the large number of trainable weights.  The next layer weights ($\mathbf{W}_{den(\text{kernelsize} \times 1)}$) are trainable and the output is passed through a sigmoid activation ($\sigma$). This ensures output values are between 0 and 1, which represent the density, for each of the coordinates in the input batch. We use Adam (\citet{kingma2014adam}) as the optimizer, with a learning rate of $2.0\times10^{-3}$ for all the experiments.

\subsection{Displacement Neural Network}
We use a neural network similar to the density neural network for approximating the displacement field. The physics-informed components shown in \citet{samaniego2020energy} are then added on the displacement output by this neural network $\textit{Disp}(\mathbf{X}_{disp})$. This can be represented as follows:
\begin{equation}
    \textit{Disp}(\mathbf{X}_{disp}) = \sin(\mathbf{X}_{disp}\mathbf{K}_{disp} +  \mathbf{b})\mathbf{W}_{disp}
\end{equation}
\ch{In each displacement network iteration, we randomly sample domain coordinates $\mathbf{X}_{disp}$}. The frequency determined by $\mathbf{K}_{disp}$ should be greater than or equal to the frequency determined by $\mathbf{K}_{den}$. This is due to the fact that if the displacement network in unable to capture and pass fine  changes in displacement  to the density network, and the density network is attempting to create very fine features, incorrect features are created and disconnections are observed in the final topology. For all our experiments, we use the same frequencies and grid sizes for $\mathbf{K}_{disp}$ and $\mathbf{K}_{den}$ and find this setting works well. Multiplying the Fourier features with $\mathbf{W}_{disp(\text{kernelsize} \times 3)}$ gives the displacements in each direction.

\subsubsection{Displacement Constraints}
Boundary conditions on displacements such as fixed sides, are implemented as hard constraints. The output of $\textit{Disp}(\mathbf{X}_{disp})$ is multiplied with a differentiable function that is 0 at the fixed boundary and 1 elsewhere. We use the exponential function for this. For example, with a cuboidal domain that has the side with all $x$ coordinates equal to zero, fixed (zero displacement in all three directions), such as in Figure \ref{fig:PINN_convergence a}, the hard constraint function takes the form $2(\frac{1}{(1+\exp(-m(c_x +0.5)))} - 0.5)$ where $c_x$ is all the $x$ coordinates in the domain, and $m$ is a constant which dictates the slope of this function. We find empirically $m = 20$ works well and use it for all our experiments. For multiple fixed sides, the displacements output by the neural network are multiplied by the functions for each fixed side.

\subsubsection{Minimum Potential Energy Loss Function}
The principle of minimum potential energy is used for approximating the displacement field as proposed in \citet{samaniego2020energy}. The neural network learns the weights that output the displacements that minimize the potential energy, and thus learns to output static equilibrium displacements. With Monte-Carlo sampling, the loss function of the displacement neural network, $L_{disp} = $ Potential Energy, for 3d problems, is defined as follows:
\begin{equation}
    L_{disp} = ISE - EW
\end{equation}
\begin{equation}
    ISE = \displaystyle\frac{V}{N} \sum^{N}_{i}(\mu\epsilon_i : \epsilon_i + \frac{\lambda (trace(\epsilon_i))^{2}}{2})
\end{equation}
\begin{equation}
    EW = \displaystyle\frac{A}{N_b}\sum^{N_b}_{i}Tu_i
\end{equation}
\\
\noindent where,\\
$ISE = $ Internal Strain Energy\\
$EW = $ External Work\\
$V = $ domain volume\\
$N = $ number of sample points in domain\\
$\mu = \frac{E}{2(1+\nu)}$, $\lambda = \frac{E\nu}{((1+\nu)(1 - 2\nu))}$\\ $E = $ Young's Modulus\\ $\nu = $ Poisson's ratio\\
$\epsilon_i = $ strain matrix at $i^{th}$ point\\
$A = $ area on which traction is applied\\
$N_b = $ number of sample points on boundary\\
$T = $ traction\\
$u_i = $ displacement at $i^{th}$ point\\
The symbol `$:$' indicates element wise multiplication, followed by a summation.

The strains are calculated using automatic differentiation in TensorFlow (\citet{Abadi2016}). We use Adam as the optimizer, with a learning rate of $5.0\times10^{-6}$ for all the experiments.

\subsection{Integration of Density and Displacement Neural Networks}
A topology optimization epoch starts by training the displacement network with randomly sampled coordinates, the corresponding current topology (found by a forward pass through the density network) and the boundary conditions. The conventional SIMP method interpolation ($E = E_{material}(\rho^{3}) $, where $\rho$ is the density) is used for obtaining the Young's modulus $E$ at each of the randomly sampled domain points in each displacement network iteration. Then, with randomly sampled coordinates, a forward pass is performed through the density network and the displacement network to get the current topology and current compliance (Internal Strain Energy) respectively, which are passed to the density network loss function. \ch{For the volume fraction constraint violation term in this loss function, we use a constant grid of domain points instead of randomly sampled points in each iteration to avoid fluctuations in the volume fraction in each iteration and thus helping stabilize the optimization process.} The density network loss function is defined as follows:
\begin{equation}
    L_{den} = \frac{c}{c_0}+\alpha(\frac{v}{V^*}-1)^2
\end{equation}
\noindent where,\\
$c = $ compliance\\
$c_0 = $ initial compliance\\
$v = $ volume fraction\\
$V^* = $ target volume fraction\\
$\alpha = $ penalty constant\\

The compliance ($c$) is a function of the densities ($\rho$) and displacements ($u$), where the displacements are also dependent on the densities. As shown in \citet{zehnder2021ntopo}, the total gradient of the compliance with respect to the densities is given by 
$
    \frac{d C}{d\rho} = \frac{\partial C}{\partial \rho} + \frac{\partial C}{\partial u}\frac{du}{d\rho} = -\frac{\partial C}{\partial \rho}
$, which needs to be incorporated while connecting the two neural networks and backpropagating the loss to the density network weights.

As shown in Algorithm \ref{alg:cap}, during each topology optimization iteration, the density network weights are updated once with a gradient descent step in Adam. Before starting the topology optimization, the displacement network is run with the initial domain as the topology, for $n_{dispinit} = 1000$ iterations for converging to static equilibrium displacements. Then in each topology optimization iteration, utilizing the concept of transfer learning as the topology does not change too drastically, we run the displacement network only for $n_{disp} = 20$ iterations. We determined the values of the $n_{dispinit}$ and $n_{disp}$ variables empirically to give the best results in the cases presented in the paper.  Though we have to increase the topology optimization iterations, this reduction of 50 times in displacement network iterations significantly reduces the computational time without compromising the compliance of the results.

We run all of our experiments on a machine with 12th Gen Intel(R) Core(TM) i7-12700 2.10 GHz processor, 16 GB of RAM and Nvidia GeForce RTX 3060 GPU.

\begin{table}[]
\small
    \centering
    \begin{tabular}{c|c|c|c}
        \toprule
        \includegraphics[width=0.17\textwidth]{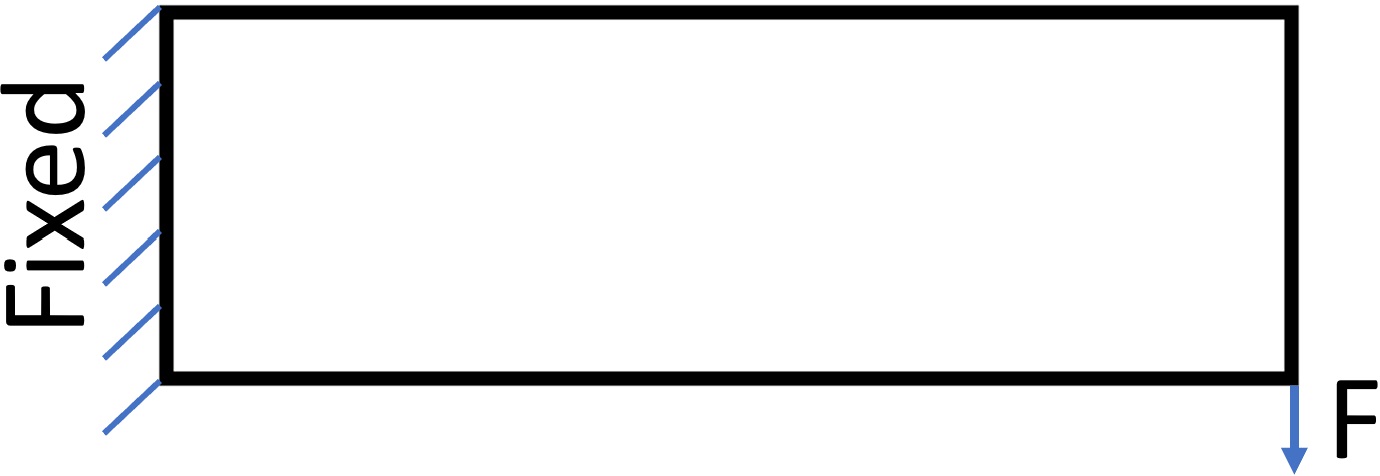}  & \includegraphics[width=0.17\textwidth]{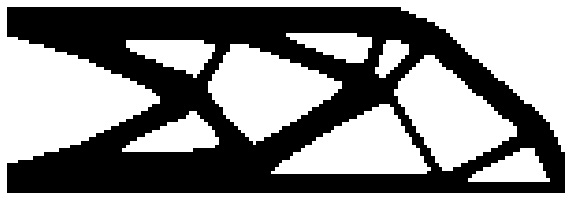}
        & \includegraphics[width=0.17\textwidth]{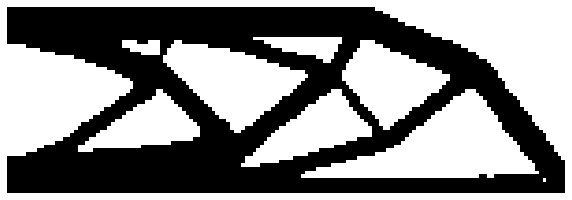}

        & \includegraphics[width=0.17\textwidth]{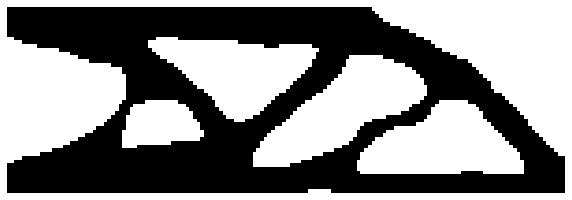}
        \\
         Boundary Conditions
         & SIMP
         & NTopo
         & DMF-TONN\\
         \midrule
         Convergence Compliance
         &$2.86\times10^{-4}$
         &$2.54\times10^{-4}$
         &$2.73\times10^{-4}$\\
         Volume Fraction Achieved
         &0.50
         &0.50
         &0.50\\
         Binary Structure Compliance
         &$2.66\times10^{-4}$
         &$2.40\times10^{-4}$
         &$2.52\times10^{-4}$\\
         Time
         &69 s
         &1465 s
         &277 s\\
        \bottomrule
    \end{tabular}
    \caption{Comparison for 2D cantilever beam problem with target volume fraction = 0.5}
    \label{tab:2dcomp05}
\end{table}

\begin{table}[]
\small
    \centering
    \begin{tabular}{c|c|c|c}
        \toprule
        \includegraphics[width=0.17\textwidth]{Images/Boundary_Conditions_2D.jpg}  & \includegraphics[width=0.17\textwidth]{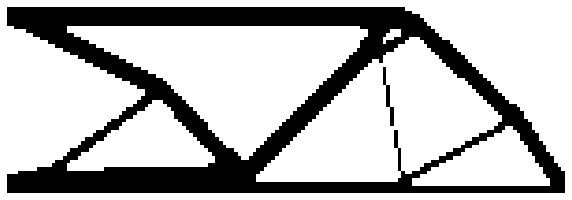}
        & \includegraphics[width=0.17\textwidth]{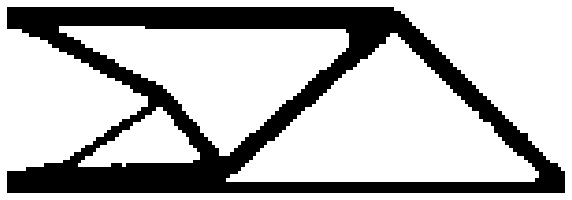}

        & \includegraphics[width=0.17\textwidth]{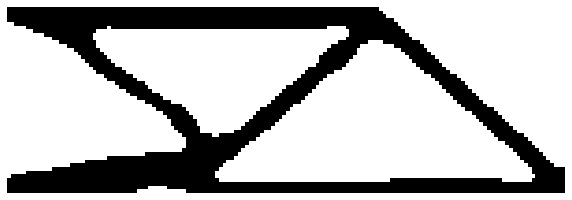}
        \\
         Boundary Conditions
         & SIMP
         & NTopo
         & DMF-TONN\\
         \midrule
         Convergence Compliance
         &$4.94\times10^{-4}$
         &$4.23\times10^{-4}$
         &$4.76\times10^{-4}$\\
         Volume Fraction Achieved
         &0.30
         &0.30
         &0.30\\
         Binary Structure Compliance
         &$4.55\times10^{-4}$
         &$3.79\times10^{-4}$
         &$4.01\times10^{-4}$\\
         Time
         &124 s
         &1461 s
         &275 s\\
         \bottomrule
    \end{tabular}
    \caption{Comparison for 2D cantilever beam problem with target volume fraction = 0.3}
    \label{tab:2dcomp03}
\end{table}

\begin{figure*}

\centering

\begin{subfigure}[t]{0.3\textwidth}
\includegraphics[width=\textwidth]{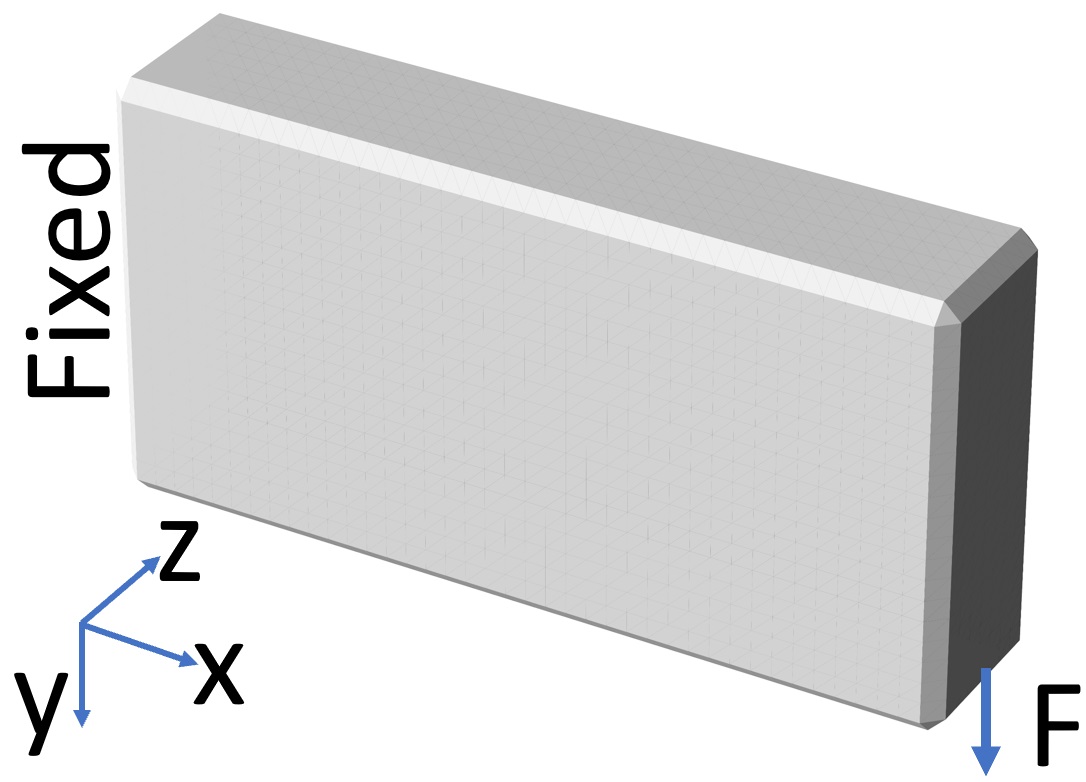}
\caption{Boundary Conditions} \label{fig:PINN_convergence a}
\end{subfigure}\hspace{0.5cm}\begin{subfigure}[t]{0.3\textwidth}
\includegraphics[width=\textwidth]{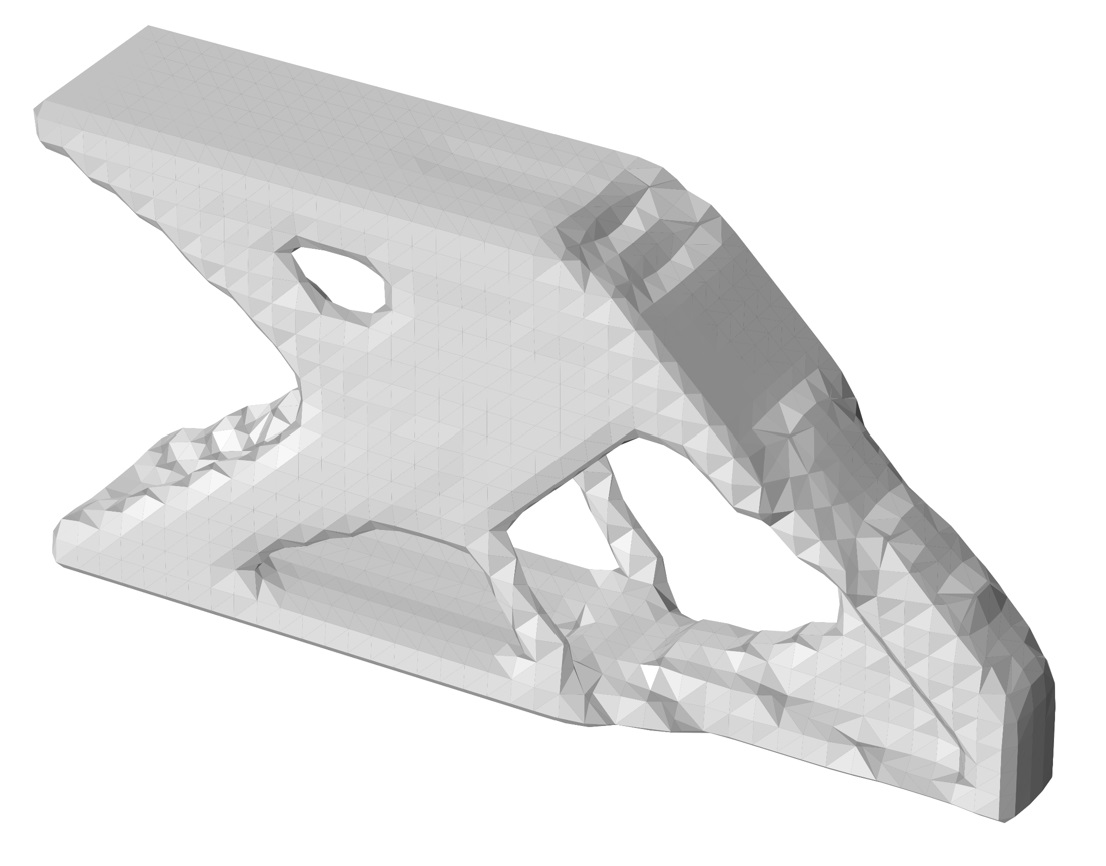}
\caption{Top3D (SIMP) Result, [7.49, 0.30]}\label{fig:PINN_convergence b}
\end{subfigure}\hspace{0.5cm}\begin{subfigure}[t]{0.3\textwidth}
\includegraphics[width=\textwidth]{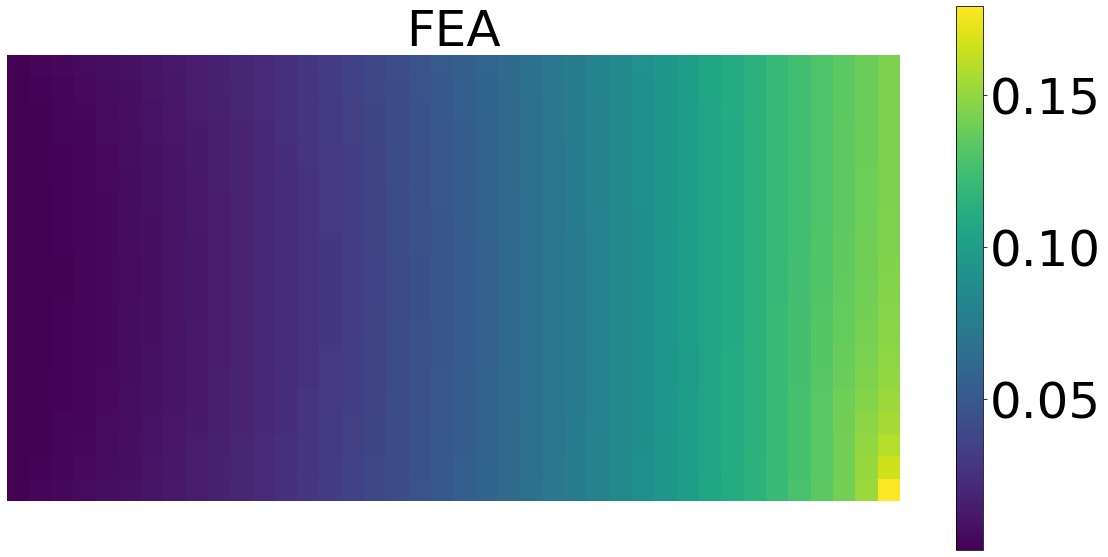}
\caption{Max Disp = 0.179mm}\label{fig:PINN_convergence c}
\end{subfigure}
\begin{subfigure}[t]{0.3\textwidth}
\includegraphics[width=\textwidth]{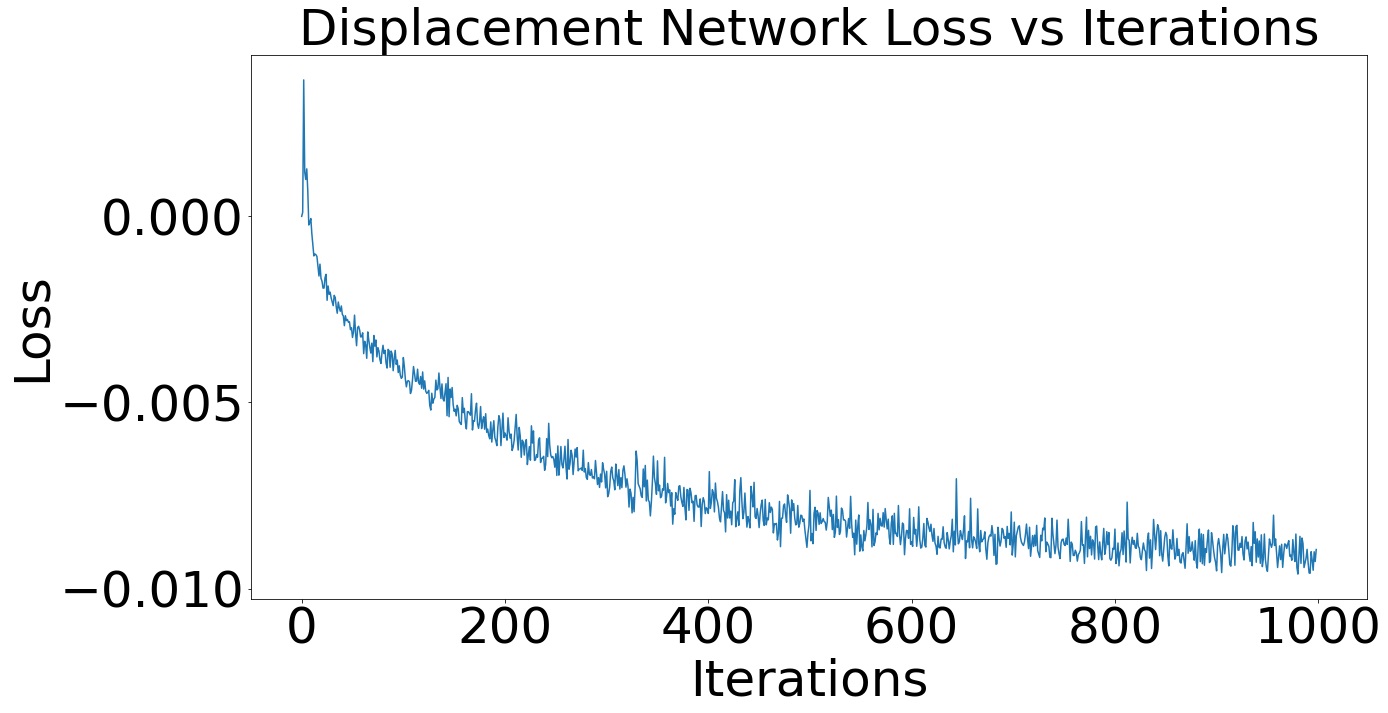}
\caption{Initial PINN Convergence}\label{fig:PINN_convergence d}
\end{subfigure}\hspace{0.5cm}\begin{subfigure}[t]{0.3\textwidth}
\includegraphics[width=\textwidth]{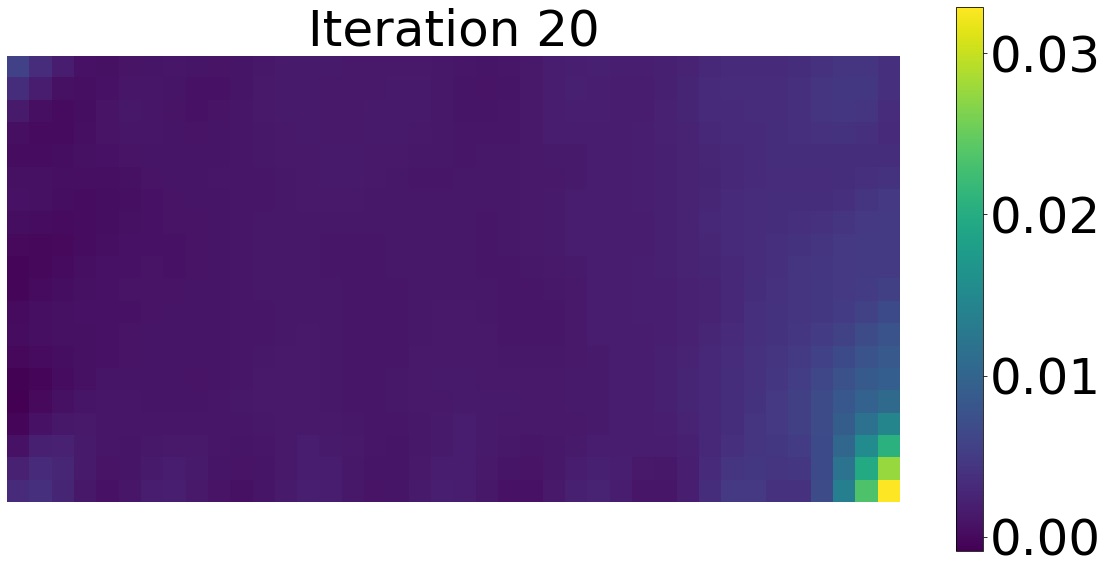}
\caption{Max Disp = 0.033mm}\label{fig:PINN_convergence e}
\end{subfigure}\hspace{0.5cm}\begin{subfigure}[t]{0.3\textwidth}
\includegraphics[width=\textwidth]{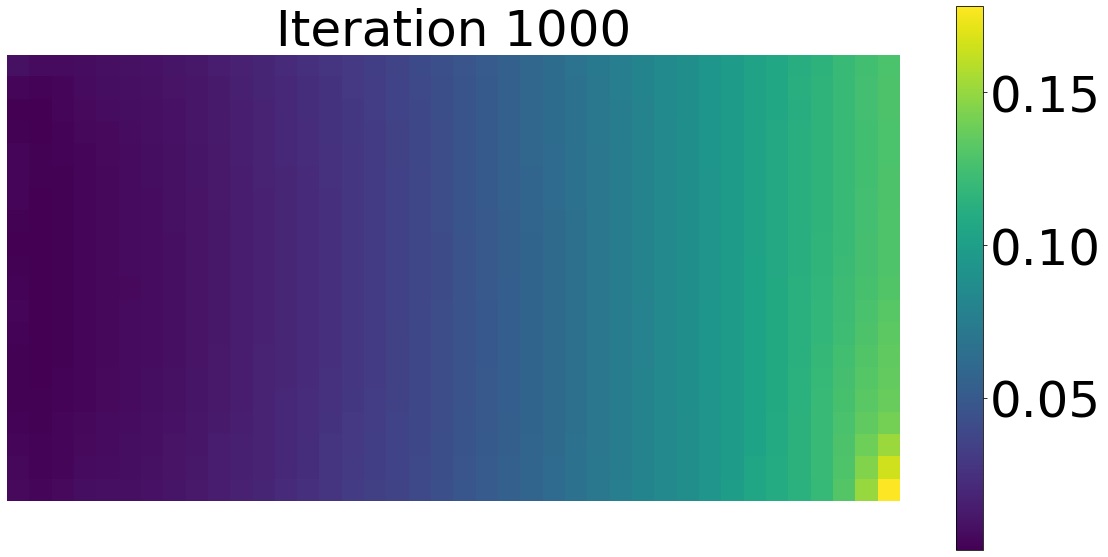}
\caption{Max Disp = 0.178mm}\label{fig:PINN_convergence f}
\end{subfigure}
\begin{subfigure}[t]{0.3\textwidth}
\includegraphics[width=\textwidth]{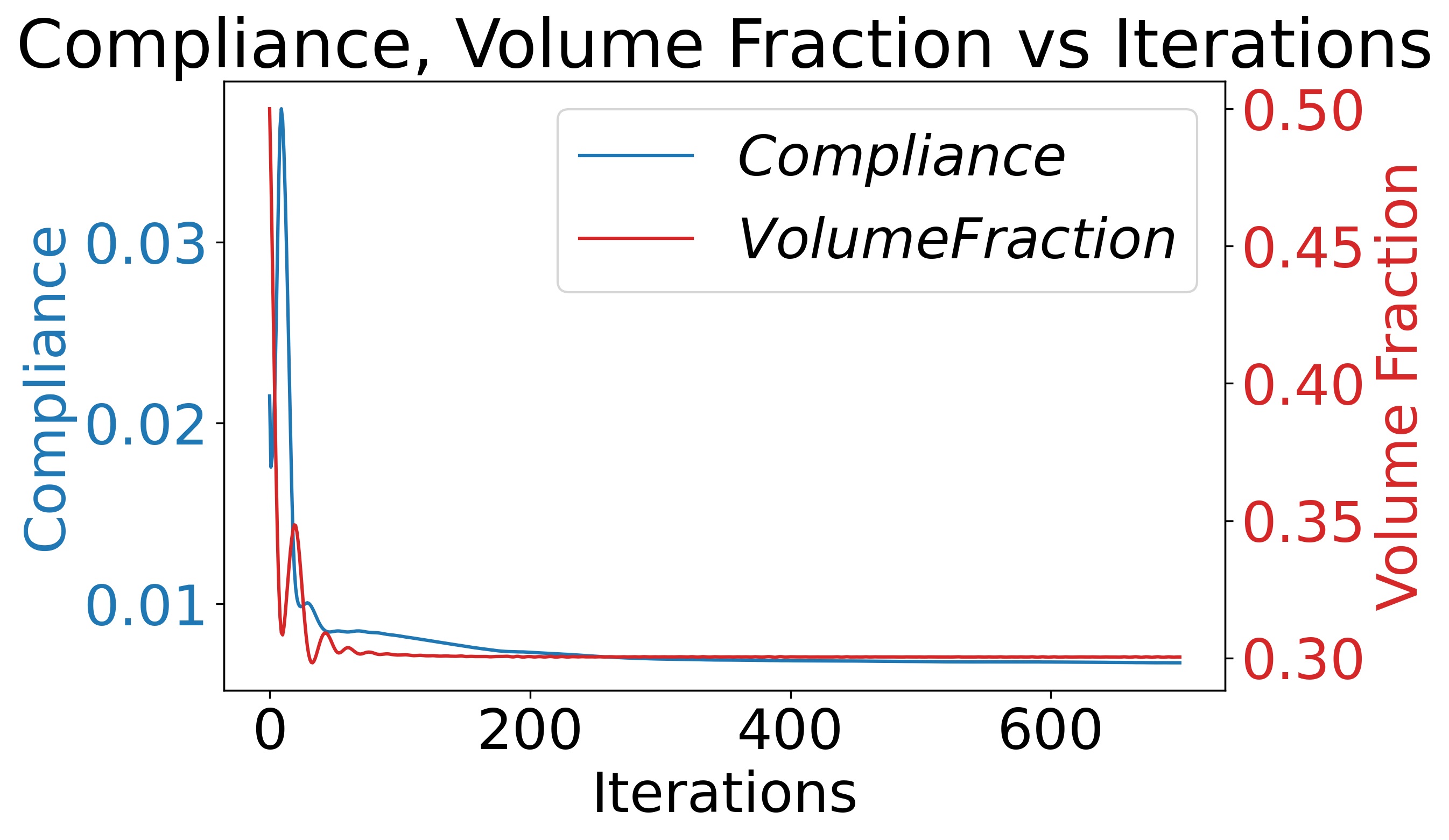}
\caption{Compliance and Volume Fraction Convergence}\label{fig:PINN_convergence g}\end{subfigure}\hspace{0.5cm}\begin{subfigure}[t]{0.3\textwidth}
\includegraphics[width=\textwidth]{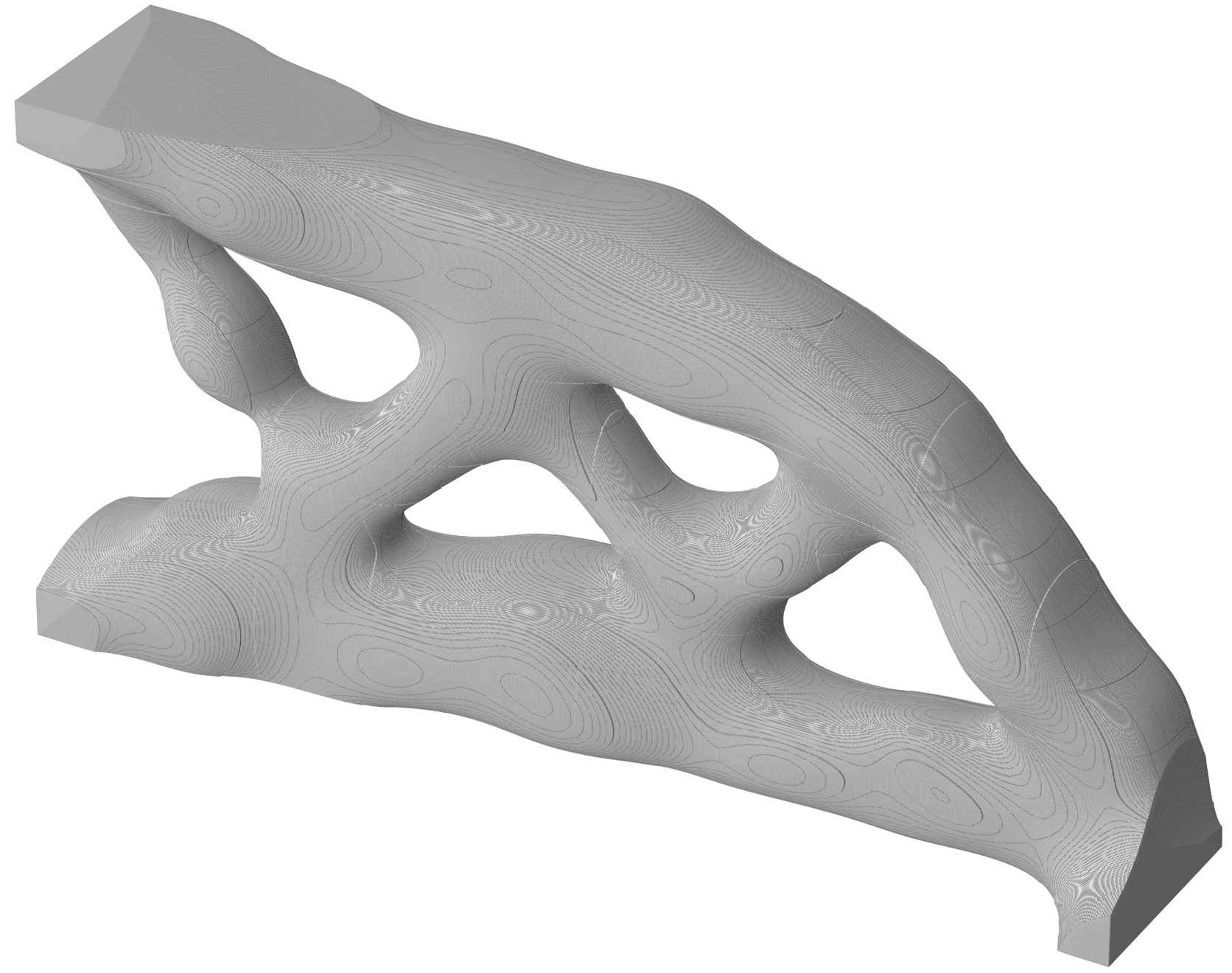}
\caption{DMF-TONN Result, [6.76, 0.30]}\label{fig:PINN_convergence h}
\end{subfigure}

\caption{Convergence history of DMF-TONN and result comparison with SIMP for 3D cantilever beam problem with target volume fraction = 0.3, Key: [compliance $\times 10^{-3}$, volume fraction]}

\label{fig:PINN_convergence}
\end{figure*}

\begin{figure*}

\centering

\begin{subfigure}[t]{0.2\textwidth}
\includegraphics[width=\textwidth]{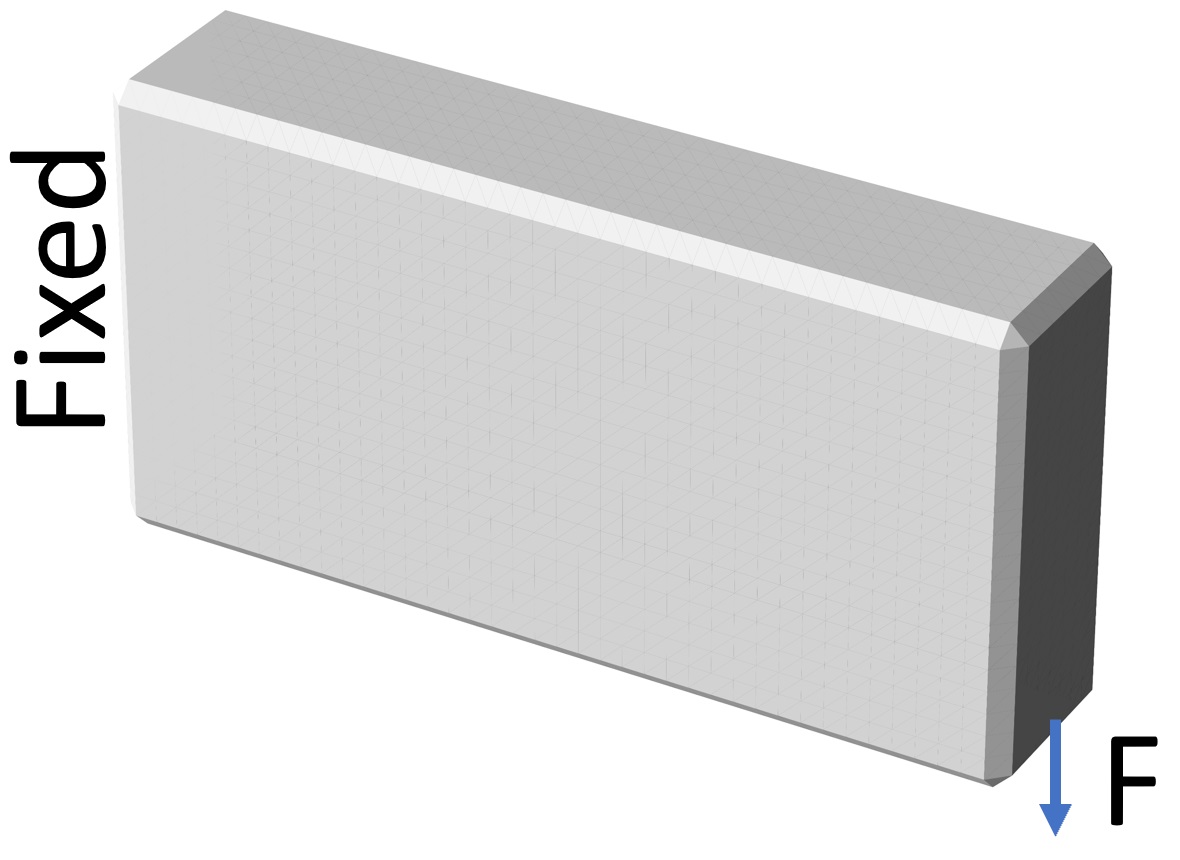}
\caption{Boundary conditions, Target volume fraction = 0.30}
\end{subfigure}\hspace{0.3cm}\begin{subfigure}[t]{0.2\textwidth}
\includegraphics[width=\textwidth]{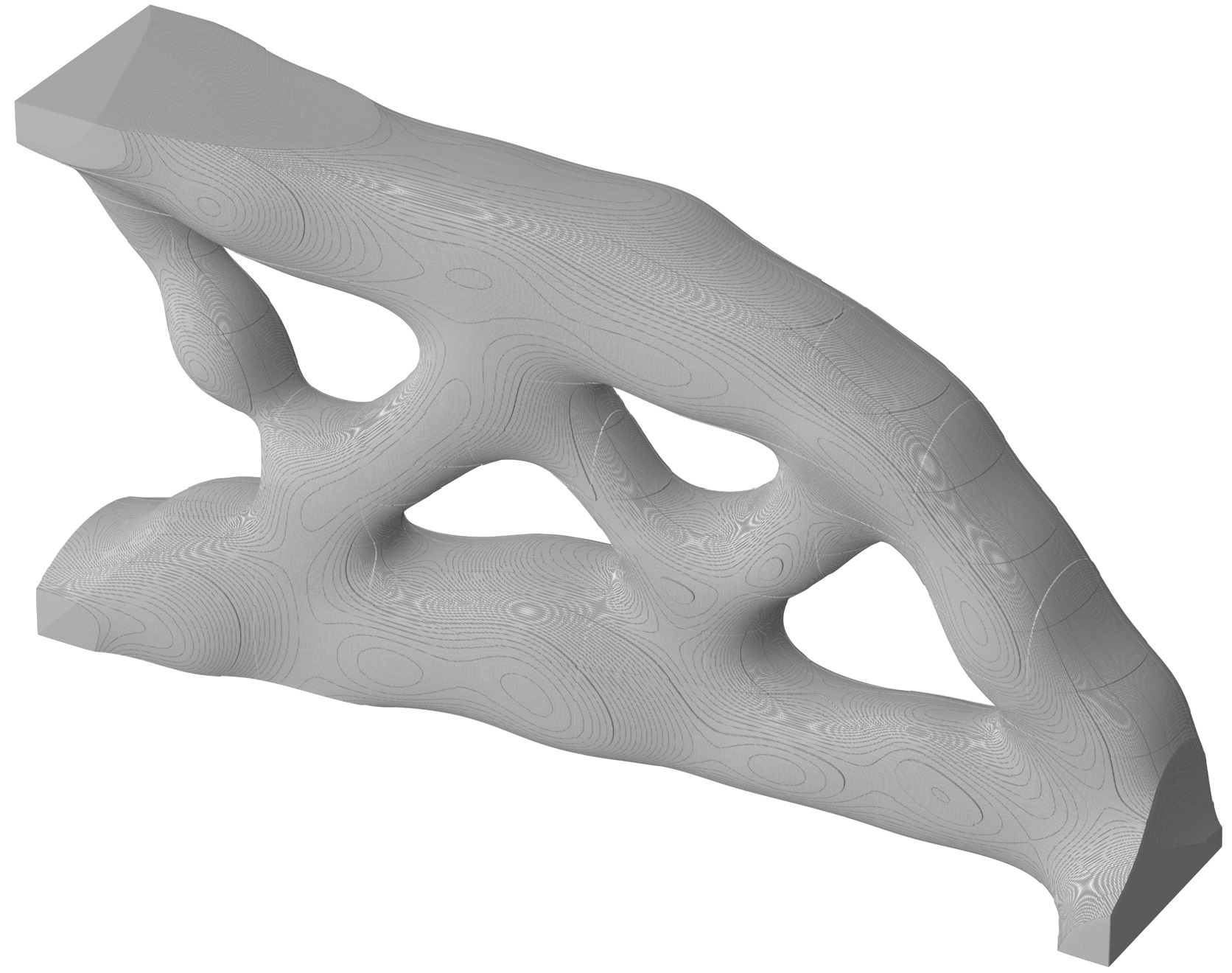}
\caption{20 PINN iterations, [6.76, 588 s, 0.30]}
\end{subfigure}\hspace{0.3cm}\begin{subfigure}[t]{0.2\textwidth}
\includegraphics[width=\textwidth]{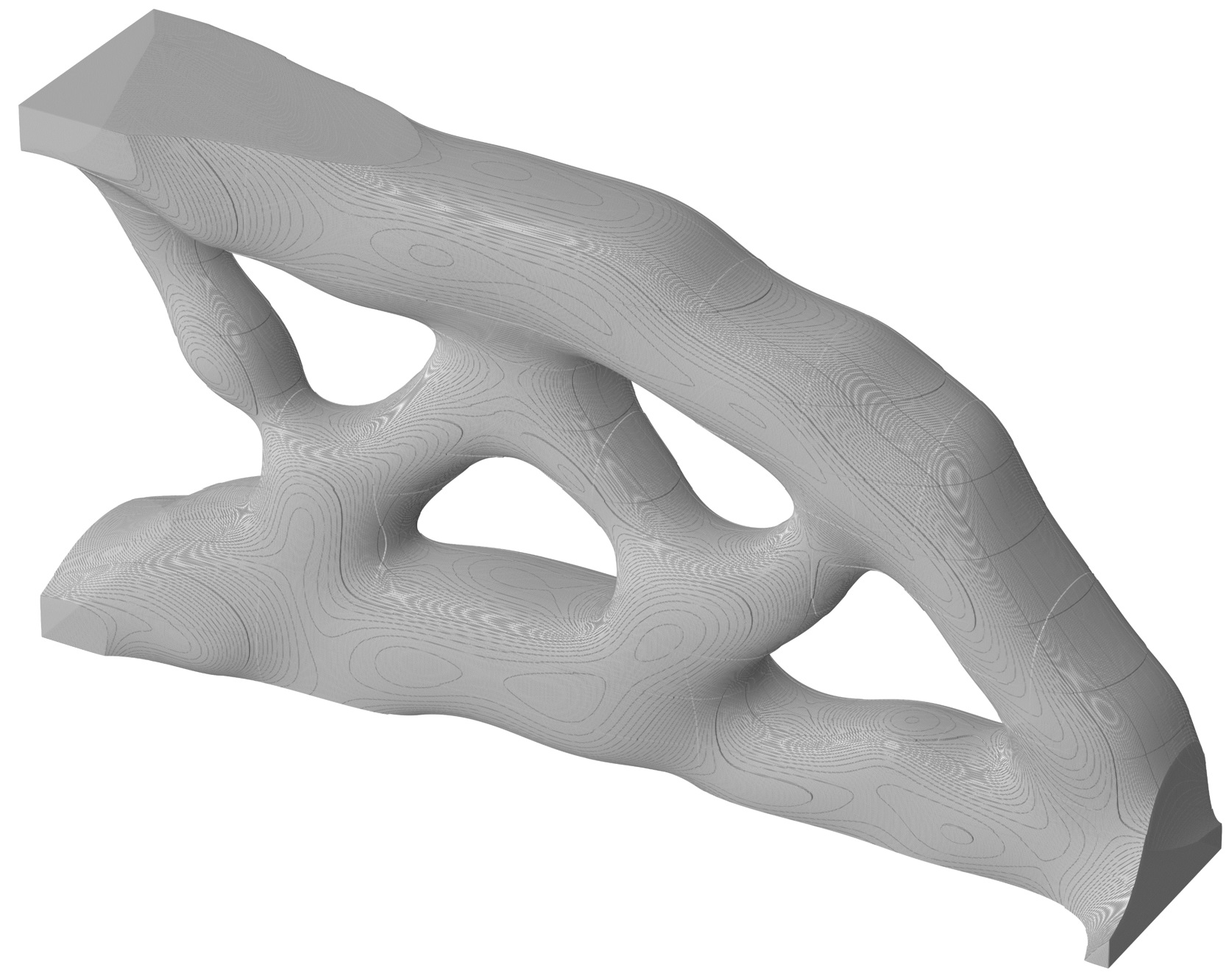}
\caption{1000 PINN iterations, [6.63, 24902 s, 0.30]}\end{subfigure}\hspace{0.3cm}\begin{subfigure}[t]{0.2\textwidth}
\includegraphics[width=\textwidth]{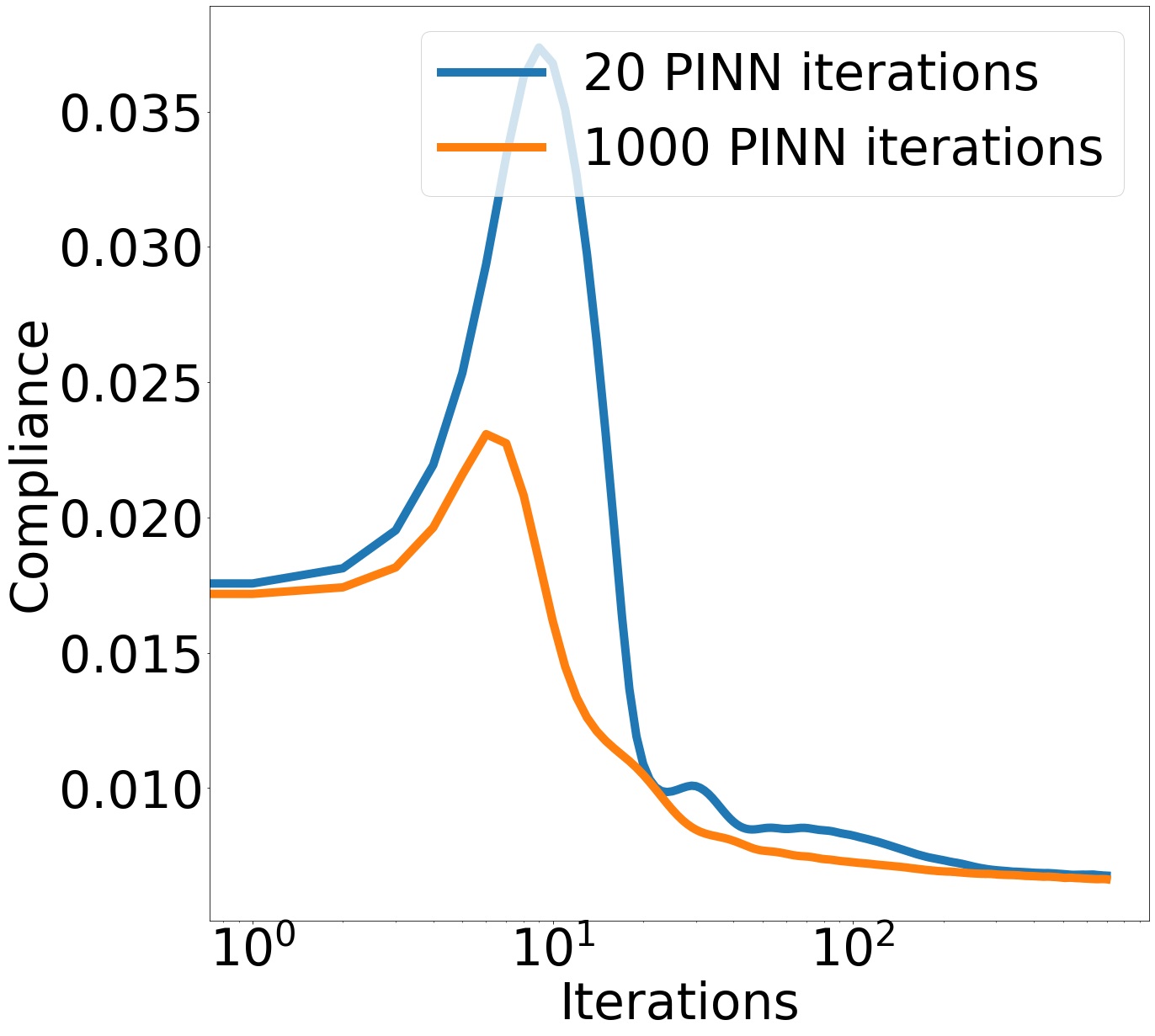}
\caption{Compliance convergence history comparison}\label{fig:ablation_study}\end{subfigure}

\caption{Transfer Learning Analysis, Key: [compliance $\times 10^{-3}$, time, volume fraction]}

\label{fig:TL1}

\end{figure*}
\begin{figure*}

\centering

\begin{subfigure}[c]{0.4\textwidth}

\includegraphics[width=\textwidth]{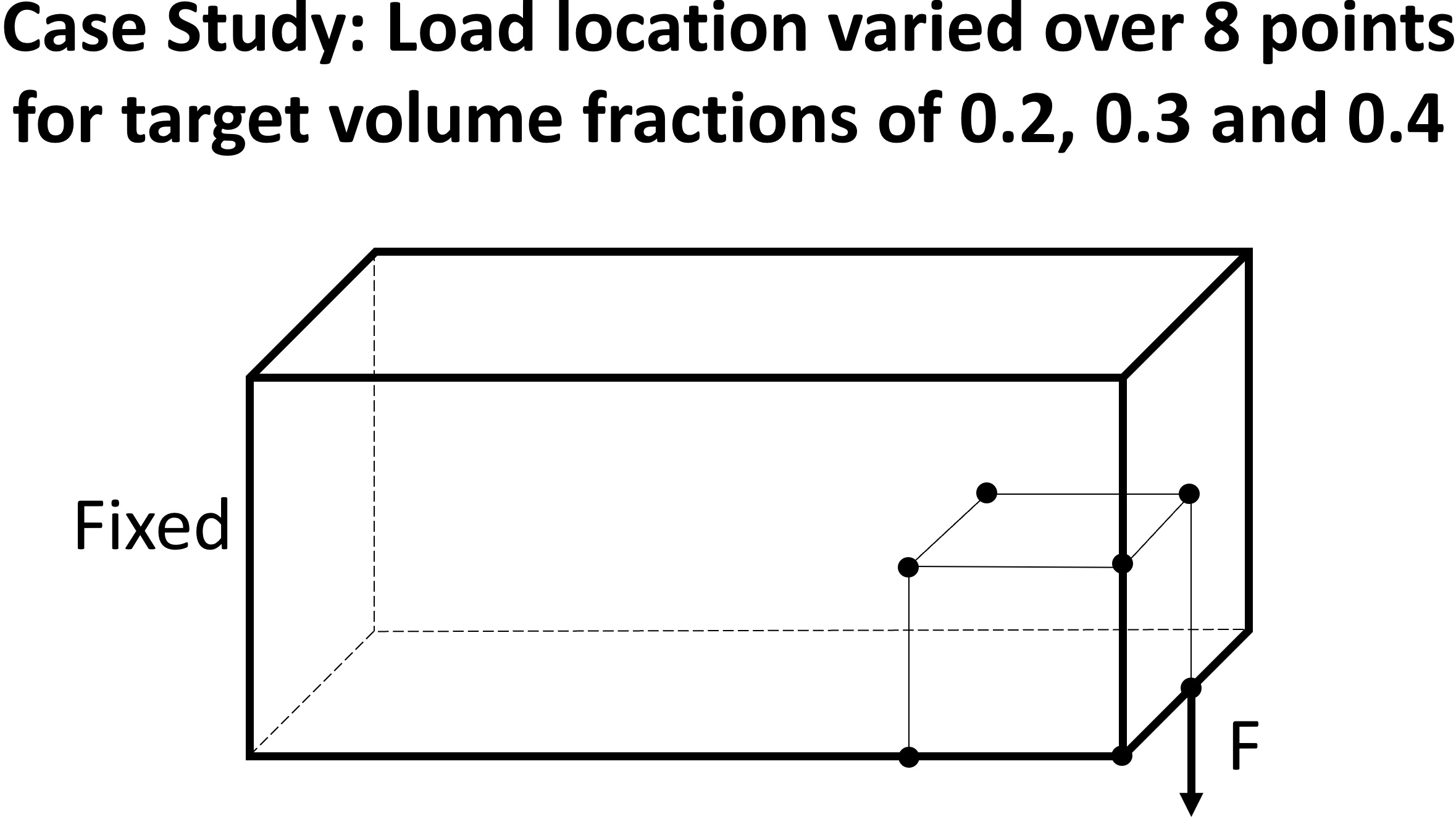}
\caption{Case study boundary conditions}
\end{subfigure}
\qquad
\begin{subfigure}[c]{0.5\textwidth}
\includegraphics[width=\textwidth]{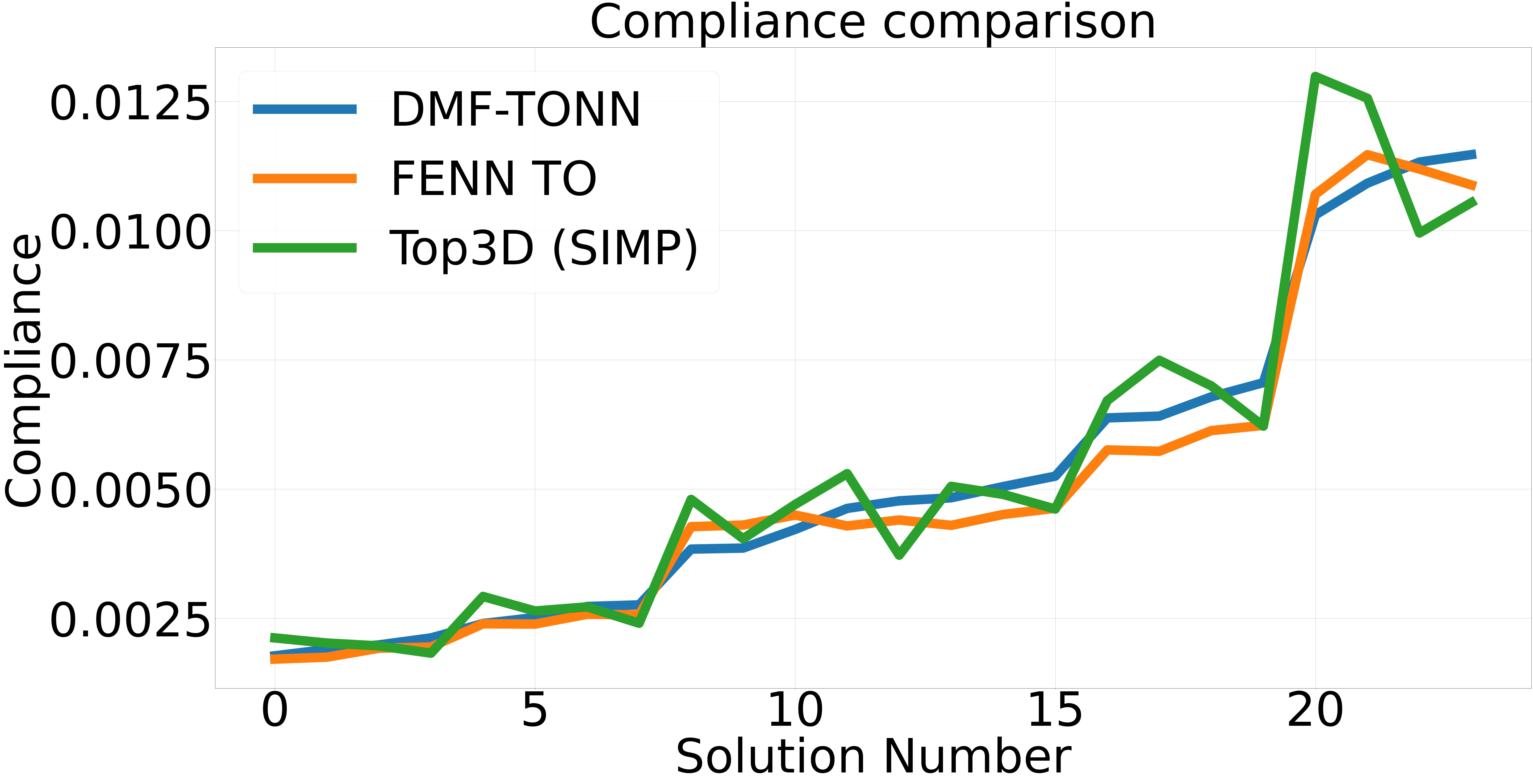}
\caption{Comparison of compliance of DMF-TONN, FENN TO and Top3D (SIMP)}
\label{fig:case_study_plot}
\end{subfigure}

\caption{Case Study of 3D Cantilever Beam Problem}

\label{fig:case_study}
\end{figure*}

\section{Results}\label{sec4}
We first compare our results with a conventional SIMP topology optimization  (\citet{andreassen2011efficient}) and NTopo (\citet{zehnder2021ntopo}) for a 2D cantilever beam problem. Then, for a 3D cantilever problem, we present the initial convergence history of the displacement network, and the convergence history of the compliance. Subsequently, we perform a case study for the 3D cantilever beam problem over varying volume fractions and load locations, comparing our results with a conventional 3D SIMP topology optimizer (\citet{liu2014efficient}), and with a method using the Fourier Features neural network for density representation and FEA for compliance calculation (similar to \citet{Chandrasekhar2021Fourier}). Then, we present an example showcasing the trade-offs in DMF-TONN and SIMP in terms of the compliance and computational time. Lastly, we validate our approach on some additional examples. \ch{We also present some examples using the same constant grid as SIMP for DMF-TONN instead of domain points that are randomly sampled in each displacement network iteration and density network iteration for compliance calculation using the trained displacement network.}

The output of our network represents an implicit function of the spatial densities. We use the marching cubes algorithm for generating the renders of the results. It should be noted that with the DMF-TONN, one can sample infinitely many points within the domain, as a continuous and differentiable function has been learned by the density network. In this paper, we use \ch{20} times the number of samples of the FEA grid size in each direction for the DMF-TONN figures for each example. On all solutions obtained by our approach, we run FEA for calculating the final compliance and use this to compare against the compliance obtained by  SIMP (using the same FE solver)  for consistency.  We ensure the degrees of freedom available are always more for SIMP than for DMF-TONN. Moreover, in SIMP, we use a density filter with  a radius  1.5 times the mesh element size in each of the presented examples, ensuring thin features and lower compliances are possible, and there is no compromise  in the SIMP results.

\subsection{Comparison of DMF-TONN, SIMP and Ntopo for 2D Cantilever Beam example}
We compare the compliances and computational times for a 2D cantilever beam example in Tables \ref{tab:2dcomp05} and \ref{tab:2dcomp03}.
We run the SIMP code (\citet{andreassen2011efficient}) with the default convergence criterion, and run NTopo for 200 iterations as shown in (\citet{zehnder2021ntopo}) for a similar 2D cantilever beam example. We run our method for $n_{opt} = 2000$ topology optimization iterations (determined empirically for 2D problems), with $n_{disp} = 20$ iterations of the displacement network in each of these topology optimization iterations. We also compare the binary (0 and 1 density values) structure compliance (ensuring the volume fraction remains the same after thresholding) which provides the actual compliance if the optimized structures were to be used in practice. We observe that though our method is slower than SIMP, it results in a mesh-free optimization with a better compliance than SIMP and a faster computational time than NTopo.

\subsection{Analysis of a 3D cantilever beam example and transfer learning}
In Figure \ref{fig:PINN_convergence}, we present the convergence history of DMF-TONN and a comparison with 3D SIMP topology optimization for an example with boundary conditions shown in Figure \ref{fig:PINN_convergence a}. We use a $40\times20\times8$ grid for the SIMP FEA, which gives 6400 design variables for the topology optimization, and $6400\times3 = 19200$ degrees of freedom (DOF) for the FEA. We use a lesser DOF model, with the number of trainable weights in both our density and displacement networks being 4096 each. We use an initial topology consisting of uniform densities of 0.5 ($\rho_{init(40\times20\times8)} = \mathbf{0.5}$) as an input for initial training of the displacement network (PINN) and Figure \ref{fig:PINN_convergence d} shows the convergence history. Figures \ref{fig:PINN_convergence e}, \ref{fig:PINN_convergence f} show the displacement field in the $y$ direction at the cross-section of the domain where the force is applied and the maximum displacement value at different iterations of the initial run of the displacement network. Figure \ref{fig:PINN_convergence c} shows the FEA displacement field at this cross section. We see that at the $1000^{th}$ iteration, the displacement network can learn a very good approximation of the FEA displacement field.

In each of the $n_{opt} = 700$ topology optimization iterations (determined empirically for 3D problems), we use only $n_{disp} = 20$ displacement network iterations and show that this is adequate for achieving results (\ref{fig:PINN_convergence h}) similar to SIMP (\ref{fig:PINN_convergence b}). Figure \ref{fig:PINN_convergence g} shows the convergence history for the topology optimization, with FEA compliance plotted for consistency with SIMP.  \ch{Maximum decrease occurs up till 300 iterations, but we continue up till 700 iterations for a more refined and better final topology. We observe similar trends in all other examples and hence use $n_{opt} = 700$ for each of the presented examples in this paper.} 
Our approach takes 588 seconds compared to 227 seconds for SIMP, but achieves mesh-free topology optimization and a better compliance than SIMP for this example.

The concept of transfer learning can be utilized with the displacement network, noting the fact that the change in topologies in each topology optimization iteration is not drastic and thus the neural network weights learned in the prior topology optimization iteration can be a good initialization for learning the weights corresponding to the static equilibrium displacement values in the next topology optimization iteration. In Figure \ref{fig:TL1}, we compare the results when using $n_{disp} = 20$ and  $n_{disp} = 1000$ displacement network iterations in each topology optimization iteration. We observe that once the displacement network is trained initially for $n_{initdisp} = 1000$ iterations before starting the topology optimization, $n_{disp} = 20$ gives similar compliance results to when using $n_{disp} = 1000$, but with a very low computational time, thus showcasing the utility of a good initialization.

\ch{Even though Figure \ref{fig:ablation_study} shows that a higher $n_{disp}$ tends to give a better objective value in lesser $n_{dens}$ (the orange curve is below the blue curve), considering the computational time $n_{disp}$ that is lower (up to a limit of 20) is better and we use $n_{disp} = 20$ for all the examples presented in this paper. If $n_{disp}$ is chosen to be very low (less than 20), then good convergence does not occur, as the displacement network has not learned much about the current topology and hence the optimization loss gradients are very inaccurate.} 

\begin{table*}[]
\small
    \centering
    \begin{tabular}{p{4cm}|p{2.5cm}|p{2.5cm}|p{2.5cm}}
        \toprule
        \includegraphics[width=0.2\textwidth]{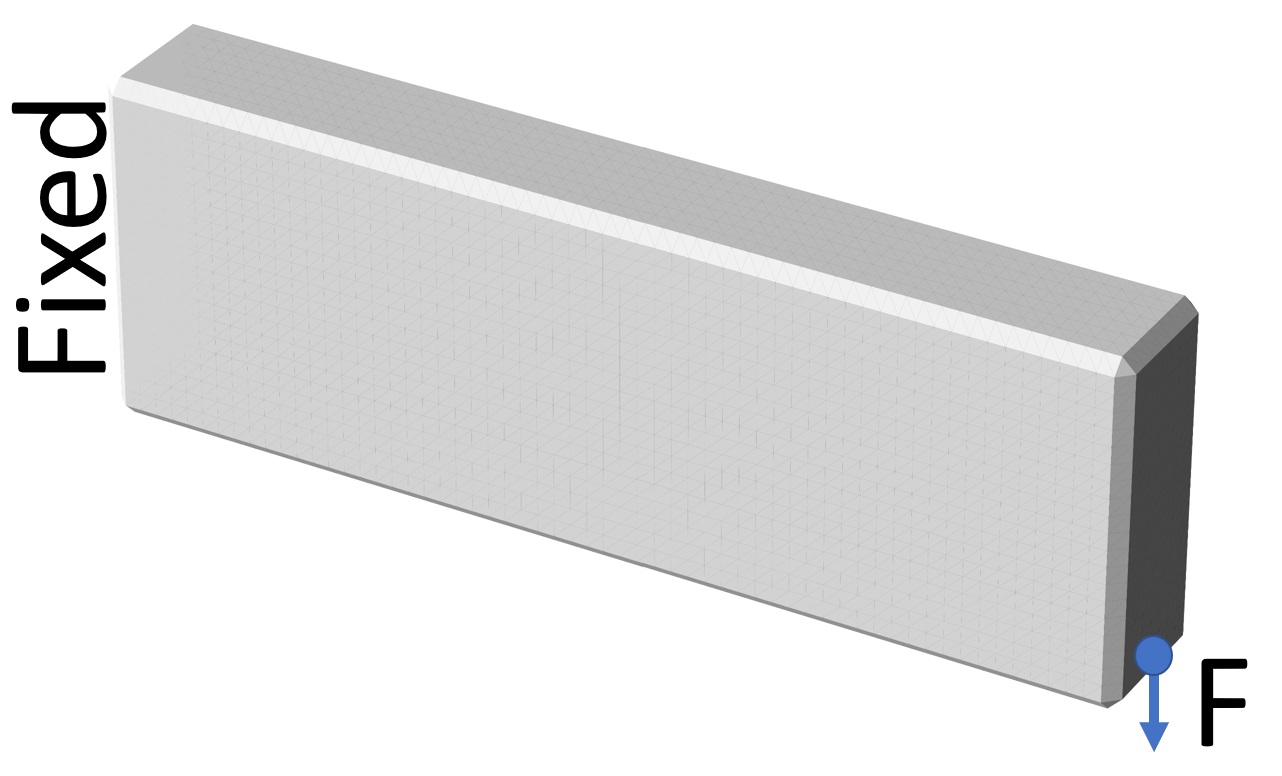}  & \includegraphics[width=0.2\textwidth]{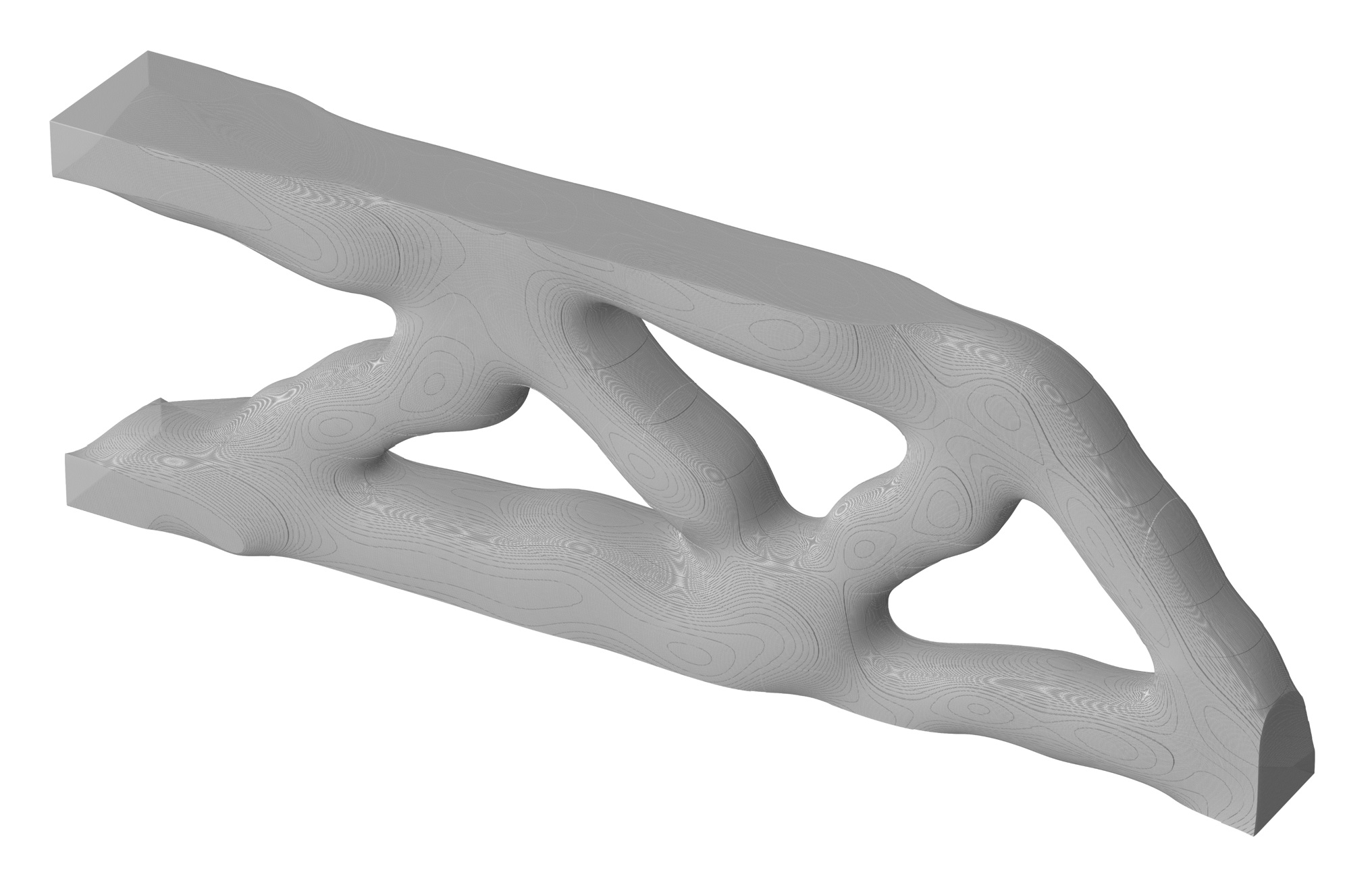}
        & \includegraphics[width=0.2\textwidth]{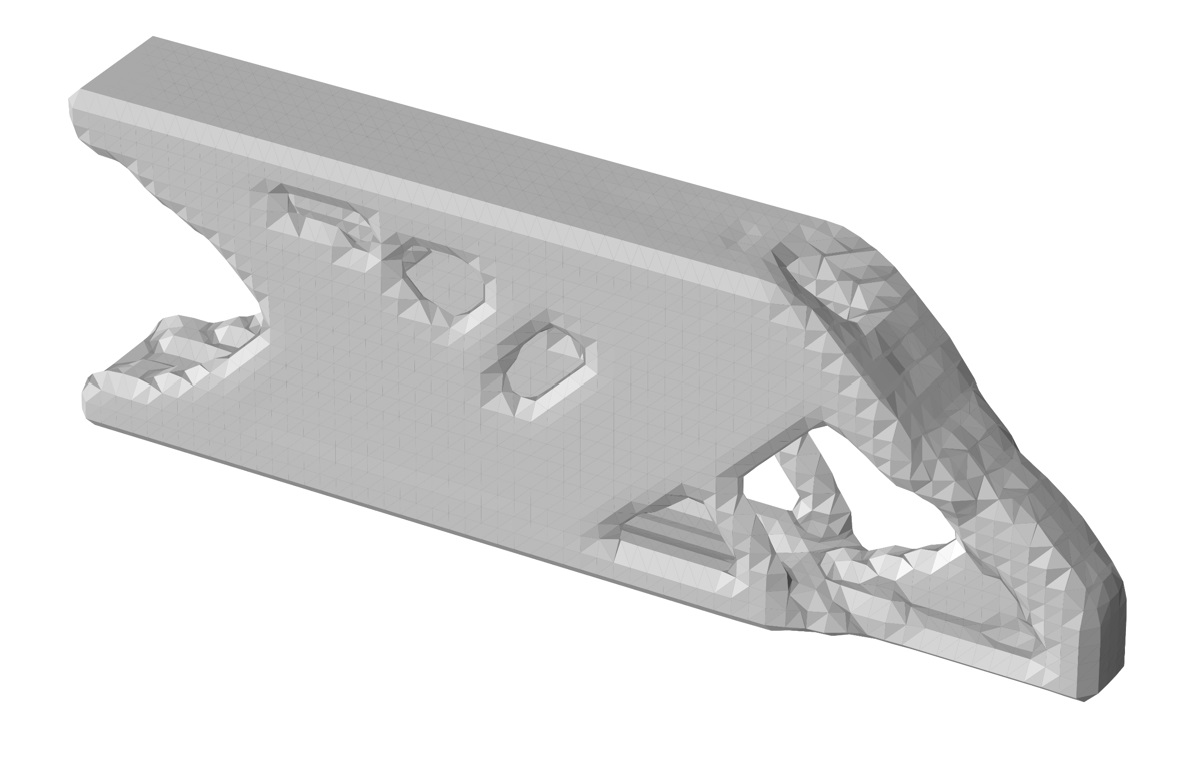}

        & \includegraphics[width=0.2\textwidth]{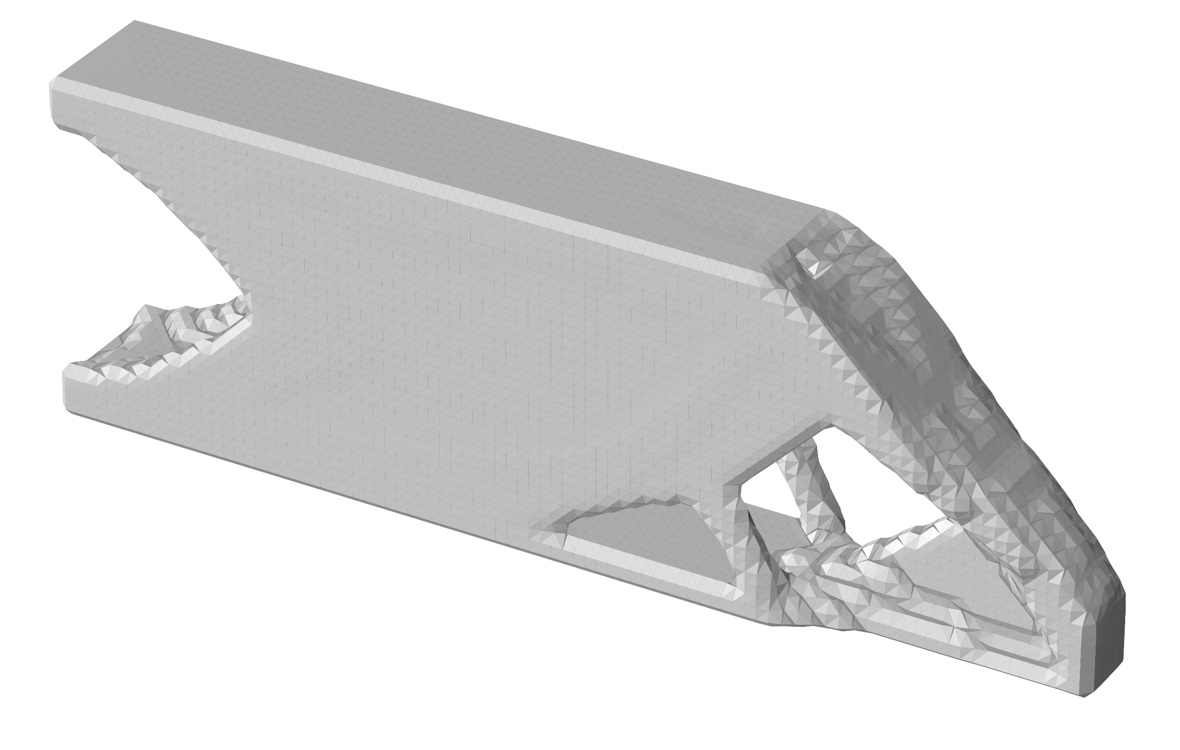}
        \\
         Boundary Conditions
         & DMF-TONN
         & SIMP ($60\times20\times8$ grid)
         & SIMP ($90\times30\times12$ grid)\\
         \midrule
         Displacement Calculation Degrees of Freedom
         &3456
         &28800
         &97200\\
         No. of Optimization Design Variables
         &3456
         &9600
         &32400\\
         Convergence Compliance
         &$2.40\times10^{-\ch{2}}$
         &$2.65\times10^{-\ch{2}}$
         &$2.38\times10^{-\ch{2}}$\\
         Volume Fraction Achieved
         &0.30
         &0.30
         &0.30\\
         Computational Time
         &517 s
         &99 s
         &807 s\\
        \bottomrule
    \end{tabular}
    \caption{Long cantilever beam with bottom load with target volume fraction = 0.3}
    \label{tab:3dcomp_long_beam}
\end{table*}

\begin{figure*}

\centering

\begin{subfigure}[t]{0.3\textwidth}
\includegraphics[width=\textwidth]{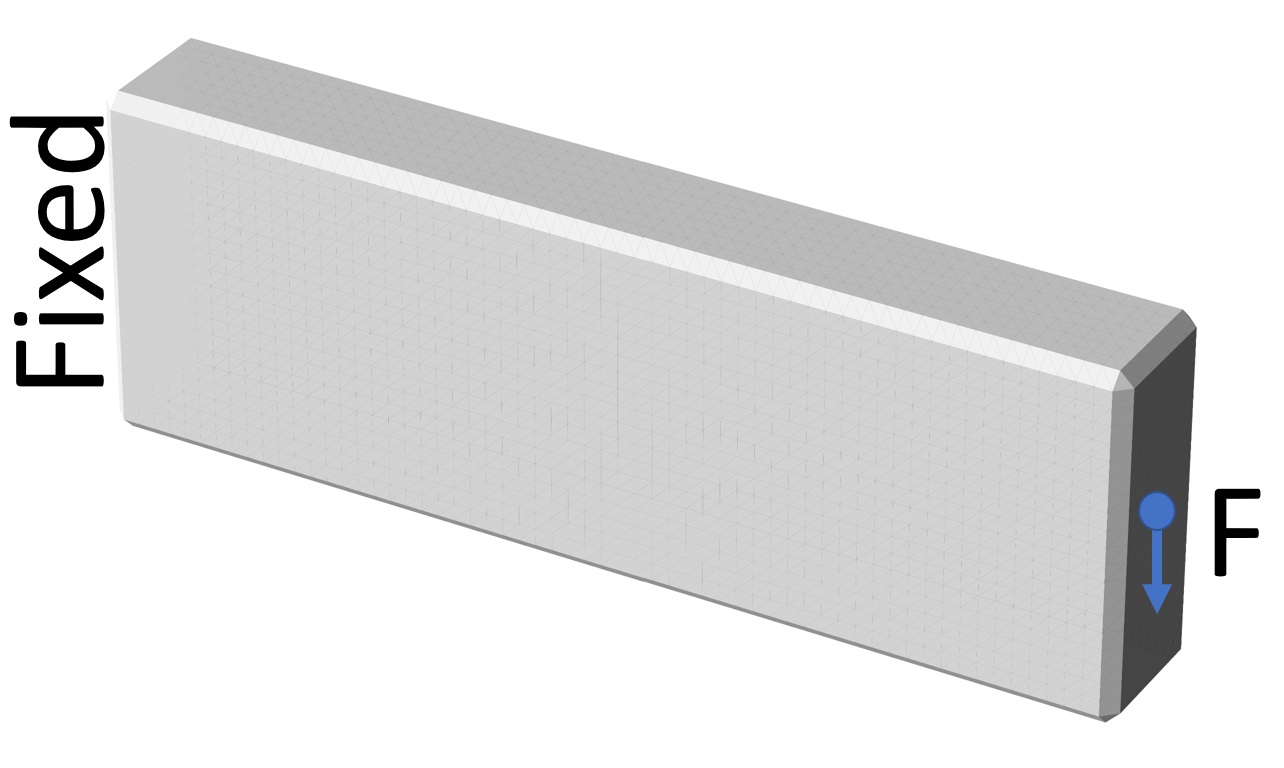}
\caption{Boundary conditions}
\end{subfigure}\hspace{0.3cm}\begin{subfigure}[t]{0.3\textwidth}
\includegraphics[width=\textwidth]{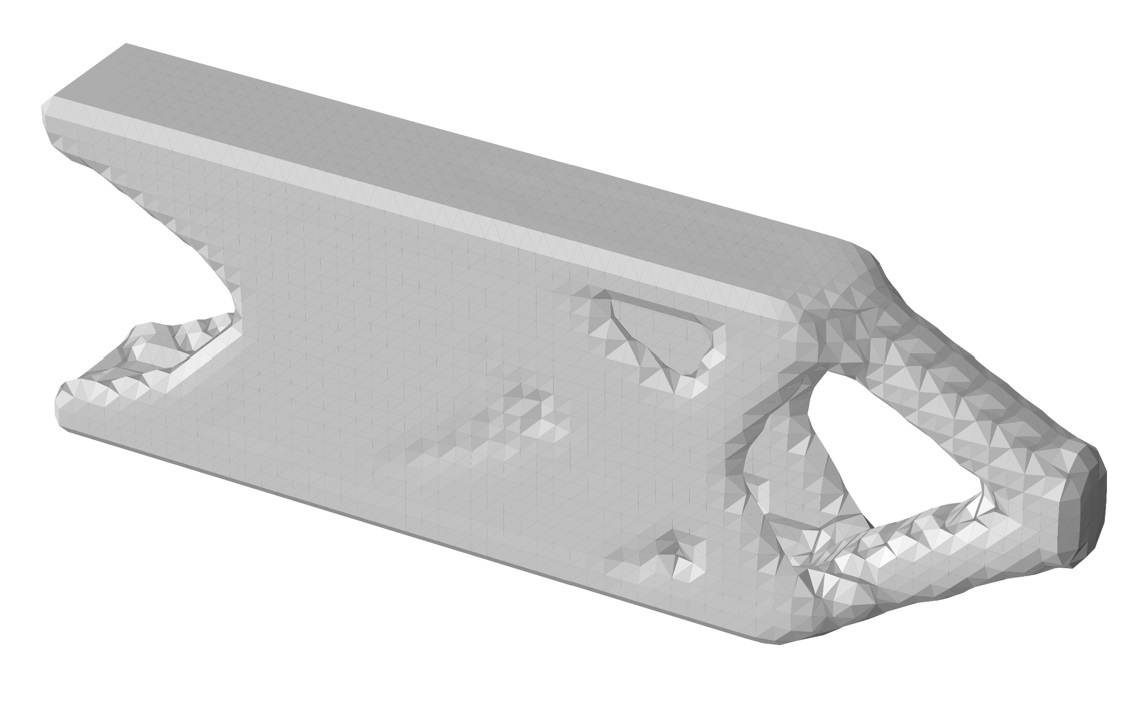}
\caption{Top3D (SIMP), [2.56, 0.30]}
\end{subfigure}\hspace{0.3cm}\begin{subfigure}[t]{0.3\textwidth}
\includegraphics[width=\textwidth]{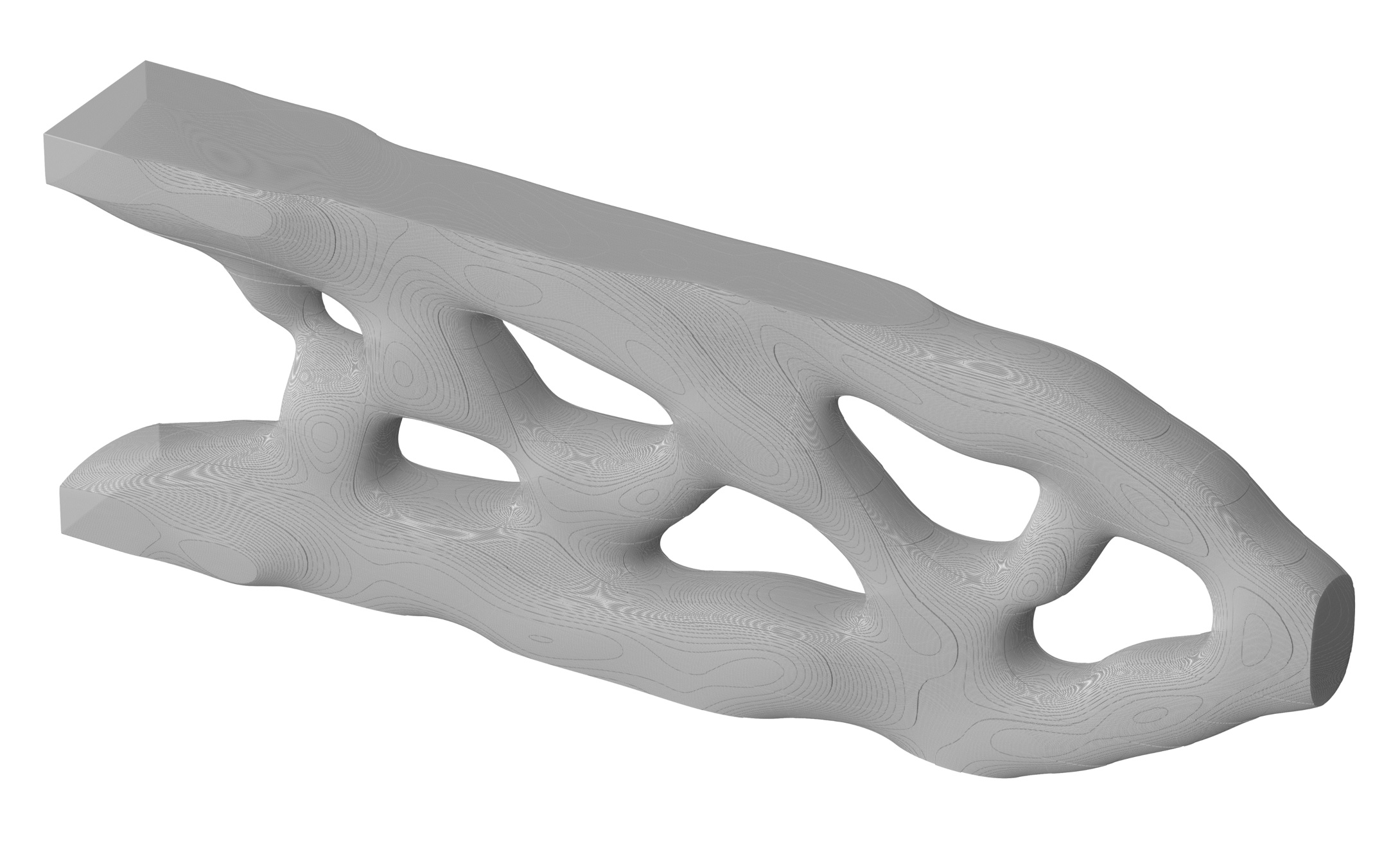}
\caption{DMF-TONN, [2.29, 0.30]}
\end{subfigure}

\caption{Long cantilever beam with center load with target volume fraction = 0.3, Key: [compliance $\times 10^{-\ch{2}}$, volume fraction]}

\label{fig:lbcenter}
\end{figure*}

\subsection{Case Study of 3D Cantilever Beam problem}
In Figure \ref{fig:case_study} we compare our fully mesh-free approach with a Finite Element Analsyis based Neural Network Topology Optimization (FENN TO) approach (i.e. using a Fourier Features neural network for representing density field and FEA for compliance calculation) and with the conventional SIMP approach. We vary the load location and target volume fraction for the 3D cantilever beam problem. All approaches are run for 700 iterations. We show the plots of the compliance of all three approaches for all these boundary conditions in Figure \ref{fig:case_study_plot} where the x-axis contains the discrete boundary conditions and y-axis represents the compliance for each of these boundary conditions. Our fully mesh-free approach achieves similar compliance values to the existing SIMP and FENN TO approaches for all the boundary conditions in this case study. The total computational times for running all the examples in this case study are 236 minutes, 96 minutes and 124 minutes for DMF-TONN, SIMP and FENN TO respectively.

\subsection{Trade-off Analysis of DMF-TONN and SIMP}\label{sec44}
In Table \ref{tab:3dcomp_long_beam}, we present the results for a right bottom end loaded cantilever beam with the ratio of 3 for the lengths of sides in the $x$ and $y$ directions (the orientation of the axes is the same as in Figure \ref{fig:PINN_convergence a}). For the Top3D (SIMP) method, we use their stated convergence criteria of 200 maximum iterations and 0.01 as tolerance of change in topology. Though the degrees of freedom are lesser for DMF-TONN, it still achieves a better compliance than SIMP with a grid of $60\times20\times8$. However, the computational time is much higher for DMF-TONN. Now, we increase the grid size of SIMP to $90\times30\times12$ for achieving a better compliance than DMF-TONN. However, as seen in Table \ref{tab:3dcomp_long_beam}, due to the 1.5 times increase in grid size in each direction, there was an exponential increase in computational time for the SIMP method and the computational time of DMF-TONN is lesser than the fine mesh SIMP. This showcases one of the advantages of the mesh-free nature of DMF-TONN, presenting interesting opportunities for tradeoffs to be explored in future research.
\begin{figure*}

\centering

\begin{subfigure}[t]{0.3\textwidth}
\includegraphics[width=\textwidth]{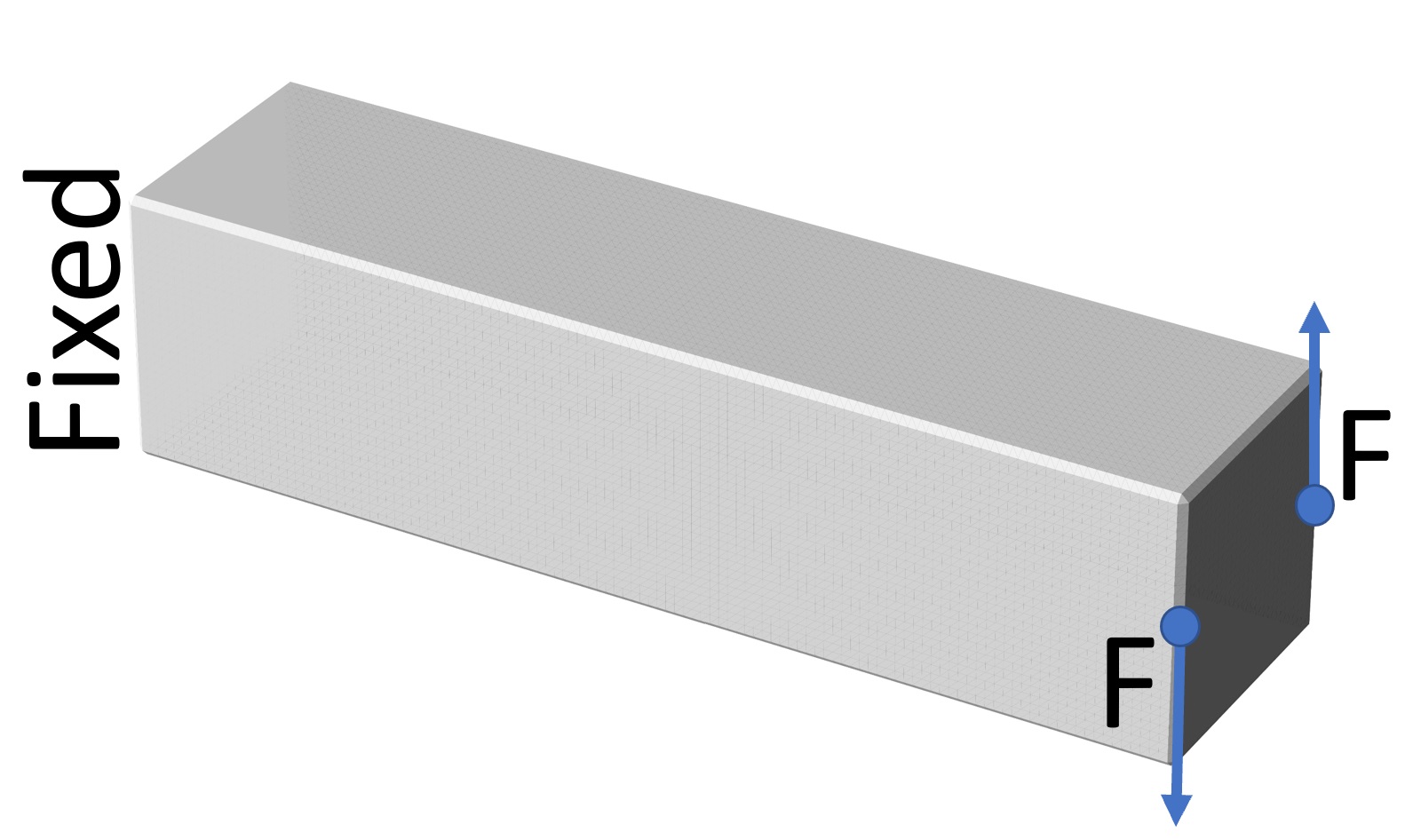}
\caption{Boundary conditions}
\end{subfigure}\hspace{0.3cm}\begin{subfigure}[t]{0.3\textwidth}
\includegraphics[width=\textwidth]{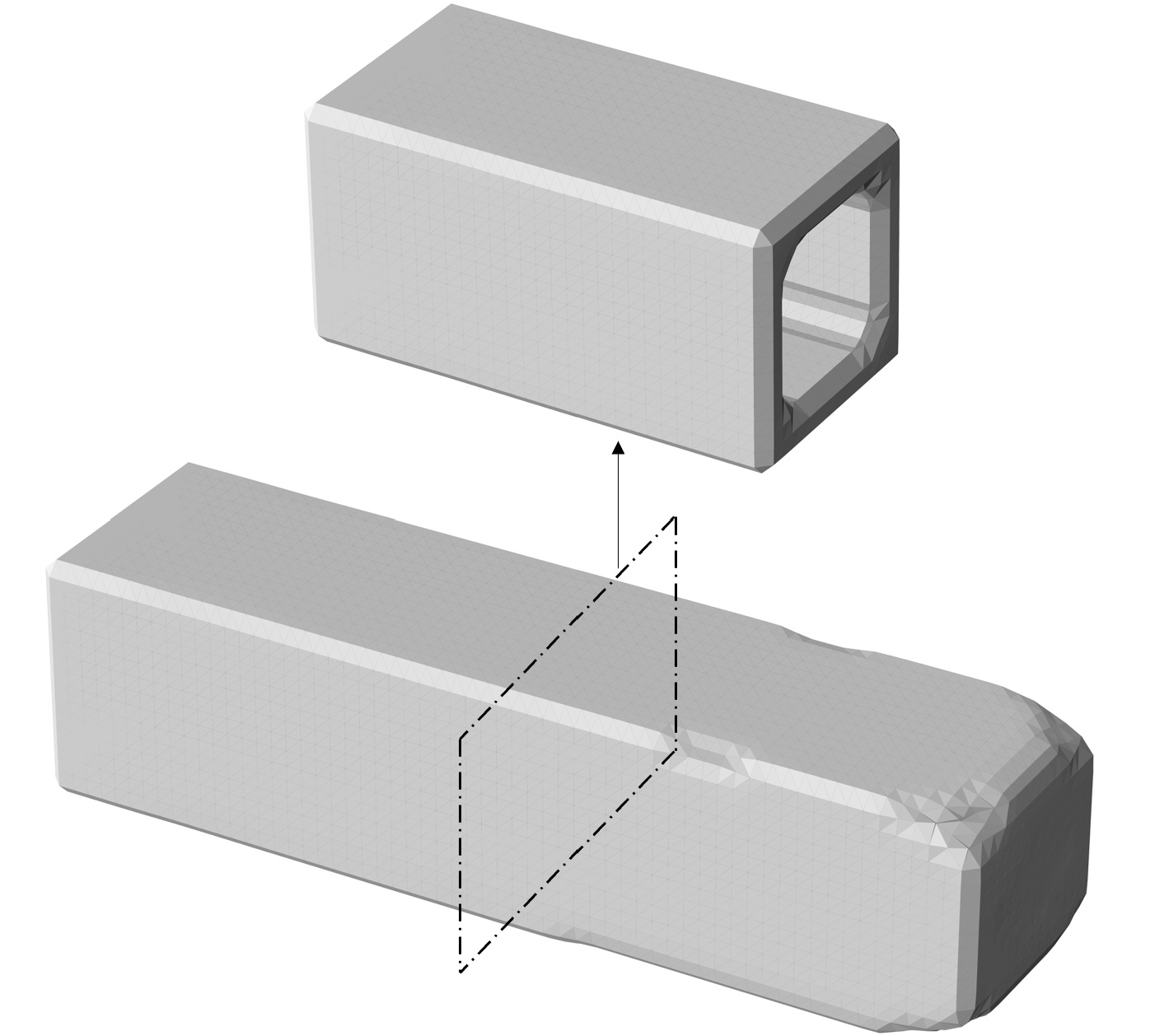}
\caption{Top3D (SIMP), [8.22, 0.50]}
\end{subfigure}\hspace{0.3cm}\begin{subfigure}[t]{0.3\textwidth}
\includegraphics[width=\textwidth]{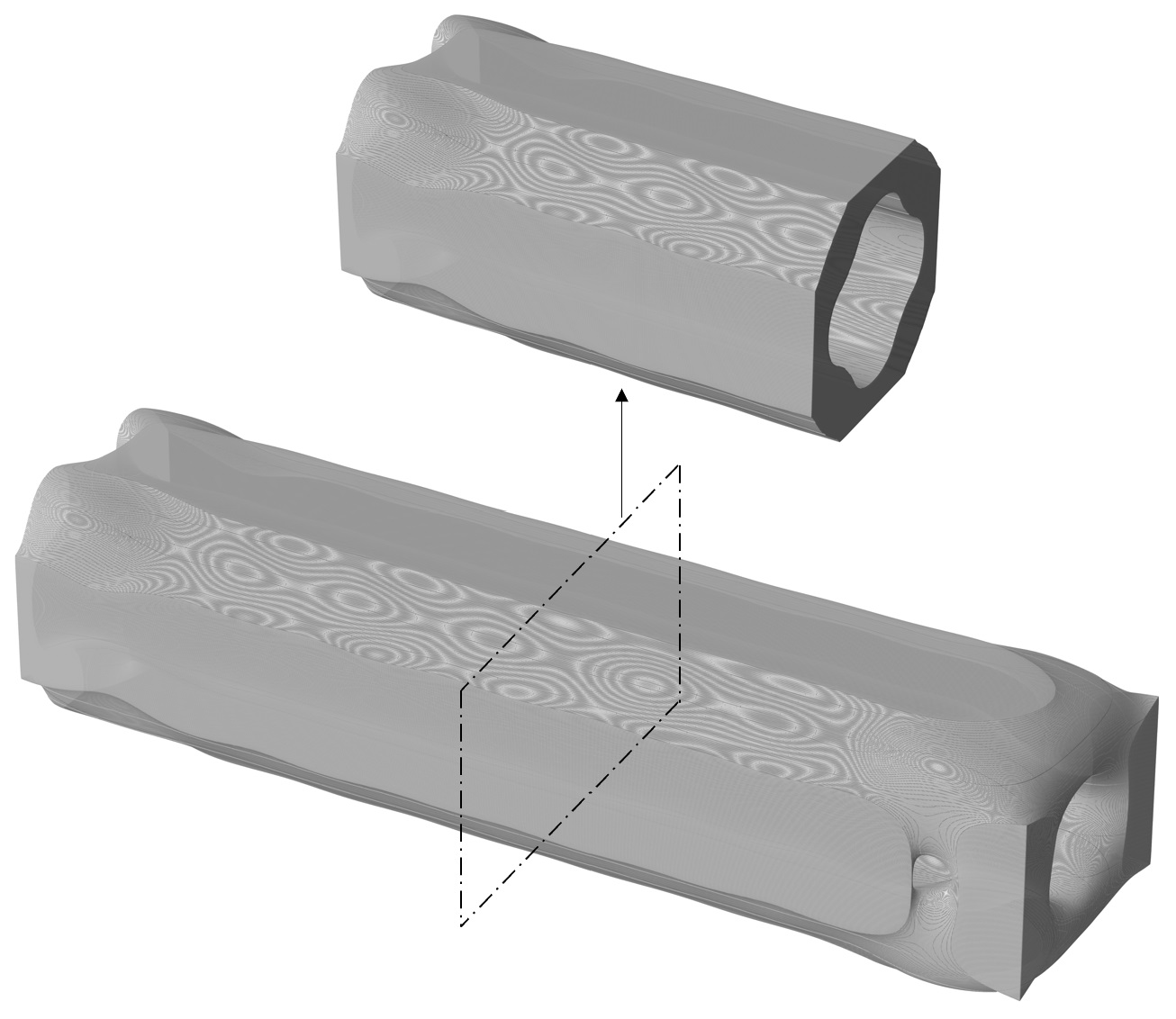}
\caption{DMF-TONN, [8.78,0.50]}
\end{subfigure}

\caption{Long cantilever beam with two loads with target volume fraction = 0.5, Key: [compliance $\times 10^{-3}$, volume fraction]}

\label{fig:torque}

\centering

\begin{subfigure}[t]{0.2\textwidth}
\includegraphics[width=\textwidth]{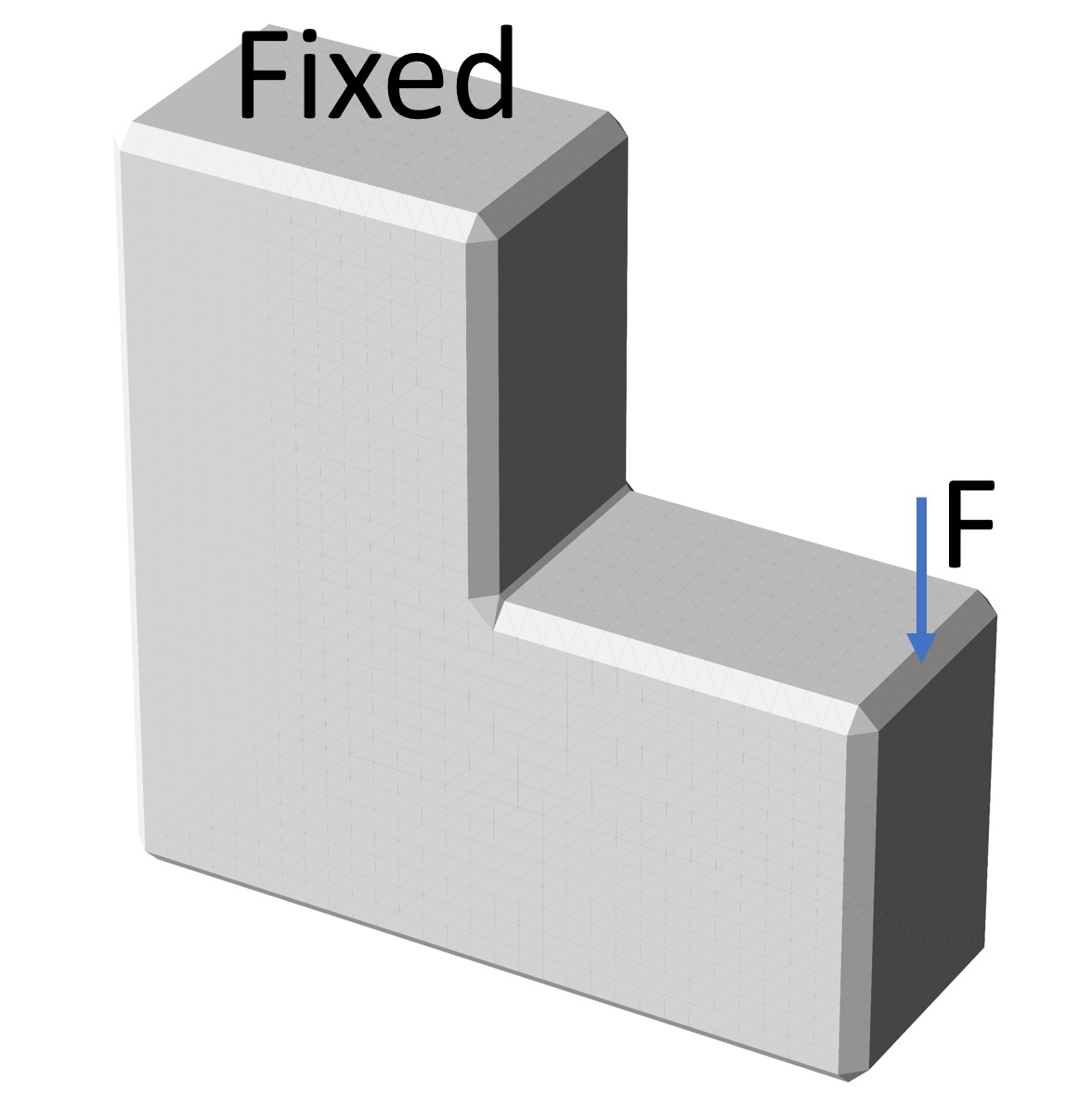}
\caption{Boundary conditions}
\end{subfigure}\hspace{0.3cm}\begin{subfigure}[t]{0.3\textwidth}
\includegraphics[width=\textwidth]{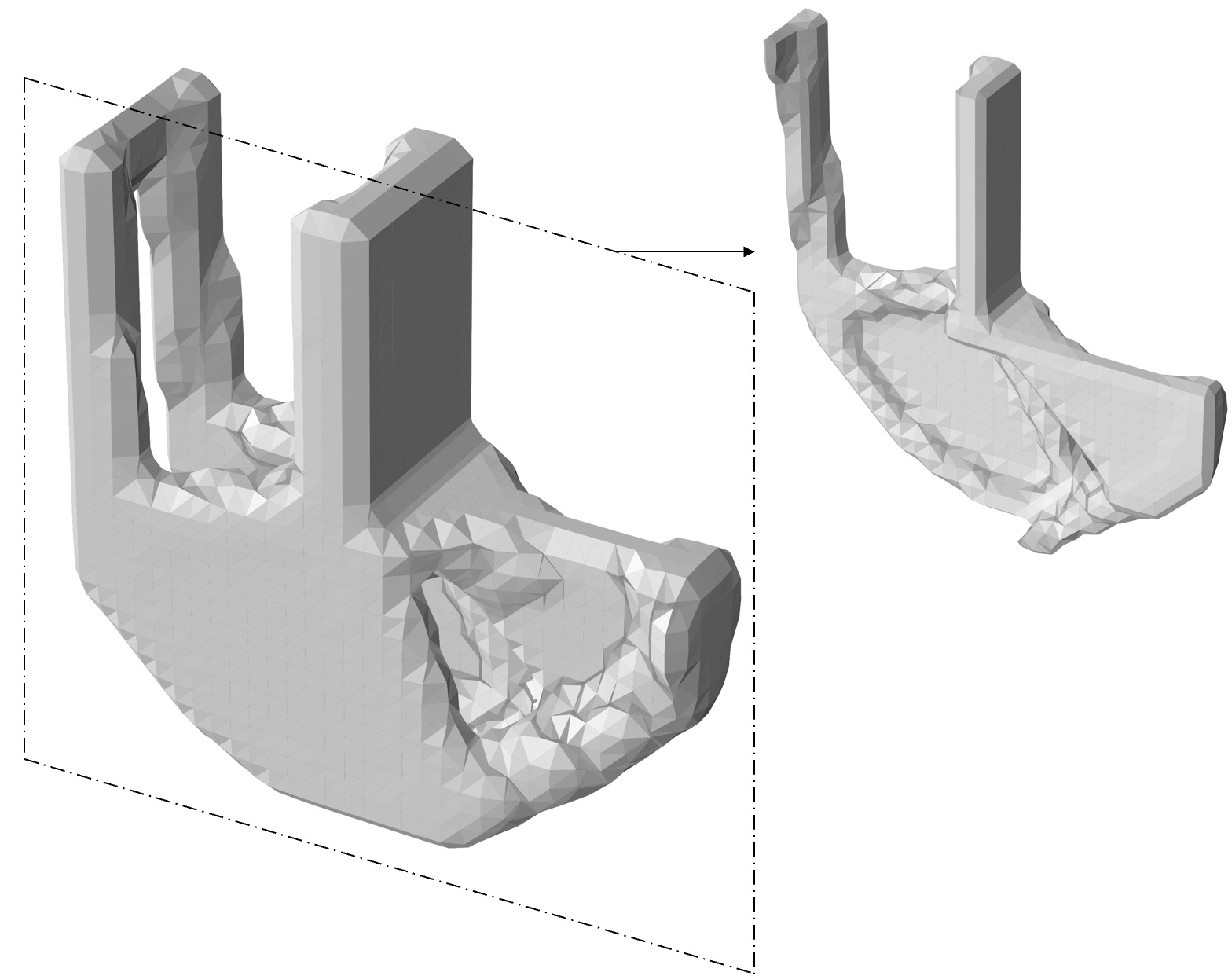}
\caption{Top3D (SIMP), [7.47, 0.20]}
\end{subfigure}\hspace{0.3cm}\begin{subfigure}[t]{0.3\textwidth}
\includegraphics[width=\textwidth]{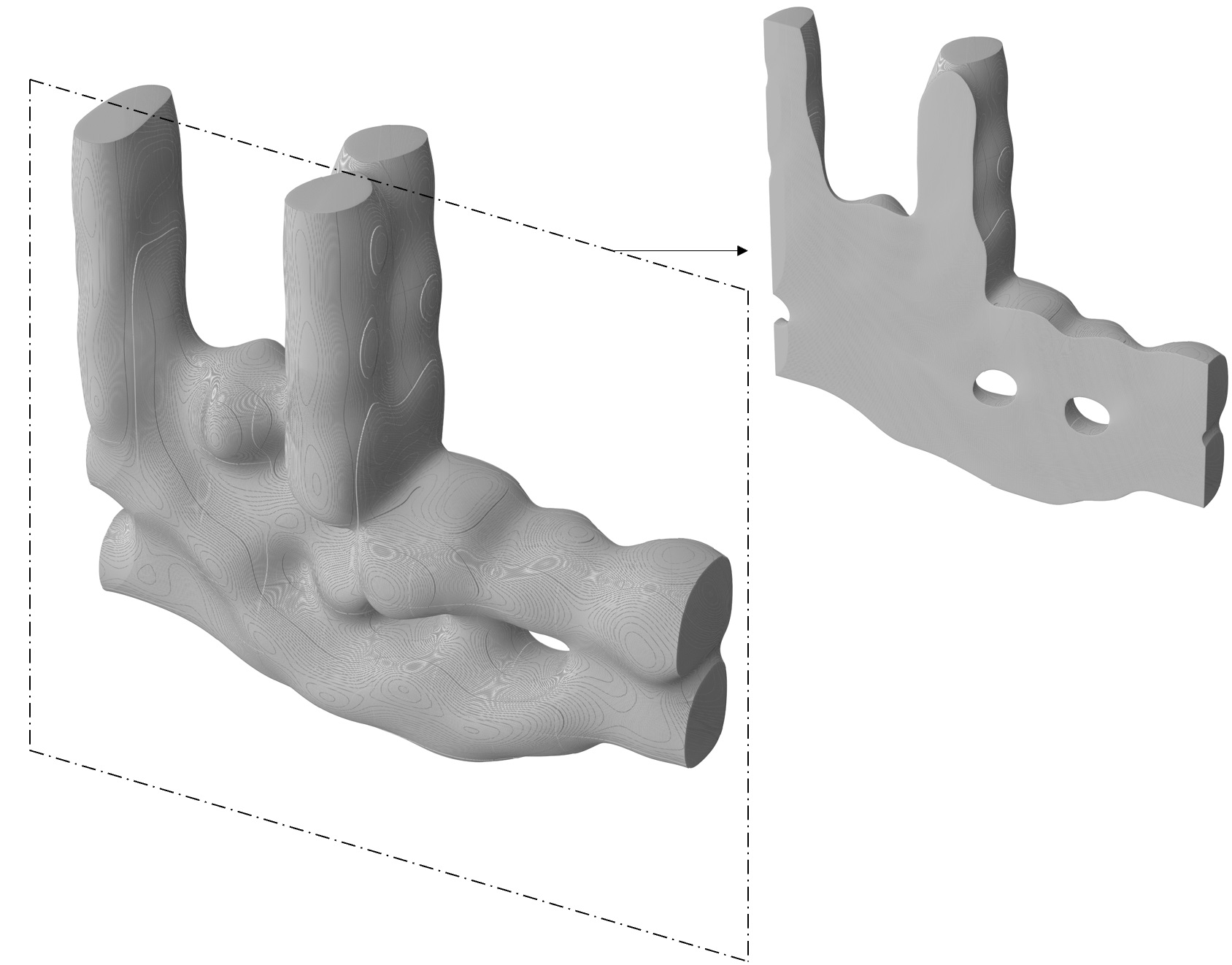}
\caption{DMF-TONN, [6.21, 0.20]}
\end{subfigure}

\caption{L-Bracket with target volume fraction = 0.2, Key: [compliance $\times 10^{-3}$, volume fraction]}

\label{fig:hook}


\centering

\begin{subfigure}[t]{0.3\textwidth}
\includegraphics[width=\textwidth]{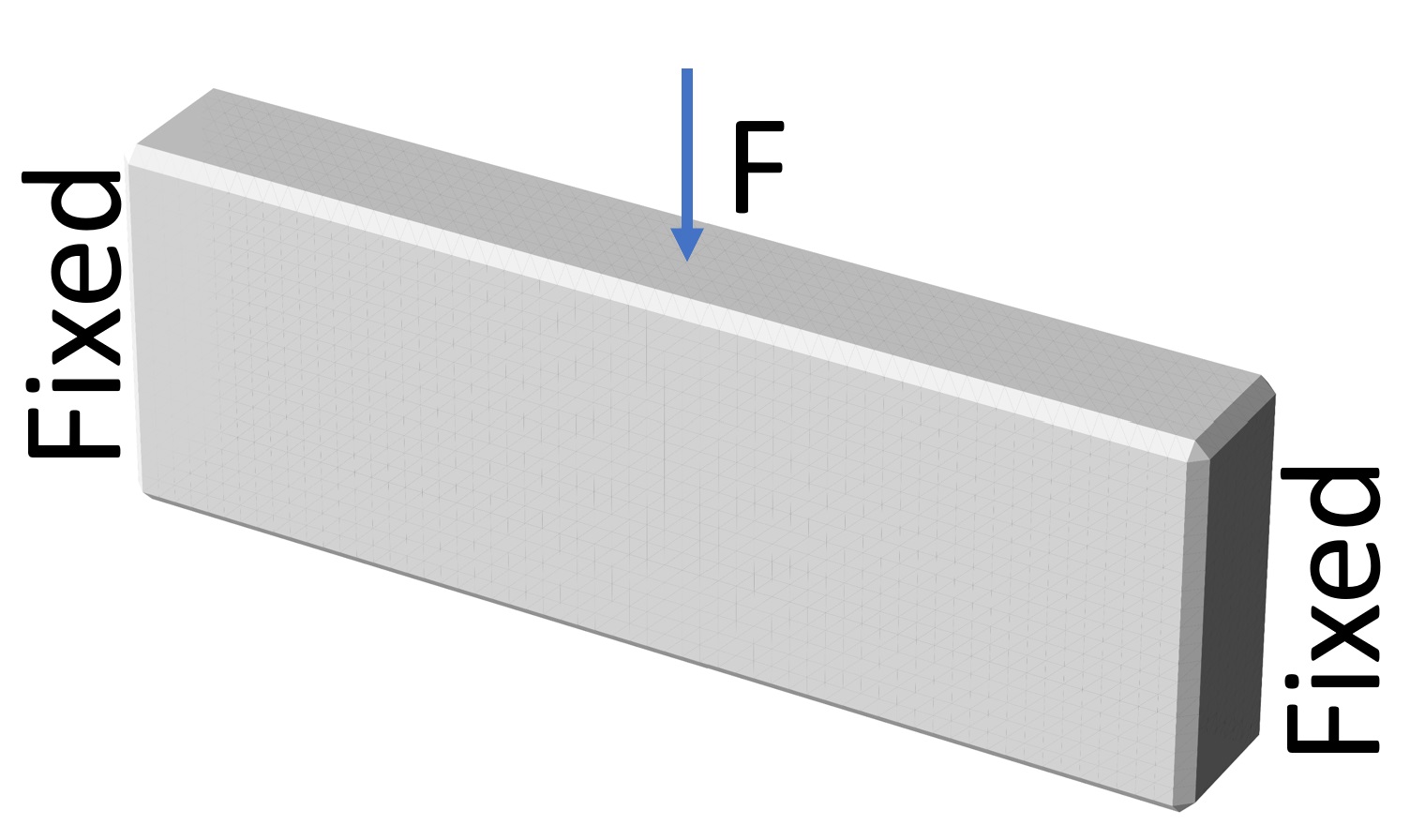}
\caption{Boundary conditions}
\end{subfigure}\hspace{0.3cm}\begin{subfigure}[t]{0.3\textwidth}
\includegraphics[width=\textwidth]{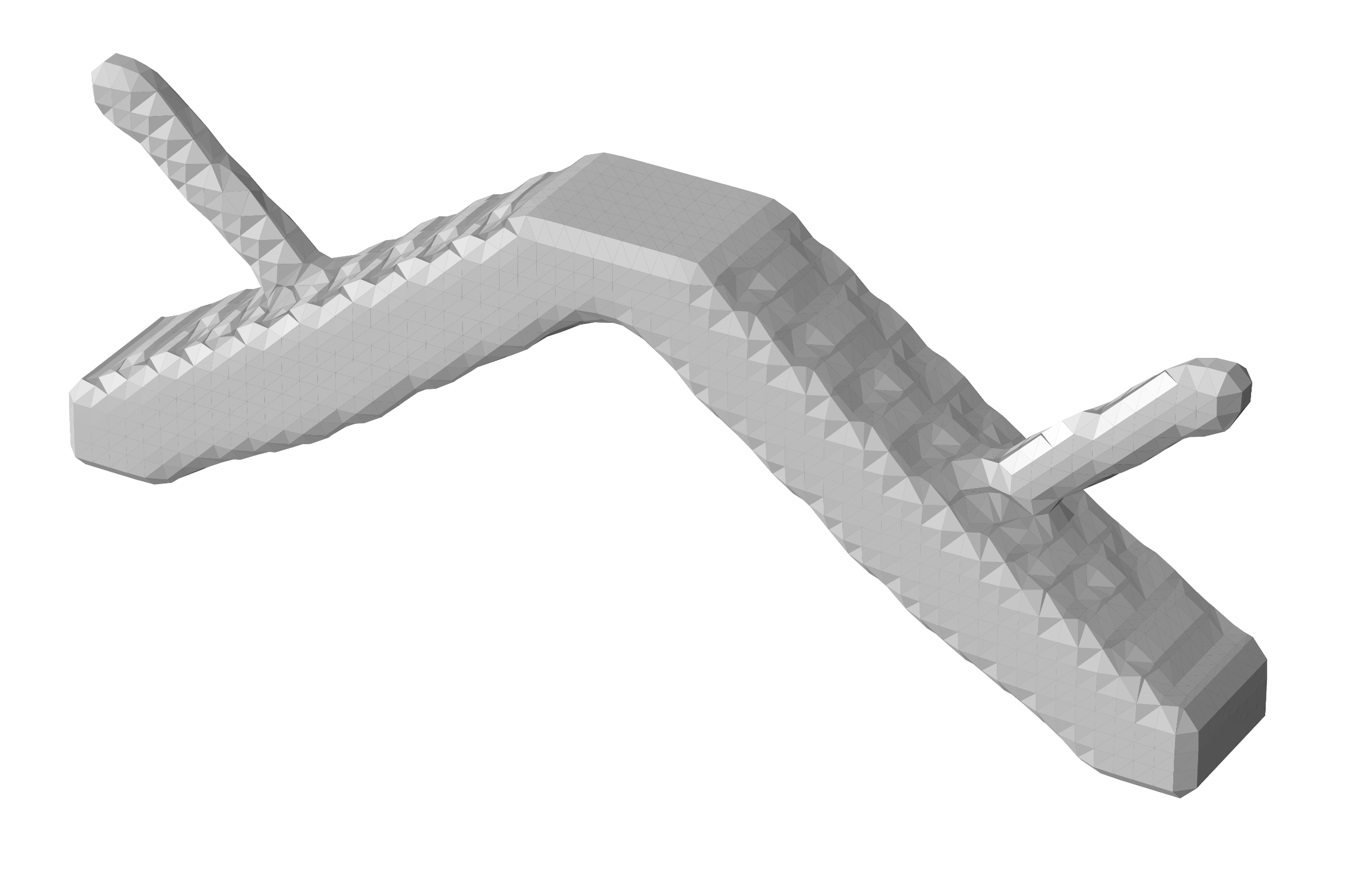}
\caption{Top3D (SIMP), [1.99, 0.30]}
\end{subfigure}\hspace{0.3cm}\begin{subfigure}[t]{0.3\textwidth}
\includegraphics[width=\textwidth]{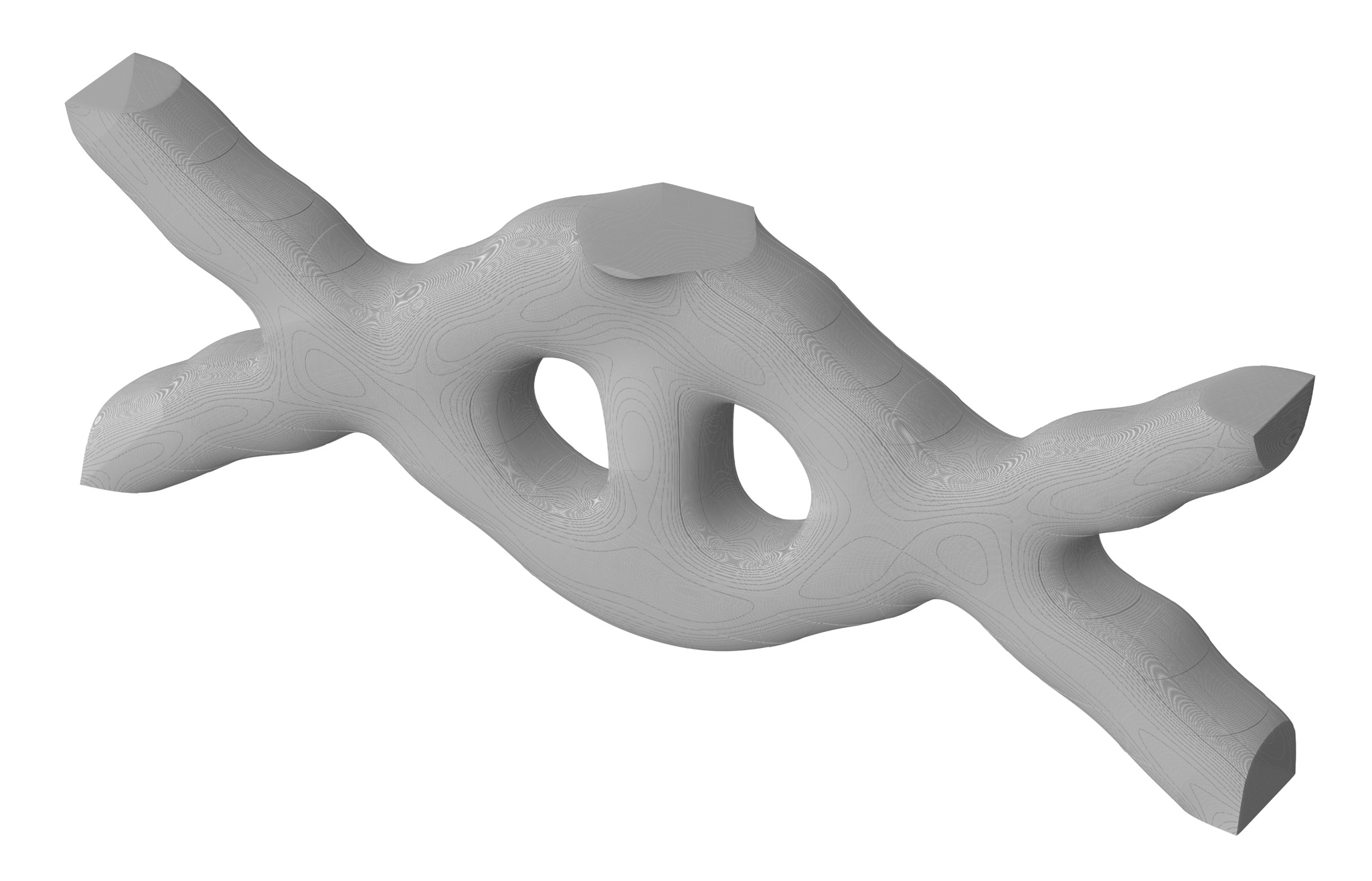}
\caption{DMF-TONN, [2.00, 0.30]}
\end{subfigure}

\caption{Bridge with target volume fraction = 0.3, Key: [compliance $\times 10^{-3}$, volume fraction]}

\label{fig:bridge}
\end{figure*}
\begin{figure*}

\centering

\begin{subfigure}[t]{0.2\textwidth}
\includegraphics[width=\textwidth]{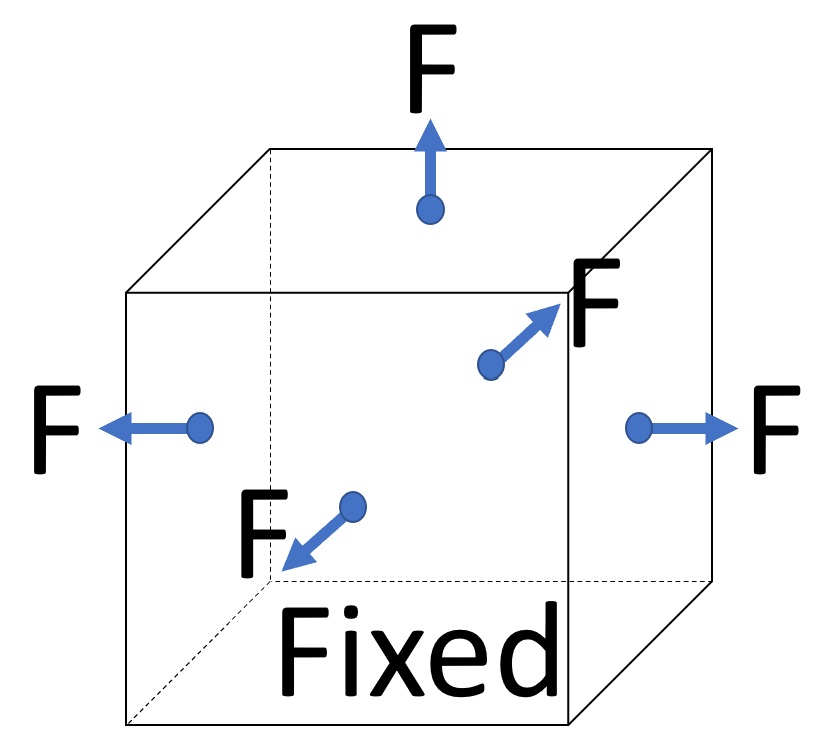}
\caption{Boundary conditions}
\end{subfigure}\hspace{0.3cm}\begin{subfigure}[t]{0.3\textwidth}
\includegraphics[width=\textwidth]{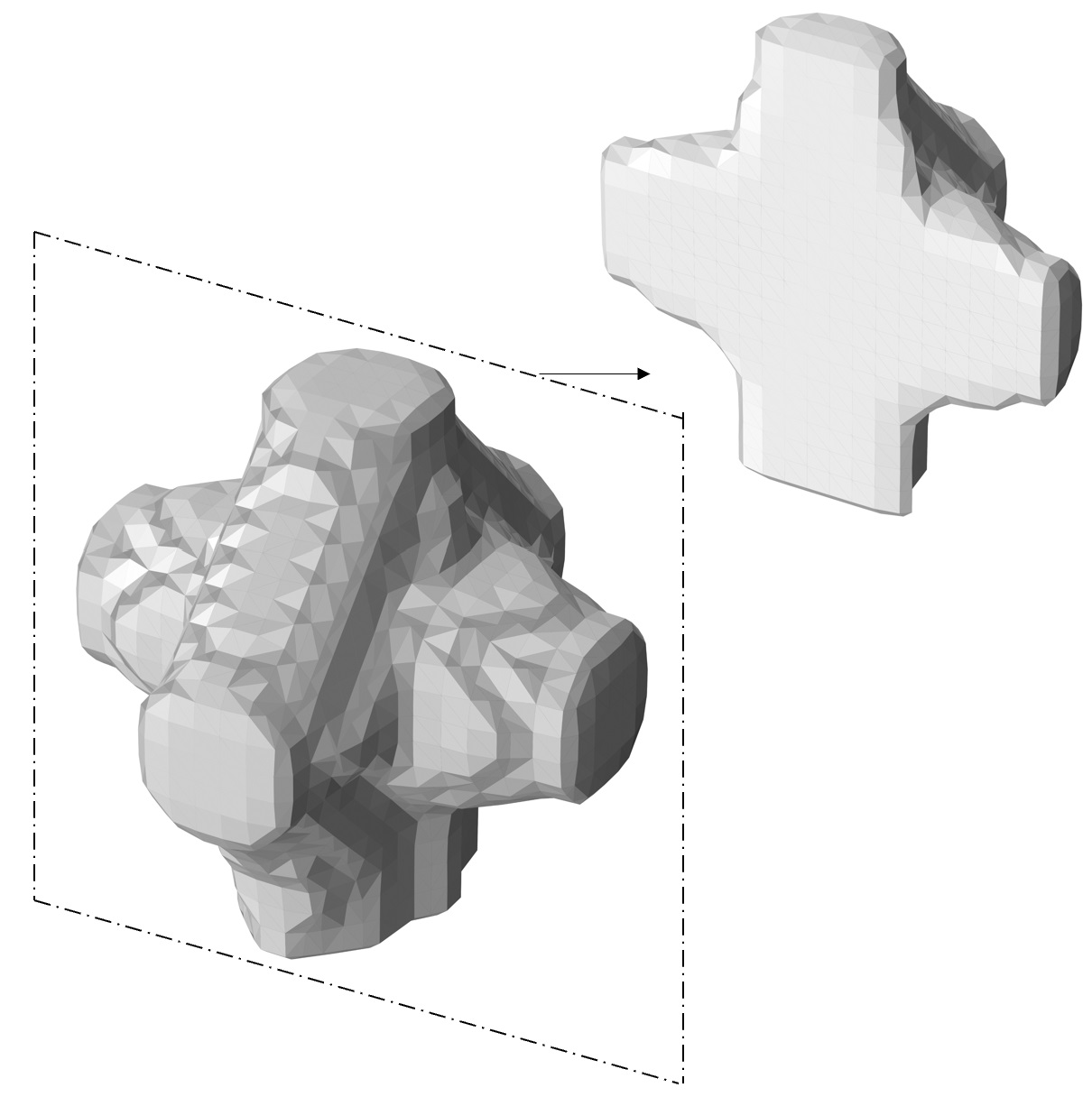}
\caption{Top3D (SIMP), [1.62, 0.30]}
\end{subfigure}\hspace{0.3cm}\begin{subfigure}[t]{0.3\textwidth}
\includegraphics[width=\textwidth]{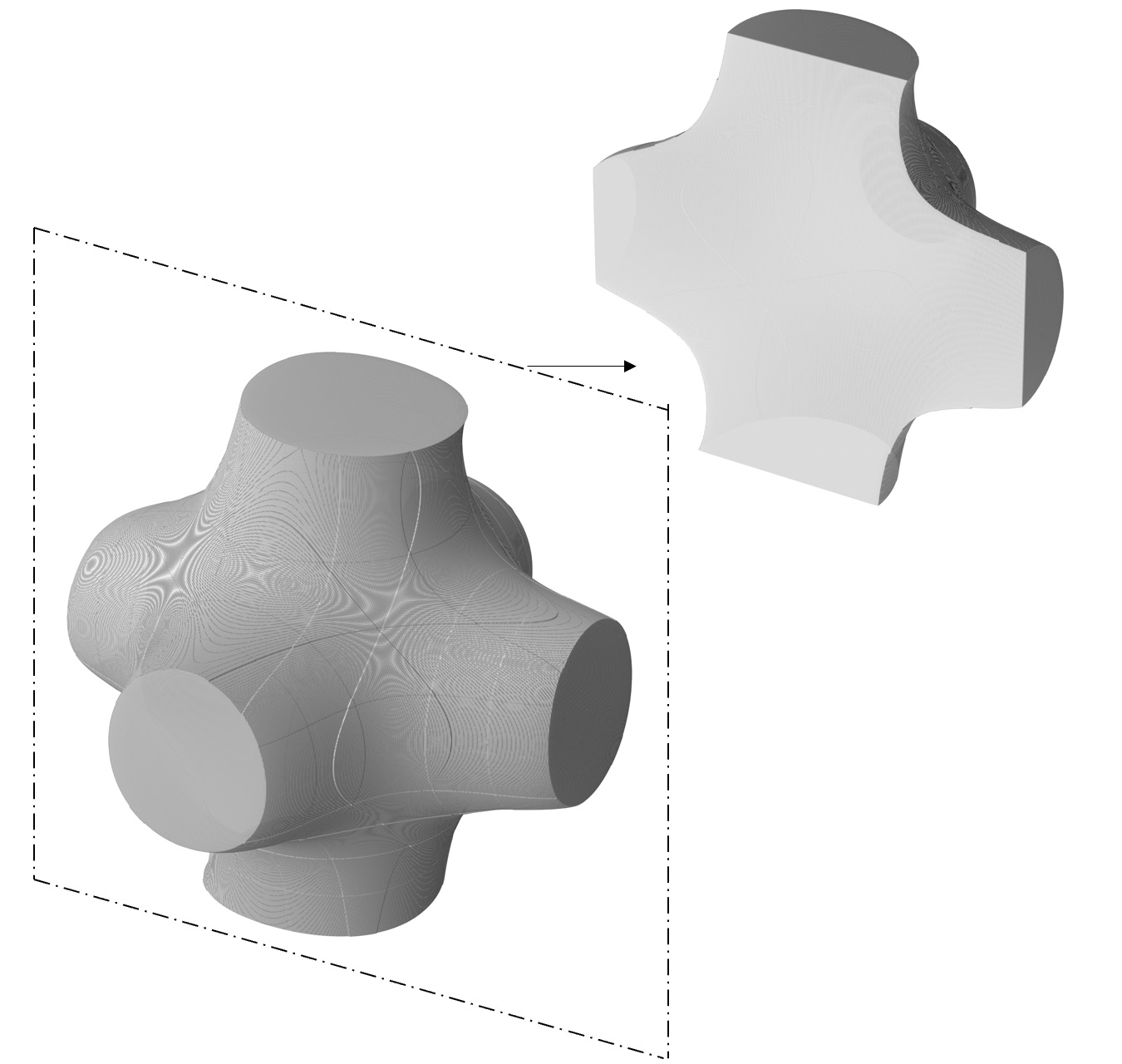}
\caption{DMF-TONN, [1.63, 0.30]}
\end{subfigure}

\caption{5 loads with target volume fraction = 0.3, Key: [compliance $\times 10^{-3}$, volume fraction]}

\label{fig:5loads}

\centering

\begin{subfigure}[t]{0.2\textwidth}
\includegraphics[width=\textwidth]{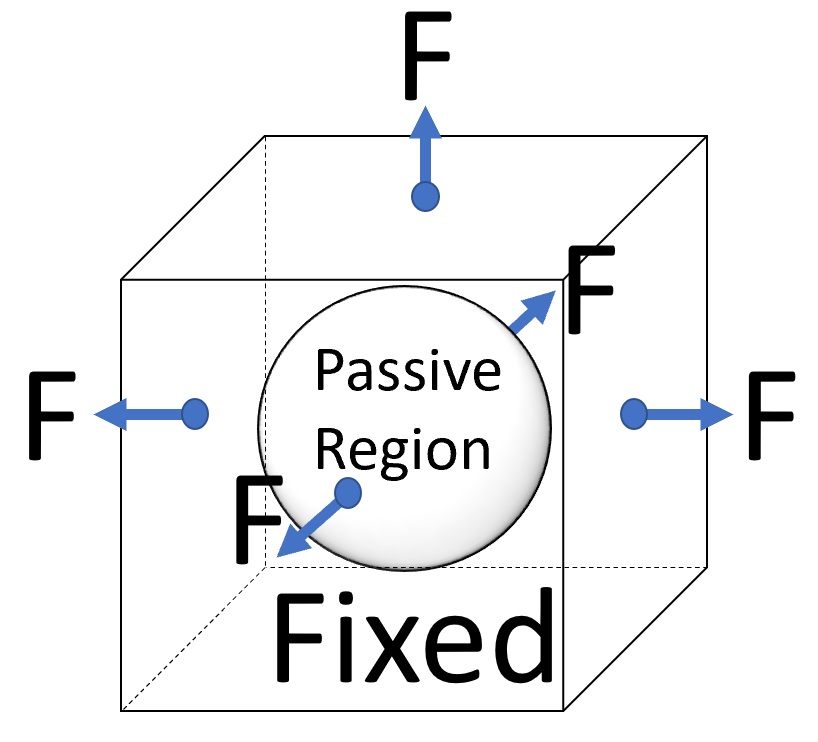}
\caption{Boundary conditions}
\end{subfigure}\hspace{0.3cm}\begin{subfigure}[t]{0.3\textwidth}
\includegraphics[width=\textwidth]{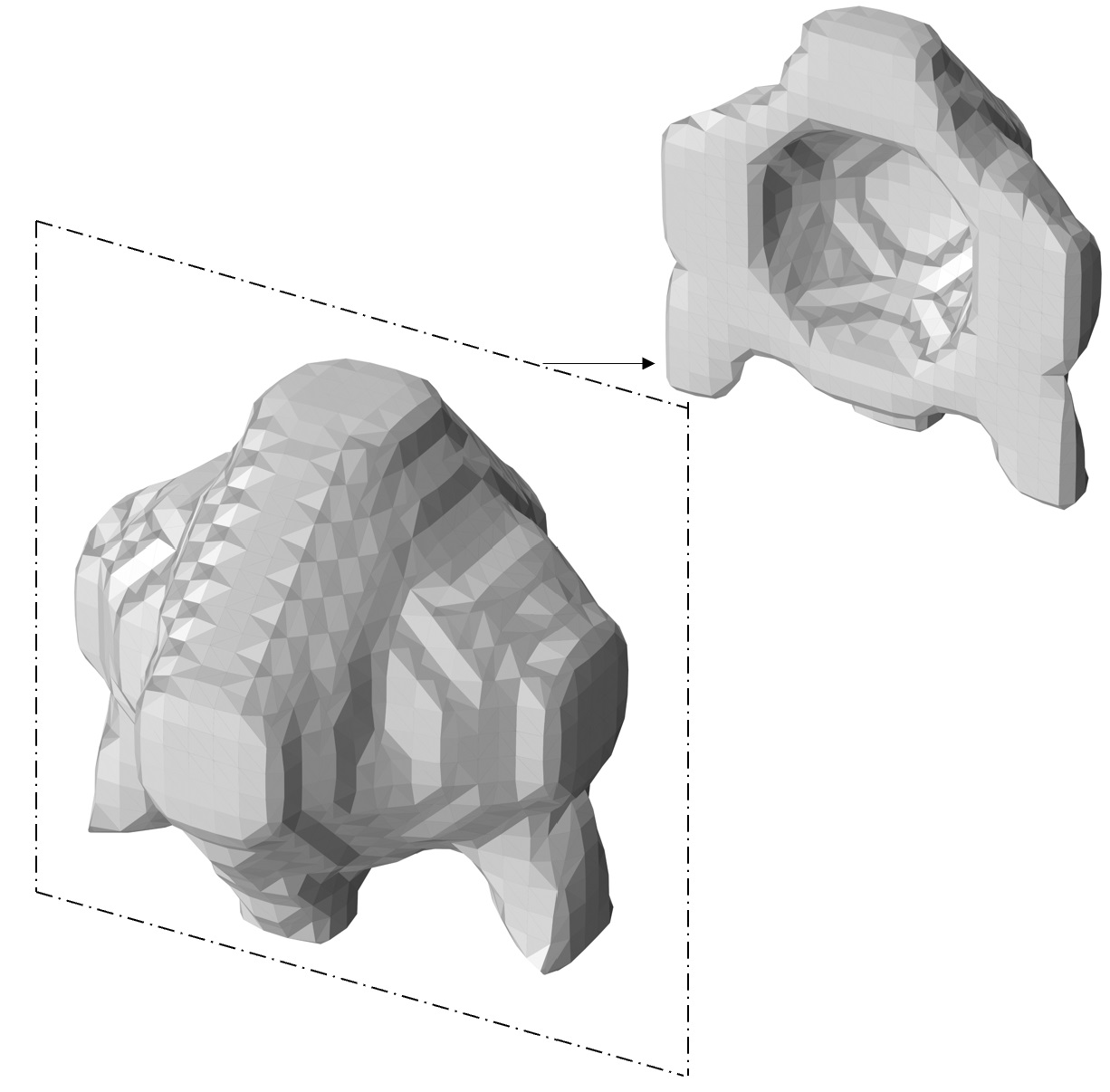}
\caption{Top3D (SIMP), [1.81, 0.30]}
\end{subfigure}\hspace{0.3cm}\begin{subfigure}[t]{0.3\textwidth}
\includegraphics[width=\textwidth]{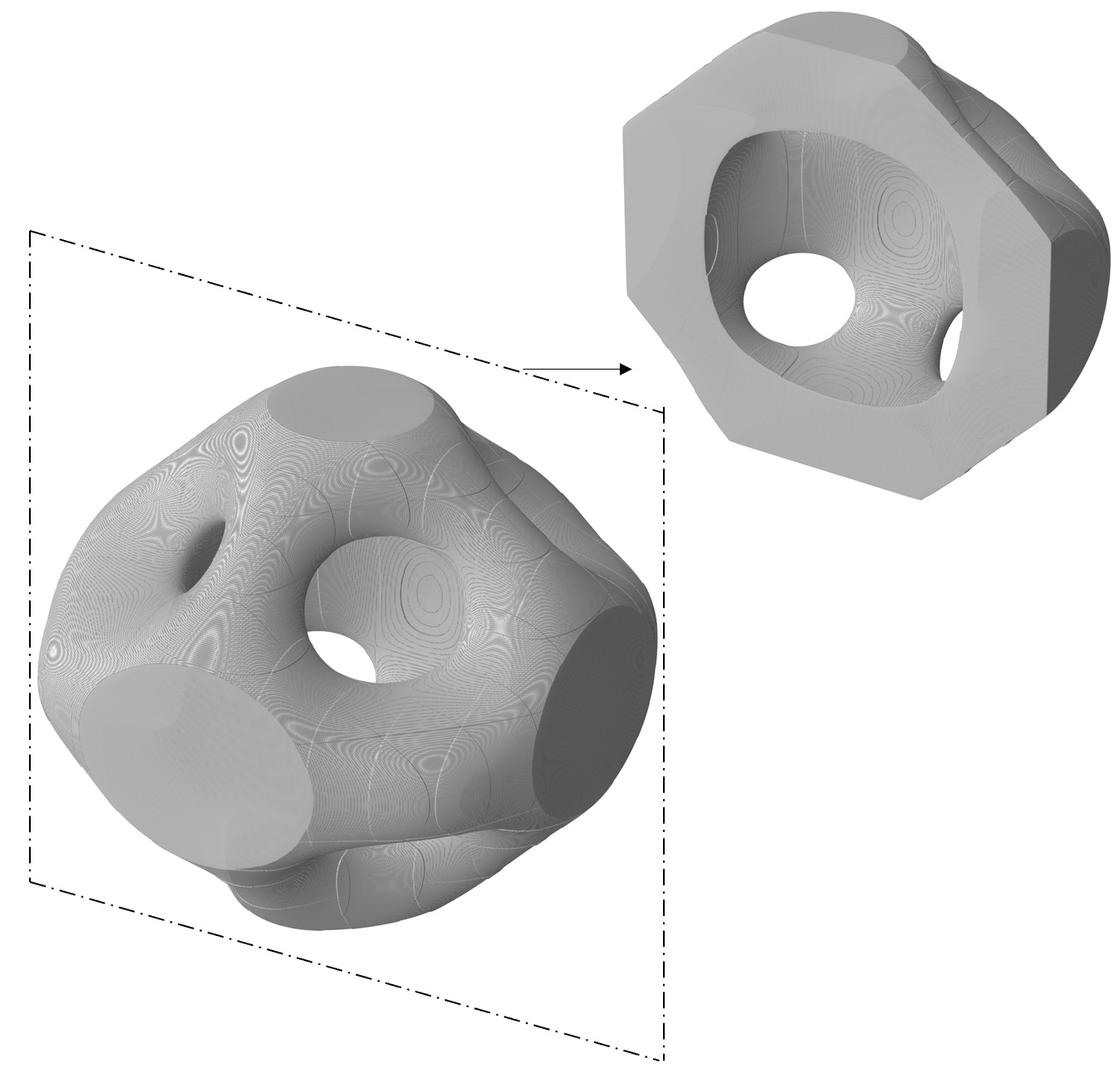}
\caption{DMF-TONN, [2.12, 0.30]}
\end{subfigure}

\caption{5 loads with passive region and target volume fraction = 0.3, Key: [compliance $\times 10^{-3}$, volume fraction]}

\label{fig:5loads passive}


\centering

\begin{subfigure}[t]{0.2\textwidth}
\includegraphics[width=\textwidth]{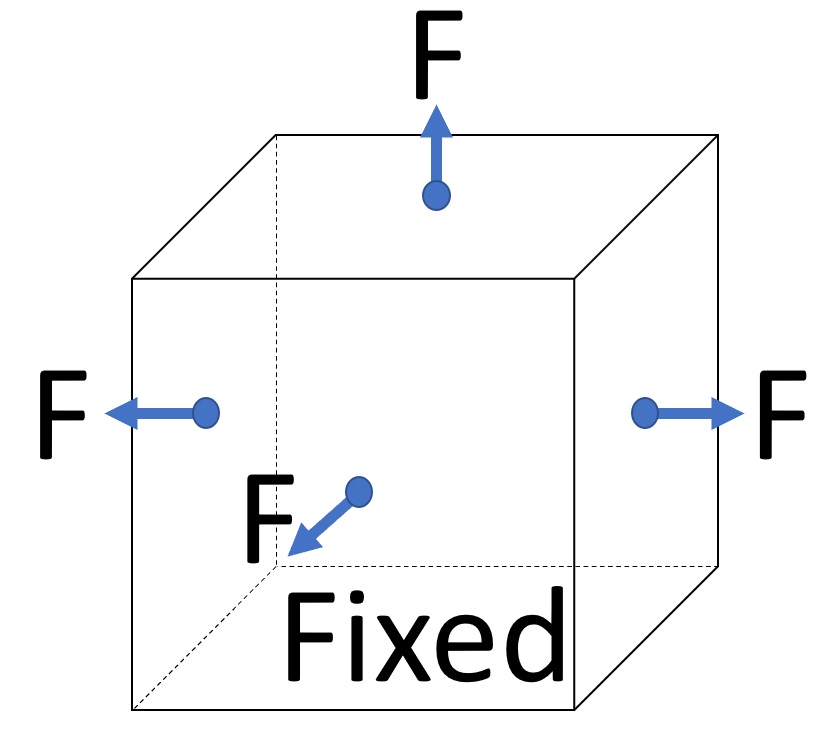}
\caption{Boundary conditions}
\end{subfigure}\hspace{0.3cm}\begin{subfigure}[t]{0.3\textwidth}
\includegraphics[width=\textwidth]{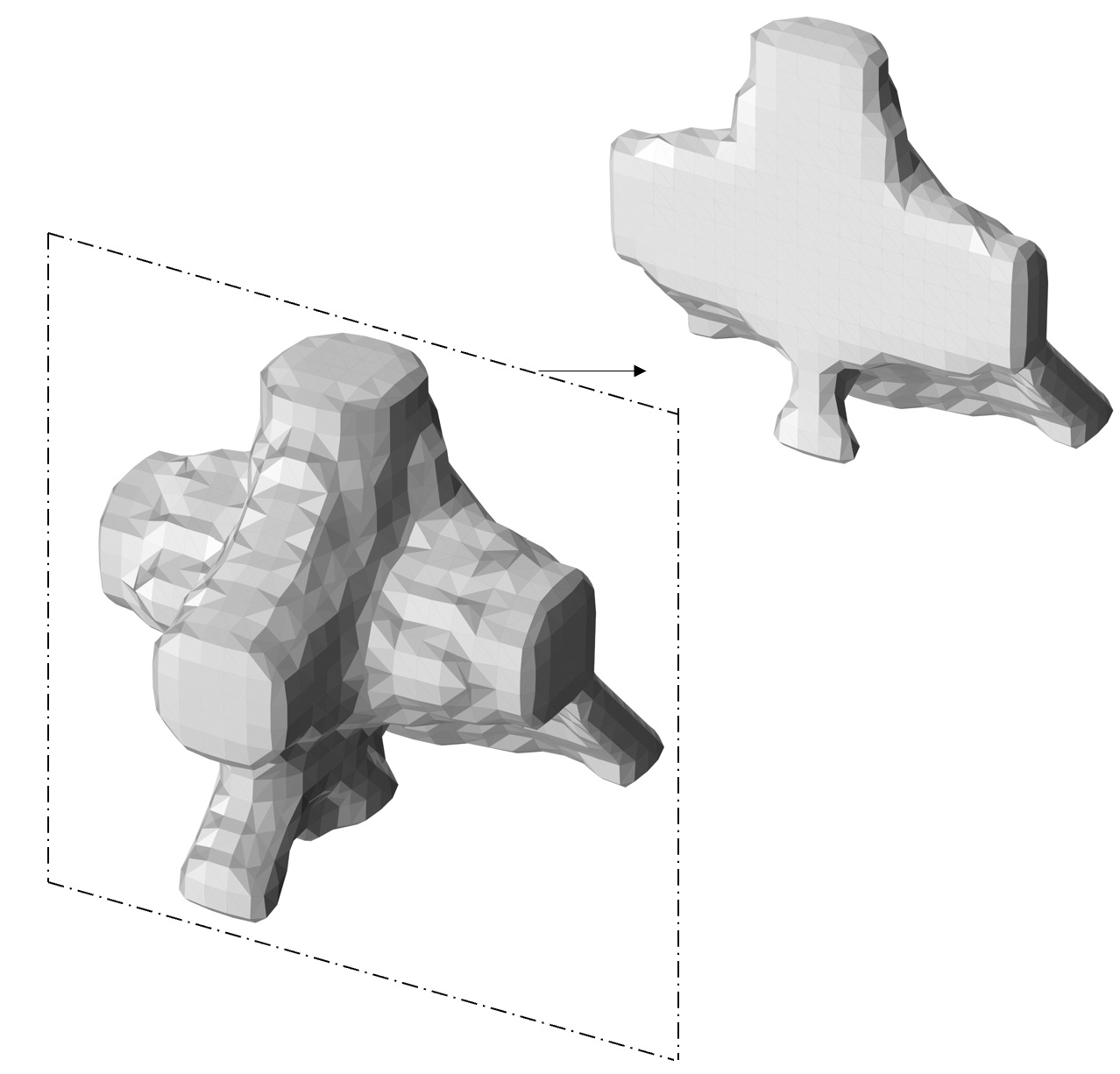}
\caption{Top3D (SIMP), [1.39, 0.30]}
\end{subfigure}\hspace{0.3cm}\begin{subfigure}[t]{0.3\textwidth}
\includegraphics[width=\textwidth]{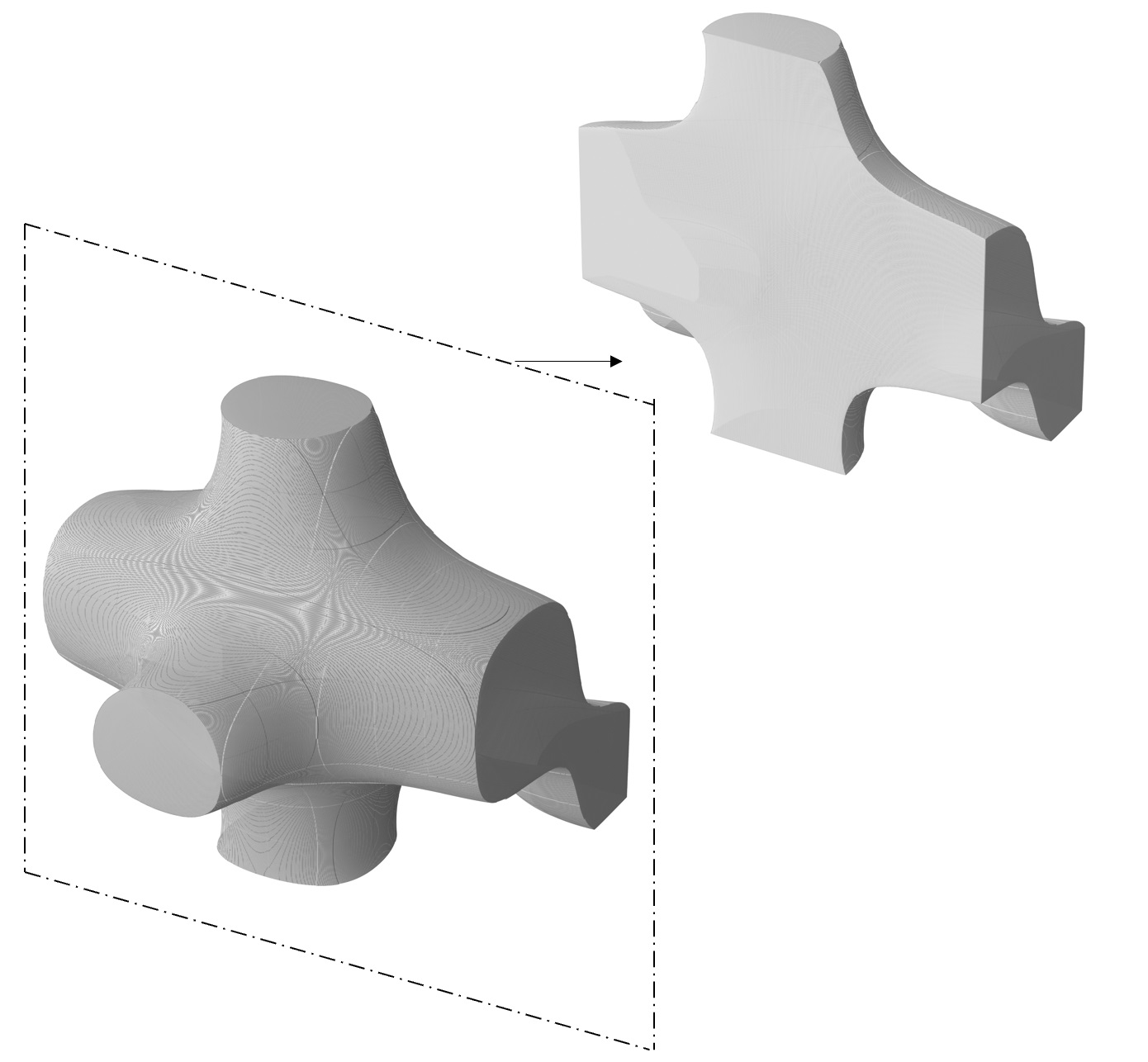}
\caption{DMF-TONN, [1.47, 0.30]}
\end{subfigure}

\caption{4 loads (unsymmetrical) with target volume fraction = 0.3, Key: [compliance $\times 10^{-3}$, volume fraction]}

\label{fig:4loads unsymm}
\end{figure*}
\begin{figure*}

\centering

\begin{subfigure}[t]{0.3\textwidth}
\includegraphics[width=\textwidth]{Images/SIMP_long_beam.jpg}
\caption{Top3D (SIMP), [2.65, 0.30, 0.540]}
\label{fig:simp_grid_dmf_tonn_long_beam_top3d}
\end{subfigure}\hspace{0.3cm}\begin{subfigure}[t]{0.3\textwidth}
\includegraphics[width=\textwidth]{Images/Long_Beam.jpg}
\caption{DMF-TONN, [2.40, 0.30, 0.739]}
\label{fig:simp_grid_dmf_tonn_long_beam_previous}
\end{subfigure}\hspace{0.3cm}\begin{subfigure}[t]{0.3\textwidth}
\includegraphics[width=\textwidth]{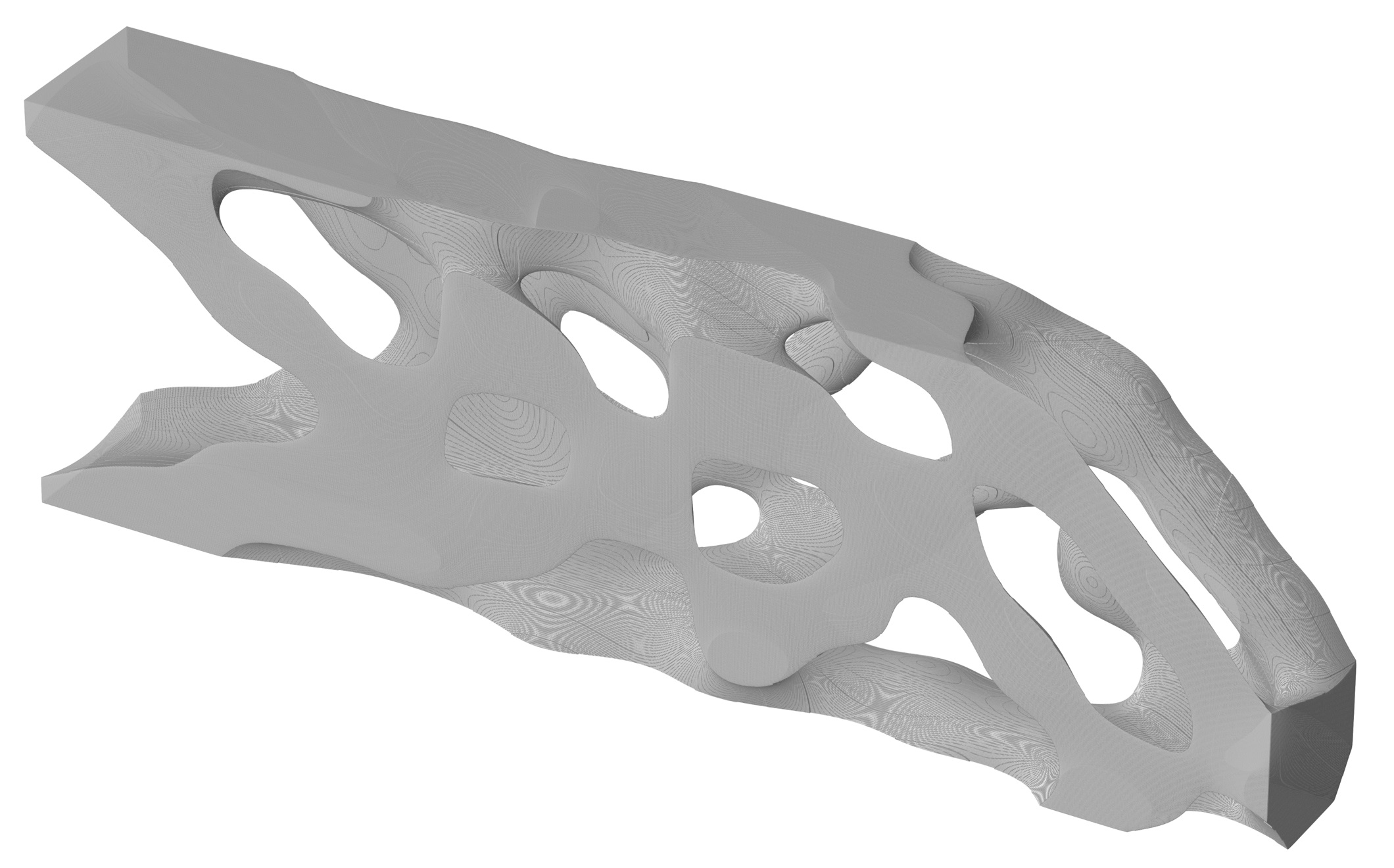}
\caption{DMF-TONN with SIMP grid, [2.34, 0.30, 1.131]}
\label{fig:simp_grid_dmf_tonn_long_beam}
\end{subfigure}

\caption{Long cantilever beam with target volume fraction = 0.3, Key: [compliance $\times 10^{-2}$, volume fraction, seconds per topology optimization iteration]}

\label{fig:lb_simp_grid}

\centering

\begin{subfigure}[t]{0.3\textwidth}
\includegraphics[width=\textwidth]{Images/SIMP_torque.jpg}
\caption{Top3D (SIMP), [8.22, 0.50, 1.325]}
\end{subfigure}\hspace{0.3cm}\begin{subfigure}[t]{0.3\textwidth}
\includegraphics[width=\textwidth]{Images/Torque.jpg}
\caption{DMF-TONN, [8.78, 0.50, 0.744]}
\end{subfigure}\hspace{0.3cm}\begin{subfigure}[t]{0.3\textwidth}
\includegraphics[width=\textwidth]{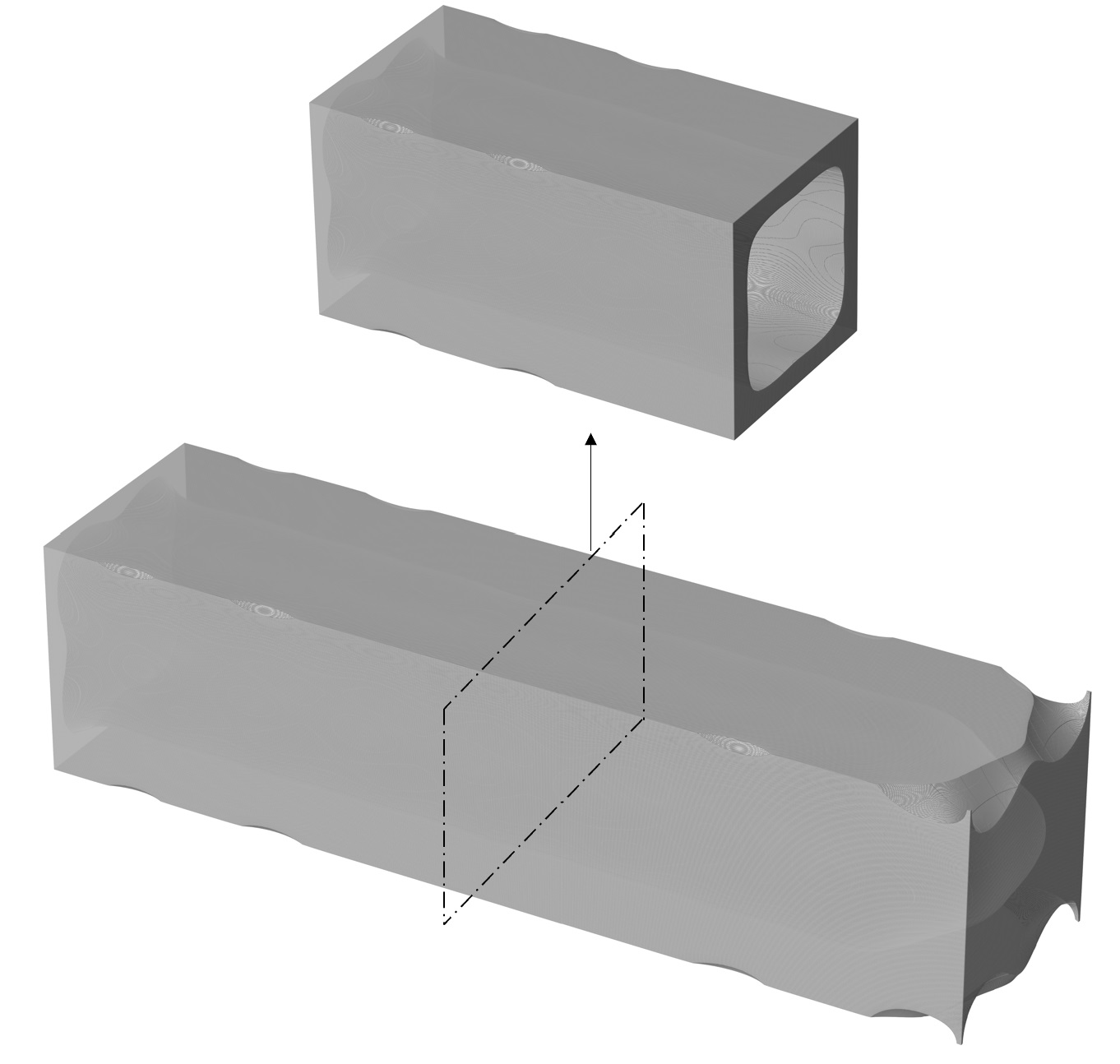}
\caption{DMF-TONN with SIMP grid, [7.77, 0.50, 1.543]}
\label{fig:simp_grid_dmf_tonn_long_beam_2loads}
\end{subfigure}

\caption{Long cantilever beam with two loads with target volume fraction = 0.5, Key: [compliance $\times 10^{-3}$, volume fraction, seconds per topology optimization iteration]}

\label{fig:lb_2loads_simp_grid}

\centering

\begin{subfigure}[t]{0.3\textwidth}
\includegraphics[width=\textwidth]{Images/SIMP_3D_CB_PINN_Iter_demo.jpg}
\caption{Top3D (SIMP), [7.49, 0.30, 0.324]}
\end{subfigure}\hspace{0.3cm}\begin{subfigure}[t]{0.3\textwidth}
\includegraphics[width=\textwidth]{Images/xPhys_Iter700_iso.jpg}
\caption{DMF-TONN, [6.76, 0.30, 0.840]}
\end{subfigure}\hspace{0.3cm}\begin{subfigure}[t]{0.3\textwidth}
\includegraphics[width=\textwidth]{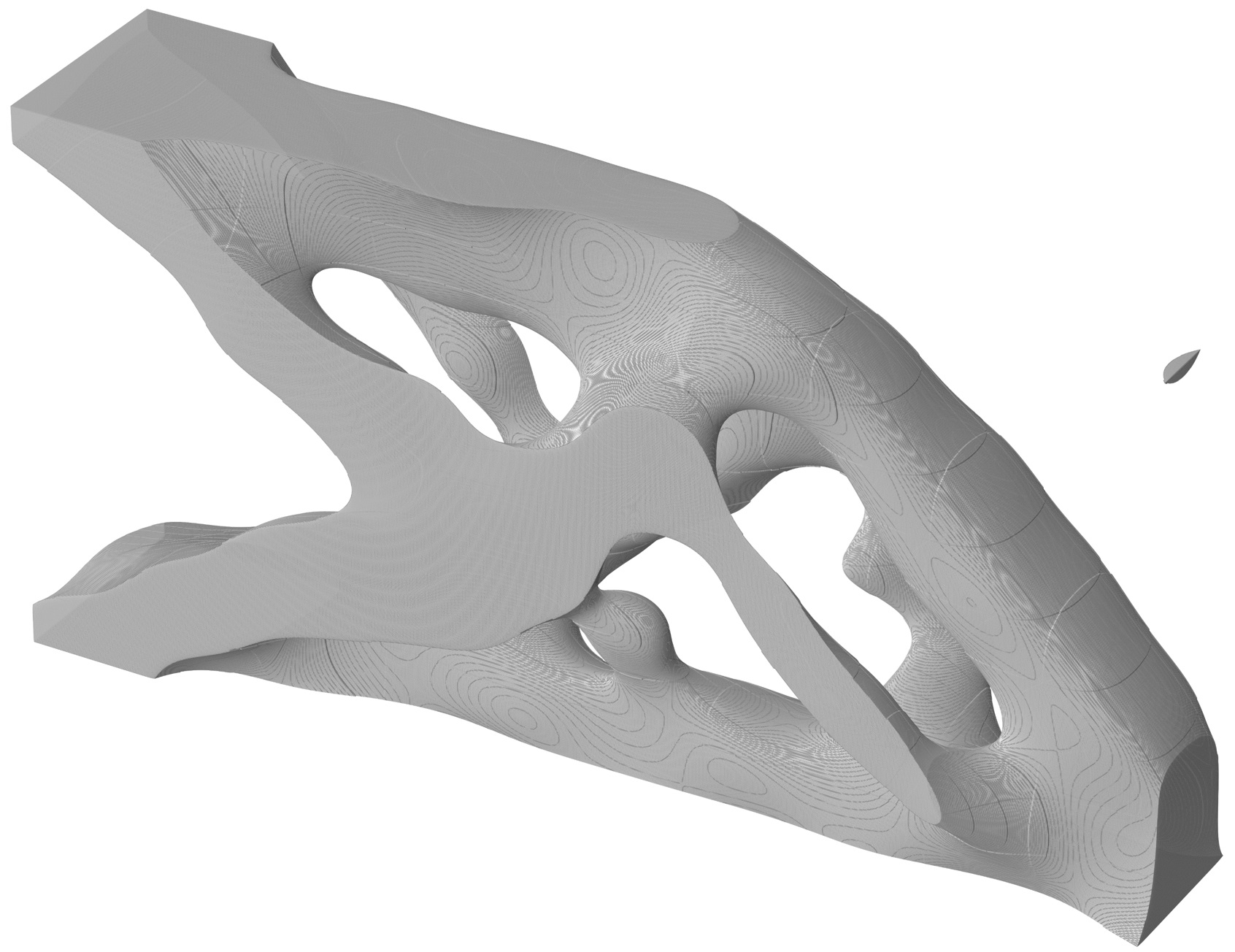}
\caption{DMF-TONN with SIMP grid, [6.57, 0.30, 0.901]}
\label{fig:simp_grid_dmf_tonn_beam}
\end{subfigure}

\caption{Cantilever beam with target volume fraction = 0.3, Key: [compliance $\times 10^{-3}$, volume fraction, seconds per topology optimization iteration]}

\label{fig:beam_simp_grid}

\end{figure*}

\subsection{Additional Examples}
In Figure \ref{fig:lbcenter}, the load acts at the center of the right side of a long cantilever beam. The compliance values obtained by DMF-TONN for these boundary conditions are better than those obtained by Top3D (SIMP) ($60\times20\times8$ grid). In Figure \ref{fig:torque}, the boundary conditions include two loads twisting a beam fixed on its one side. The ideal topology should be a hollowed-out beam, and DMF-TONN correctly outputs a similar topology. For this example, we observe that the compliance of the SIMP ($60\times15\times15$ grid) result is better than that of DMF-TONN. In Figure \ref{fig:hook} we present a case with a  passive (non-design) region is present in the upper right quarter of the domain. This condition is enforced using an additional constraint violation objective for the density network, where if the density network outputs density values close to 1 in the passive region, the loss is penalized. For this L-Bracket example, the compliance of the result obtained by DMF-TONN is better than that obtained by SIMP ($30\times30\times10$ grid). In Figure \ref{fig:bridge}, both left and right sides are fixed and the load acts at the center of the beam. The compliances of the results obtained by SIMP ($60\times20\times8$ grid) and DMF-TONN are similar for this example. In Figure \ref{fig:5loads}, the boundary conditions consist of a cube fixed at the bottom and a load acting at the center of each of its remaining sides. In Figure \ref{fig:5loads passive}, there is a passive region (sphere) at the center of the cube, and hence the topology has to form around this sphere. In Figure \ref{fig:4loads unsymm}, unlike Figure \ref{fig:5loads}, the loading is unsymmetric, i.e there are 4 loads and one side does not have a load acting on it. For the boundary conditions in Figures \ref{fig:5loads}, \ref{fig:5loads passive} and \ref{fig:4loads unsymm}, SIMP \ch{($20\times20\times20$ grid)} achieves better compliance than DMF-TONN. Internal analysis has shown that DMF-TONN relies on several hyperparameters, including the batch-size, the kernel range and size and the learning rate. Finding the optimum values of these hyperparameters, such that DMF-TONN always (for all boundary conditions and target volume fractions) gives better compliance than SIMP, is something we do not study exhaustively here, and is a topic for future work.

We use a Young's Modulus (E) $= 1000 N/mm^2$, Poisson's ratio of 0.3, Force $= 0.1N$ and we normalize the domain with the longest side $= 1 mm$ for all examples. We use 6000 randomly sampled domain points as the batch in each iteration \ch{of the displacement network (making sure the force vector location points and fixed boundary condition points are included) and during each iteration of density network for compliance calculation. For the volume fraction constraint violation calculation, we use a constant grid same as the SIMP method grid used in each example (to avoid fluctuations in volume fraction).} We use a kernel grid  size of 16 in all three directions and set upper and lower bounds of the kernel values to 35 and -35 respectively. We find this hyper-parameters setting works well for most problems, unless when the domain has a markedly skewed size ratio, such as in the long beam and bridge examples. In those cases, one has to accordingly adjust the kernel size in the different directions. In these examples, we use 24 kernel grid points in the longest direction, and 12 kernel grid points in the other directions, and an upper and lower bounds of the kernel values of 45 and -45 respectively, for the results presented. For the examples in Figures \ref{fig:5loads}, \ref{fig:5loads passive} and \ref{fig:4loads unsymm}, we use a smaller value of lower and upper bound (10) of kernel values for the results presented, as higher values tend to give unnecessary artifacts in the final topology. The computational time for DMF-TONN for each of the presented examples is less than 800 seconds. For each of the presented examples, the degrees of freedom  are always more for SIMP than for DMF-TONN.

\ch{\subsection{Using same grid points as SIMP for DMF-TONN}
We showcase a few examples using the same grid points as SIMP for input to DMF-TONN instead of domain points that are randomly sampled in each displacement network iteration and density network iteration for compliance calculation using the trained displacement network. For the long cantilever beam example (Figure \ref{fig:lb_simp_grid}), where the SIMP grid has 9600 points, we get a better compliance (Figure \ref{fig:simp_grid_dmf_tonn_long_beam}) than when using the randomly sampled points for DMF-TONN with batch size equal to 6000 for compliance calculation (as used in all previous examples up till now) (Figure \ref{fig:simp_grid_dmf_tonn_long_beam_previous}). Note that the randomly sampled DMF-TONN compliance is already better than the SIMP compliance (Figure \ref{fig:simp_grid_dmf_tonn_long_beam_top3d}). For the long cantilever beam with two loads example (Figure \ref{fig:lb_2loads_simp_grid}), the DMF-TONN result had worse compliance than SIMP, but now after using the SIMP grid in DMF-TONN, it achieves a far better compliance than SIMP. This better compliance when using the same grid points as SIMP in DMF-TONN can be attributed to the increased number of points (9600 instead of 6000 in the long cantilever beam problem and 13500 instead of 6000 in long cantilever beam with two loads problem). However, it should be noted as the batch size of DMF-TONN now increased, the computation time increased by 205s and 571s respectively for the two examples. Also, sometimes we do not get a better result with a SIMP grid even if the number of points is higher, as seen in the cantilever beam example (Figure \ref{fig:beam_simp_grid}), which has better compliance than DMF-TONN but a dangling feature is present. We can attribute this dangling feature to the fact that random sampling in each topology optimization iteration as well as displacement network iteration is not present for the compliance calculation and all the hyperparameters of DMF-TONN (such as the neural network kernel size and frequency) are tuned for random sampling with a batch size of 6000.

In each of the examples presented in this paper, we use a SIMP grid that always has a higher number of points than the DMF-TONN batch size of 6000, to satisfy the ratio of lengths of sides while making sure SIMP never has any disadvantage due to lower number of points. From the preliminary analysis presented in this section we conclude that using the SIMP grid as input to DMF-TONN, we may get a better result than the original DMF-TONN in terms of compliance, but there is an increase in the computation time and there may be issues such as dangling features. The domain coordinate sampling method used in DMF-TONN plays a crucial role in the results and detailed analysis and experimentation of different sampling methods is a good direction for future work.}

\section{Conclusion}\label{sec5}
We show that using directly connected
displacement field estimation and density field estimation neural networks is indeed an effective approach for mesh-free topology optimization. We verify through various examples that DMF-TONN, which involves using just one gradient descent step of the density network in each topology optimization epoch without any sensitivity filtering or density filtering leads to comparable results to conventional topology optimization. We significantly reduce the computational time compared to prior related works and also explore the trade-offs between DMF-TONN and SIMP for a cantilever beam example, showcasing the advantage of the mesh-free nature of DMF-TONN.

There are several limitations observed currently with DMF-TONN. The first one concerns the kernels used in the density and displacement neural network. The kernel grid has to be scaled according to the domain size if there exist sides with ratio of lengths greater than or equal to 3. Moreover, we observed that for low target volume fractions (less than 0.2), the kernel is not able to capture the required features and the optimization does not converge. Also, there are multiple hyperparameters involved in DMF-TONN, and an exhaustive study for finding their optimum values is a topic for future work. 

\ch{To remove the meshing process in the complex inverse design problem of topology optimization, we devise DMF-TONN and show that it works well for various 3D problems and opens up avenues taking advantage of the mesh-free nature for seamless integration with post-processing software, or performing optimizations where FEA may be unsuitable.} Future work involves improving the robustness of the kernel, extending the approach to complex problems, \ch{comparisons with state-of-the-art irregular grid topology optimization software}, and experimenting with and analyzing the effect and advantages of different domain coordinate sampling methods.

\section{Funding and Competing interests}
No funding was received to assist with the preparation of this manuscript. The authors have no competing interests to declare that are relevant to the content of this article.


\begin{thebibliography}{35}
\ifx \bisbn   \undefined \def \bisbn  #1{ISBN #1}\fi
\ifx \binits  \undefined \def \binits#1{#1}\fi
\ifx \bauthor  \undefined \def \bauthor#1{#1}\fi
\ifx \batitle  \undefined \def \batitle#1{#1}\fi
\ifx \bjtitle  \undefined \def \bjtitle#1{#1}\fi
\ifx \bvolume  \undefined \def \bvolume#1{\textbf{#1}}\fi
\ifx \byear  \undefined \def \byear#1{#1}\fi
\ifx \bissue  \undefined \def \bissue#1{#1}\fi
\ifx \bfpage  \undefined \def \bfpage#1{#1}\fi
\ifx \blpage  \undefined \def \blpage #1{#1}\fi
\ifx \burl  \undefined \def \burl#1{\textsf{#1}}\fi
\ifx \doiurl  \undefined \def \doiurl#1{\url{https://doi.org/#1}}\fi
\ifx \betal  \undefined \def \betal{\textit{et al.}}\fi
\ifx \binstitute  \undefined \def \binstitute#1{#1}\fi
\ifx \binstitutionaled  \undefined \def \binstitutionaled#1{#1}\fi
\ifx \bctitle  \undefined \def \bctitle#1{#1}\fi
\ifx \beditor  \undefined \def \beditor#1{#1}\fi
\ifx \bpublisher  \undefined \def \bpublisher#1{#1}\fi
\ifx \bbtitle  \undefined \def \bbtitle#1{#1}\fi
\ifx \bedition  \undefined \def \bedition#1{#1}\fi
\ifx \bseriesno  \undefined \def \bseriesno#1{#1}\fi
\ifx \blocation  \undefined \def \blocation#1{#1}\fi
\ifx \bsertitle  \undefined \def \bsertitle#1{#1}\fi
\ifx \bsnm \undefined \def \bsnm#1{#1}\fi
\ifx \bsuffix \undefined \def \bsuffix#1{#1}\fi
\ifx \bparticle \undefined \def \bparticle#1{#1}\fi
\ifx \barticle \undefined \def \barticle#1{#1}\fi
\bibcommenthead
\ifx \bconfdate \undefined \def \bconfdate #1{#1}\fi
\ifx \botherref \undefined \def \botherref #1{#1}\fi
\ifx \url \undefined \def \url#1{\textsf{#1}}\fi
\ifx \bchapter \undefined \def \bchapter#1{#1}\fi
\ifx \bbook \undefined \def \bbook#1{#1}\fi
\ifx \bcomment \undefined \def \bcomment#1{#1}\fi
\ifx \oauthor \undefined \def \oauthor#1{#1}\fi
\ifx \citeauthoryear \undefined \def \citeauthoryear#1{#1}\fi
\ifx \endbibitem  \undefined \def \endbibitem {}\fi
\ifx \bconflocation  \undefined \def \bconflocation#1{#1}\fi
\ifx \arxivurl  \undefined \def \arxivurl#1{\textsf{#1}}\fi
\csname PreBibitemsHook\endcsname

\bibitem[\protect\citeauthoryear{Bends{\o}e}{1989}]{bendsoe1989optimal}
\begin{barticle}
\bauthor{\bsnm{Bends{\o}e}, \binits{M.P.}}:
\batitle{Optimal shape design as a material distribution problem}.
\bjtitle{Structural optimization}
\bvolume{1},
\bfpage{193}--\blpage{202}
(\byear{1989})
\end{barticle}
\endbibitem

\bibitem[\protect\citeauthoryear{Zhou and Rozvany}{1991}]{zhou1991coc}
\begin{barticle}
\bauthor{\bsnm{Zhou}, \binits{M.}},
\bauthor{\bsnm{Rozvany}, \binits{G.}}:
\batitle{The coc algorithm, part ii: Topological, geometrical and generalized shape optimization}.
\bjtitle{Computer methods in applied mechanics and engineering}
\bvolume{89}(\bissue{1-3}),
\bfpage{309}--\blpage{336}
(\byear{1991})
\end{barticle}
\endbibitem

\bibitem[\protect\citeauthoryear{Chandrasekhar and Suresh}{2021}]{Chandrasekhar2021}
\begin{botherref}
\oauthor{\bsnm{Chandrasekhar}, \binits{A.}},
\oauthor{\bsnm{Suresh}, \binits{K.}}:
Tounn: Topology optimization using neural networks.
Structural and Multidisciplinary Optimization
\textbf{63}
(2021)
\doiurl{10.1007/s00158-020-02748-4}
\end{botherref}
\endbibitem

\bibitem[\protect\citeauthoryear{Samaniego et~al.}{2020}]{samaniego2020energy}
\begin{barticle}
\bauthor{\bsnm{Samaniego}, \binits{E.}},
\bauthor{\bsnm{Anitescu}, \binits{C.}},
\bauthor{\bsnm{Goswami}, \binits{S.}},
\bauthor{\bsnm{Nguyen-Thanh}, \binits{V.M.}},
\bauthor{\bsnm{Guo}, \binits{H.}},
\bauthor{\bsnm{Hamdia}, \binits{K.}},
\bauthor{\bsnm{Zhuang}, \binits{X.}},
\bauthor{\bsnm{Rabczuk}, \binits{T.}}:
\batitle{An energy approach to the solution of partial differential equations in computational mechanics via machine learning: Concepts, implementation and applications}.
\bjtitle{Computer Methods in Applied Mechanics and Engineering}
\bvolume{362},
\bfpage{112790}
(\byear{2020})
\end{barticle}
\endbibitem

\bibitem[\protect\citeauthoryear{Zehnder et~al.}{2021}]{zehnder2021ntopo}
\begin{barticle}
\bauthor{\bsnm{Zehnder}, \binits{J.}},
\bauthor{\bsnm{Li}, \binits{Y.}},
\bauthor{\bsnm{Coros}, \binits{S.}},
\bauthor{\bsnm{Thomaszewski}, \binits{B.}}:
\batitle{Ntopo: Mesh-free topology optimization using implicit neural representations}.
\bjtitle{Advances in Neural Information Processing Systems}
\bvolume{34},
\bfpage{10368}--\blpage{10381}
(\byear{2021})
\end{barticle}
\endbibitem

\bibitem[\protect\citeauthoryear{Bends{\o}e and Kikuchi}{1988}]{bendsoe1988generating}
\begin{barticle}
\bauthor{\bsnm{Bends{\o}e}, \binits{M.P.}},
\bauthor{\bsnm{Kikuchi}, \binits{N.}}:
\batitle{Generating optimal topologies in structural design using a homogenization method}.
\bjtitle{Computer methods in applied mechanics and engineering}
\bvolume{71}(\bissue{2}),
\bfpage{197}--\blpage{224}
(\byear{1988})
\end{barticle}
\endbibitem

\bibitem[\protect\citeauthoryear{Allaire et~al.}{2002}]{allaire2002level}
\begin{barticle}
\bauthor{\bsnm{Allaire}, \binits{G.}},
\bauthor{\bsnm{Jouve}, \binits{F.}},
\bauthor{\bsnm{Toader}, \binits{A.-M.}}:
\batitle{A level-set method for shape optimization}.
\bjtitle{Comptes Rendus Mathematique}
\bvolume{334}(\bissue{12}),
\bfpage{1125}--\blpage{1130}
(\byear{2002})
\end{barticle}
\endbibitem

\bibitem[\protect\citeauthoryear{Wang et~al.}{2003}]{wang2003level}
\begin{barticle}
\bauthor{\bsnm{Wang}, \binits{M.Y.}},
\bauthor{\bsnm{Wang}, \binits{X.}},
\bauthor{\bsnm{Guo}, \binits{D.}}:
\batitle{A level set method for structural topology optimization}.
\bjtitle{Computer methods in applied mechanics and engineering}
\bvolume{192}(\bissue{1-2}),
\bfpage{227}--\blpage{246}
(\byear{2003})
\end{barticle}
\endbibitem

\bibitem[\protect\citeauthoryear{Woldseth et~al.}{2022}]{woldseth2022use}
\begin{barticle}
\bauthor{\bsnm{Woldseth}, \binits{R.V.}},
\bauthor{\bsnm{Aage}, \binits{N.}},
\bauthor{\bsnm{B{\ae}rentzen}, \binits{J.A.}},
\bauthor{\bsnm{Sigmund}, \binits{O.}}:
\batitle{On the use of artificial neural networks in topology optimisation}.
\bjtitle{Structural and Multidisciplinary Optimization}
\bvolume{65}(\bissue{10}),
\bfpage{294}
(\byear{2022})
\end{barticle}
\endbibitem

\bibitem[\protect\citeauthoryear{Raissi et~al.}{2019}]{raissi2019physics}
\begin{barticle}
\bauthor{\bsnm{Raissi}, \binits{M.}},
\bauthor{\bsnm{Perdikaris}, \binits{P.}},
\bauthor{\bsnm{Karniadakis}, \binits{G.E.}}:
\batitle{Physics-informed neural networks: A deep learning framework for solving forward and inverse problems involving nonlinear partial differential equations}.
\bjtitle{Journal of Computational physics}
\bvolume{378},
\bfpage{686}--\blpage{707}
(\byear{2019})
\end{barticle}
\endbibitem

\bibitem[\protect\citeauthoryear{Nguyen-Thanh et~al.}{2020}]{nguyen2020deep}
\begin{barticle}
\bauthor{\bsnm{Nguyen-Thanh}, \binits{V.M.}},
\bauthor{\bsnm{Zhuang}, \binits{X.}},
\bauthor{\bsnm{Rabczuk}, \binits{T.}}:
\batitle{A deep energy method for finite deformation hyperelasticity}.
\bjtitle{European Journal of Mechanics-A/Solids}
\bvolume{80},
\bfpage{103874}
(\byear{2020})
\end{barticle}
\endbibitem

\bibitem[\protect\citeauthoryear{Sitzmann et~al.}{2020}]{sitzmann2020implicit}
\begin{barticle}
\bauthor{\bsnm{Sitzmann}, \binits{V.}},
\bauthor{\bsnm{Martel}, \binits{J.}},
\bauthor{\bsnm{Bergman}, \binits{A.}},
\bauthor{\bsnm{Lindell}, \binits{D.}},
\bauthor{\bsnm{Wetzstein}, \binits{G.}}:
\batitle{Implicit neural representations with periodic activation functions}.
\bjtitle{Advances in Neural Information Processing Systems}
\bvolume{33},
\bfpage{7462}--\blpage{7473}
(\byear{2020})
\end{barticle}
\endbibitem

\bibitem[\protect\citeauthoryear{Tancik et~al.}{2020}]{tancik2020fourier}
\begin{barticle}
\bauthor{\bsnm{Tancik}, \binits{M.}},
\bauthor{\bsnm{Srinivasan}, \binits{P.}},
\bauthor{\bsnm{Mildenhall}, \binits{B.}},
\bauthor{\bsnm{Fridovich-Keil}, \binits{S.}},
\bauthor{\bsnm{Raghavan}, \binits{N.}},
\bauthor{\bsnm{Singhal}, \binits{U.}},
\bauthor{\bsnm{Ramamoorthi}, \binits{R.}},
\bauthor{\bsnm{Barron}, \binits{J.}},
\bauthor{\bsnm{Ng}, \binits{R.}}:
\batitle{Fourier features let networks learn high frequency functions in low dimensional domains}.
\bjtitle{Advances in Neural Information Processing Systems}
\bvolume{33},
\bfpage{7537}--\blpage{7547}
(\byear{2020})
\end{barticle}
\endbibitem

\bibitem[\protect\citeauthoryear{Banga et~al.}{2018}]{banga20183d}
\begin{botherref}
\oauthor{\bsnm{Banga}, \binits{S.}},
\oauthor{\bsnm{Gehani}, \binits{H.}},
\oauthor{\bsnm{Bhilare}, \binits{S.}},
\oauthor{\bsnm{Patel}, \binits{S.}},
\oauthor{\bsnm{Kara}, \binits{L.}}:
3d topology optimization using convolutional neural networks.
arXiv preprint arXiv:1808.07440
(2018)
\end{botherref}
\endbibitem

\bibitem[\protect\citeauthoryear{Yu et~al.}{2019}]{yu2019deep}
\begin{barticle}
\bauthor{\bsnm{Yu}, \binits{Y.}},
\bauthor{\bsnm{Hur}, \binits{T.}},
\bauthor{\bsnm{Jung}, \binits{J.}},
\bauthor{\bsnm{Jang}, \binits{I.G.}}:
\batitle{Deep learning for determining a near-optimal topological design without any iteration}.
\bjtitle{Structural and Multidisciplinary Optimization}
\bvolume{59}(\bissue{3}),
\bfpage{787}--\blpage{799}
(\byear{2019})
\end{barticle}
\endbibitem

\bibitem[\protect\citeauthoryear{Nakamura and Suzuki}{2020}]{nakamura2020deep}
\begin{botherref}
\oauthor{\bsnm{Nakamura}, \binits{K.}},
\oauthor{\bsnm{Suzuki}, \binits{Y.}}:
Deep learning-based topological optimization for representing a user-specified design area.
arXiv preprint arXiv:2004.05461
(2020)
\end{botherref}
\endbibitem

\bibitem[\protect\citeauthoryear{Nie et~al.}{2021}]{nie2021topologygan}
\begin{botherref}
\oauthor{\bsnm{Nie}, \binits{Z.}},
\oauthor{\bsnm{Lin}, \binits{T.}},
\oauthor{\bsnm{Jiang}, \binits{H.}},
\oauthor{\bsnm{Kara}, \binits{L.B.}}:
Topologygan: Topology optimization using generative adversarial networks based on physical fields over the initial domain.
Journal of Mechanical Design
\textbf{143}(3)
(2021)
\end{botherref}
\endbibitem

\bibitem[\protect\citeauthoryear{Behzadi and Ilie{\c{s}}}{2021}]{behzadi2021real}
\begin{barticle}
\bauthor{\bsnm{Behzadi}, \binits{M.M.}},
\bauthor{\bsnm{Ilie{\c{s}}}, \binits{H.T.}}:
\batitle{Real-time topology optimization in 3d via deep transfer learning}.
\bjtitle{Computer-Aided Design}
\bvolume{135},
\bfpage{103014}
(\byear{2021})
\end{barticle}
\endbibitem

\bibitem[\protect\citeauthoryear{Maz{\'e} and Ahmed}{2022}]{maze2022diffusion}
\begin{botherref}
\oauthor{\bsnm{Maz{\'e}}, \binits{F.}},
\oauthor{\bsnm{Ahmed}, \binits{F.}}:
Diffusion models beat gans on topology optimization
(2022)
\end{botherref}
\endbibitem

\bibitem[\protect\citeauthoryear{White et~al.}{2019}]{white2019multiscale}
\begin{barticle}
\bauthor{\bsnm{White}, \binits{D.A.}},
\bauthor{\bsnm{Arrighi}, \binits{W.J.}},
\bauthor{\bsnm{Kudo}, \binits{J.}},
\bauthor{\bsnm{Watts}, \binits{S.E.}}:
\batitle{Multiscale topology optimization using neural network surrogate models}.
\bjtitle{Computer Methods in Applied Mechanics and Engineering}
\bvolume{346},
\bfpage{1118}--\blpage{1135}
(\byear{2019})
\end{barticle}
\endbibitem

\bibitem[\protect\citeauthoryear{Chandrasekhar and Suresh}{2021a}]{Chandrasekhar2021Fourier}
\begin{botherref}
\oauthor{\bsnm{Chandrasekhar}, \binits{A.}},
\oauthor{\bsnm{Suresh}, \binits{K.}}:
Length scale control in topology optimization using fourier enhanced neural networks.
CoRR
\textbf{abs/2109.01861}
(2021)
{\href{https://arxiv.org/abs/2109.01861}{{2109.01861}}}
\end{botherref}
\endbibitem

\bibitem[\protect\citeauthoryear{Chandrasekhar and Suresh}{2021b}]{Chandrasekhar2021MM}
\begin{botherref}
\oauthor{\bsnm{Chandrasekhar}, \binits{A.}},
\oauthor{\bsnm{Suresh}, \binits{K.}}:
Multi-material topology optimization using neural networks.
CAD Computer Aided Design
\textbf{136}
(2021)
\doiurl{10.1016/j.cad.2021.103017}
\end{botherref}
\endbibitem

\bibitem[\protect\citeauthoryear{Deng and To}{2020}]{deng2020topology}
\begin{barticle}
\bauthor{\bsnm{Deng}, \binits{H.}},
\bauthor{\bsnm{To}, \binits{A.C.}}:
\batitle{Topology optimization based on deep representation learning (drl) for compliance and stress-constrained design}.
\bjtitle{Computational Mechanics}
\bvolume{66}(\bissue{2}),
\bfpage{449}--\blpage{469}
(\byear{2020})
\end{barticle}
\endbibitem

\bibitem[\protect\citeauthoryear{Zhang et~al.}{2023}]{zhang2023topology}
\begin{barticle}
\bauthor{\bsnm{Zhang}, \binits{Z.}},
\bauthor{\bsnm{Yao}, \binits{W.}},
\bauthor{\bsnm{Li}, \binits{Y.}},
\bauthor{\bsnm{Zhou}, \binits{W.}},
\bauthor{\bsnm{Chen}, \binits{X.}}:
\batitle{Topology optimization via implicit neural representations}.
\bjtitle{Computer Methods in Applied Mechanics and Engineering}
\bvolume{411},
\bfpage{116052}
(\byear{2023})
\end{barticle}
\endbibitem

\bibitem[\protect\citeauthoryear{Hoyer et~al.}{2019}]{hoyer2019neural}
\begin{botherref}
\oauthor{\bsnm{Hoyer}, \binits{S.}},
\oauthor{\bsnm{Sohl-Dickstein}, \binits{J.}},
\oauthor{\bsnm{Greydanus}, \binits{S.}}:
Neural reparameterization improves structural optimization.
arXiv preprint arXiv:1909.04240
(2019)
\end{botherref}
\endbibitem

\bibitem[\protect\citeauthoryear{Chen et~al.}{2023a}]{chen2023concurrent}
\begin{botherref}
\oauthor{\bsnm{Chen}, \binits{H.}},
\oauthor{\bsnm{Joglekar}, \binits{A.}},
\oauthor{\bsnm{Whitefoot}, \binits{K.S.}},
\oauthor{\bsnm{Kara}, \binits{L.B.}}:
Concurrent build direction, part segmentation, and topology optimization for additive manufacturing using neural networks.
Journal of Mechanical Design
(2023)
\end{botherref}
\endbibitem

\bibitem[\protect\citeauthoryear{Chen et~al.}{2023b}]{raychen2023idetc}
\begin{bchapter}
\bauthor{\bsnm{Chen}, \binits{H.}},
\bauthor{\bsnm{Joglekar}, \binits{A.}},
\bauthor{\bsnm{Kara}, \binits{L.B.}}:
\bctitle{Topology optimization using neural networks with conditioning field initialization for improved efficiency}.
In: \bbtitle{ASME 2023 International Design Engineering Technical Conferences and Computers and Information in Engineering and Conference}
(\byear{2023}).
\bcomment{American Society of Mechanical Engineers}
\end{bchapter}
\endbibitem

\bibitem[\protect\citeauthoryear{He et~al.}{2022}]{he2022deep}
\begin{botherref}
\oauthor{\bsnm{He}, \binits{J.}},
\oauthor{\bsnm{Chadha}, \binits{C.}},
\oauthor{\bsnm{Kushwaha}, \binits{S.}},
\oauthor{\bsnm{Koric}, \binits{S.}},
\oauthor{\bsnm{Abueidda}, \binits{D.}},
\oauthor{\bsnm{Jasiuk}, \binits{I.}}:
Deep energy method in topology optimization applications.
Acta Mechanica,
1--15
(2022)
\end{botherref}
\endbibitem

\bibitem[\protect\citeauthoryear{Jeong et~al.}{2023}]{jeong2023physics}
\begin{barticle}
\bauthor{\bsnm{Jeong}, \binits{H.}},
\bauthor{\bsnm{Bai}, \binits{J.}},
\bauthor{\bsnm{Batuwatta-Gamage}, \binits{C.}},
\bauthor{\bsnm{Rathnayaka}, \binits{C.}},
\bauthor{\bsnm{Zhou}, \binits{Y.}},
\bauthor{\bsnm{Gu}, \binits{Y.}}:
\batitle{A physics-informed neural network-based topology optimization (pinnto) framework for structural optimization}.
\bjtitle{Engineering Structures}
\bvolume{278},
\bfpage{115484}
(\byear{2023})
\end{barticle}
\endbibitem

\bibitem[\protect\citeauthoryear{Lu et~al.}{2021}]{lu2021physics}
\begin{barticle}
\bauthor{\bsnm{Lu}, \binits{L.}},
\bauthor{\bsnm{Pestourie}, \binits{R.}},
\bauthor{\bsnm{Yao}, \binits{W.}},
\bauthor{\bsnm{Wang}, \binits{Z.}},
\bauthor{\bsnm{Verdugo}, \binits{F.}},
\bauthor{\bsnm{Johnson}, \binits{S.G.}}:
\batitle{Physics-informed neural networks with hard constraints for inverse design}.
\bjtitle{SIAM Journal on Scientific Computing}
\bvolume{43}(\bissue{6}),
\bfpage{1105}--\blpage{1132}
(\byear{2021})
\end{barticle}
\endbibitem

\bibitem[\protect\citeauthoryear{Mai et~al.}{2023}]{mai2023physics}
\begin{botherref}
\oauthor{\bsnm{Mai}, \binits{H.T.}},
\oauthor{\bsnm{Mai}, \binits{D.D.}},
\oauthor{\bsnm{Kang}, \binits{J.}},
\oauthor{\bsnm{Lee}, \binits{J.}},
\oauthor{\bsnm{Lee}, \binits{J.}}:
Physics-informed neural energy-force network: a unified solver-free numerical simulation for structural optimization.
Engineering with Computers,
1--24
(2023)
\end{botherref}
\endbibitem

\bibitem[\protect\citeauthoryear{Kingma and Ba}{2014}]{kingma2014adam}
\begin{botherref}
\oauthor{\bsnm{Kingma}, \binits{D.P.}},
\oauthor{\bsnm{Ba}, \binits{J.}}:
Adam: A method for stochastic optimization.
arXiv preprint arXiv:1412.6980
(2014)
\end{botherref}
\endbibitem

\bibitem[\protect\citeauthoryear{Abadi et~al.}{2016}]{Abadi2016}
\begin{botherref}
\oauthor{\bsnm{Abadi}, \binits{M.}},
\oauthor{\bsnm{Agarwal}, \binits{A.}},
\oauthor{\bsnm{Barham}, \binits{P.}},
\oauthor{\bsnm{Brevdo}, \binits{E.}},
\oauthor{\bsnm{Chen}, \binits{Z.}},
\oauthor{\bsnm{Citro}, \binits{C.}},
\oauthor{\bsnm{Corrado}, \binits{G.S.}},
\oauthor{\bsnm{Davis}, \binits{A.}},
\oauthor{\bsnm{Dean}, \binits{J.}},
\oauthor{\bsnm{Devin}, \binits{M.}},
\oauthor{\bsnm{Ghemawat}, \binits{S.}},
\oauthor{\bsnm{Goodfellow}, \binits{I.}},
\oauthor{\bsnm{Harp}, \binits{A.}},
\oauthor{\bsnm{Irving}, \binits{G.}},
\oauthor{\bsnm{Isard}, \binits{M.}},
\oauthor{\bsnm{Jia}, \binits{Y.}},
\oauthor{\bsnm{Jozefowicz}, \binits{R.}},
\oauthor{\bsnm{Kaiser}, \binits{L.}},
\oauthor{\bsnm{Kudlur}, \binits{M.}},
\oauthor{\bsnm{Levenberg}, \binits{J.}},
\oauthor{\bsnm{Mane}, \binits{D.}},
\oauthor{\bsnm{Monga}, \binits{R.}},
\oauthor{\bsnm{Moore}, \binits{S.}},
\oauthor{\bsnm{Murray}, \binits{D.}},
\oauthor{\bsnm{Olah}, \binits{C.}},
\oauthor{\bsnm{Schuster}, \binits{M.}},
\oauthor{\bsnm{Shlens}, \binits{J.}},
\oauthor{\bsnm{Steiner}, \binits{B.}},
\oauthor{\bsnm{Sutskever}, \binits{I.}},
\oauthor{\bsnm{Talwar}, \binits{K.}},
\oauthor{\bsnm{Tucker}, \binits{P.}},
\oauthor{\bsnm{Vanhoucke}, \binits{V.}},
\oauthor{\bsnm{Vasudevan}, \binits{V.}},
\oauthor{\bsnm{Viegas}, \binits{F.}},
\oauthor{\bsnm{Vinyals}, \binits{O.}},
\oauthor{\bsnm{Warden}, \binits{P.}},
\oauthor{\bsnm{Wattenberg}, \binits{M.}},
\oauthor{\bsnm{Wicke}, \binits{M.}},
\oauthor{\bsnm{Yu}, \binits{Y.}},
\oauthor{\bsnm{Zheng}, \binits{X.}}:
Tensorflow: Large-scale machine learning on heterogeneous distributed systems
(2016)
\end{botherref}
\endbibitem

\bibitem[\protect\citeauthoryear{Andreassen et~al.}{2011}]{andreassen2011efficient}
\begin{barticle}
\bauthor{\bsnm{Andreassen}, \binits{E.}},
\bauthor{\bsnm{Clausen}, \binits{A.}},
\bauthor{\bsnm{Schevenels}, \binits{M.}},
\bauthor{\bsnm{Lazarov}, \binits{B.S.}},
\bauthor{\bsnm{Sigmund}, \binits{O.}}:
\batitle{Efficient topology optimization in matlab using 88 lines of code}.
\bjtitle{Structural and Multidisciplinary Optimization}
\bvolume{43},
\bfpage{1}--\blpage{16}
(\byear{2011})
\end{barticle}
\endbibitem

\bibitem[\protect\citeauthoryear{Liu and Tovar}{2014}]{liu2014efficient}
\begin{barticle}
\bauthor{\bsnm{Liu}, \binits{K.}},
\bauthor{\bsnm{Tovar}, \binits{A.}}:
\batitle{An efficient 3d topology optimization code written in matlab}.
\bjtitle{Structural and Multidisciplinary Optimization}
\bvolume{50},
\bfpage{1175}--\blpage{1196}
(\byear{2014})
\end{barticle}
\endbibitem

\end{thebibliography}


\end{document}